%% file: ms.tex
\RequirePackage{etoolbox}
\csdef{input@path}%
{%
 {sty/},
 {img/},
}
\documentclass{elsevierbook}

\input{CCPScolors}

\usepackage{tensor}
\usepackage[ruled,vlined]{algorithm2e}
\usepackage{booktabs} 
\usepackage{optidef}
\usepackage{arydshln}
\usepackage{ifthen}

\DeclareMathAlphabet{\mathantt}{OT1}{antt}{li}{it}
\DeclareMathAlphabet{\mathpzc}{OT1}{pzc}{m}{it}
\let\emptyset\varnothing

\usepackage{tikz}
\usetikzlibrary{shapes,arrows}
\usetikzlibrary{arrows,calc,positioning,bending}
\usetikzlibrary{decorations.markings}

\newcommand{\GE}{\mathpzc{G}}

\newcommand{\mapS}[3]{#1 \colon #2 \to #3 }

\newcommand{\fun}[1]{\mathantt{#1}}
\newcommand{\dfun}[1]{\dot{\mathantt{#1}}}

\newtheorem{remark}{Remark}[section]

\usepackage{fullpage}
\usepackage{optidef}
\usepackage{caption}
\usepackage{subcaption}

\usepackage[doi=false,isbn=false,url=false,giveninits=true, eprint=false, maxbibnames=99]{biblatex}
\AtEveryBibitem{\clearfield{month}}
\usepackage{breakcites} 
\usepackage{tabularx}
\bibliography{references}

\begin{document}

  \include{aboutauthors}

  
\renewcommand{\thefootnote}{\fnsymbol{footnote}}

\chapter*{Machine Learning for Process Control of (Bio)Chemical Processes}

Andreas Himmel$^\dagger$, Janine Matschek$^\ddagger$, Rudolph Kok$^\ddagger$, Bruno Morabito$^\ddagger$, Hoang Hai Nguyen$^\dagger$, Rolf Findeisen$^{\dagger,}$\footnote{rolf.findeisen@iat.tu-darmstadt.de}\\[1ex]
$^\dagger$Technical University of Darmstadt, Control and Cyber–Physical Systems Laboratory, Darmstadt, Germany
$^\ddagger$Otto von Guericke University Magdeburg, Systems Theory and Automatic Control Laboratory, Magdeburg, Germany
 \\[1ex]
 
\begin{abstract}
The control of manufacturing processes must satisfy high quality and efficiency requirements while meeting safety requirements. A broad spectrum of monitoring and control strategies, such as model- and optimization-based controllers, are utilized to address these issues. 
Driven by rising demand for flexible yet energy and resource-efficient operations existing approaches are challenged due to high uncertainties and changes. Machine learning algorithms are becoming increasingly important in tackling these challenges, especially due to the growing amount of available data. The ability for automatic adaptation and learning from human operators offer new opportunities to increase efficiency yet provide flexible operation. Combining machine learning algorithms with safe or robust controls offers novel reliable operation methods. This chapter highlights ways to fuse machine learning and control for the safe and improved operation of chemical and biochemical processes. 
We outline and summarize both - learning models for control and learning the control components. We offer a general overview, including a literature review, to provide a guideline for utilizing machine learning techniques in control structures.   
\end{abstract}

\begin{keywords}
\kwd{Machine Learning}
\kwd{Control}
\kwd{(Bio)Chemical Processes}
\kwd{Production Systems}
\kwd{Predictive Control}
\kwd{Neural Networks}
\kwd{Gaussian Processes}
\end{keywords}

\def\MLterm{ml-oracle}
\setcounter{chapter}{1}
\section{Introduction}\label{MLcontrolmanufacturing}
\input{Sections/1_Introduction}

\section{Process Setup}
\label{sec:Process_setup} 
\input{Sections/2_Structure_of_manufacturing_process}

\section{Generic Description of Machine Learning Approaches}
\label{sec:Generator}

\input{Sections/3_Generator}

\section{Plant Modeling via Machine Learning for Monitoring and Control}
\label{sec:ml4model}

\input{Sections/4_ML_within_the_model}

\section{Controller Design via Machine Learning}
\label{sec:ML_Controller}
\input{Sections/5_ML_for_the_control_law}

\section{Summary and Outlook}
\label{sec:Summary}
\input{Sections/6_Summary_outlook}

\newpage
\Backmatter
\printbibliography
\end{document}

%% file: CCPScolors.tex
\definecolor{TUDCCPSturquoise1}{RGB}{49,133,156}
\definecolor{TUDCCPSturquoise2}{RGB}{219,238,244}

\definecolor{TUDCCPSblue1}{RGB}{23,55,94}
\definecolor{TUDCCPSblue2}{RGB}{31,73,125}
\definecolor{CCPSblue1}{RGB}{37,64,97}
\definecolor{CCPSblue2}{RGB}{56,96,146}
\definecolor{CCPSblue3}{RGB}{220,230,242}

\definecolor{CCPSgreen1}{RGB}{67,80,58}
\definecolor{CCPSgreen2}{RGB}{170,171,128}
\definecolor{CCPSgreen3}{RGB}{215,228,189}

\definecolor{CCPSred1}{RGB}{99,37,35}
\definecolor{CCPSred2}{RGB}{149,55,53}
\definecolor{CCPSred3}{RGB}{242,220,219}

\definecolor{CCPSgray1}{RGB}{127,127,127}
\definecolor{CCPSgray2}{RGB}{191,191,191}
\definecolor{CCPSgray23}{RGB}{217,217,217}
\definecolor{CCPSgray3}{RGB}{242,242,242}

\definecolor{darktangerine}{rgb}{1.0, 0.66, 0.07}
\definecolor{CCPSorange1}{rgb}{0.72, 0.25, 0.05}
\definecolor{CCPSorange2}{rgb}{0.82, 0.41, 0.12}
\definecolor{CCPSorange3}{rgb}{1.0, 0.8, 0.6}
\definecolor{CCPSorange4}{RGB}{236,233,212}

\definecolor{MPIblue}{RGB}{51,165,195}
\definecolor{MPIgrey}{RGB}{135,135,141}
\definecolor{MPIgreen}{RGB}{0,118,117}
\definecolor{MPIred}{RGB}{120,0,75}
\definecolor{MPIredBound}{rgb}{0.75, 0, 0}

\definecolor{MPIsand}{RGB}{236,233,212}
\definecolor{MPItext}{RGB}{56,60,60}
\definecolor{MPIbluetext}{RGB}{23,161,193}

\definecolor{MPIbrown}{RGB}{139,69,19}
\definecolor{MPIorange}{RGB}{255,204,153}
\definecolor{MPIyellow}{RGB}{255,255,104}
\definecolor{MPIbazaar}{rgb}{0.6, 0.47, 0.48}
\definecolor{MPIburgund}{rgb}{0.5, 0.0, 0.13}
\definecolor{MPIcadet}{rgb}{0.33, 0.41, 0.47}
\definecolor{MPIcoal}{rgb}{0.21, 0.27, 0.31}
\definecolor{MPImagenta}{rgb}{0.55, 0.0, 0.55}
\definecolor{MPIindigo}{rgb}{0.29, 0.0, 0.51}
\definecolor{MPImauve}{rgb}{0.4, 0.19, 0.28}
\definecolor{MPIpine}{rgb}{0.25, 0.28, 0.2}

\definecolor{IFATblue1}{RGB}{37,64,97}
\definecolor{IFATblue2}{RGB}{56,96,146}
\definecolor{IFATblue3}{RGB}{220,230,242}

\definecolor{IFATgreen1}{RGB}{67,80,58}
\definecolor{IFATgreen2}{RGB}{170,171,128}
\definecolor{IFATgreen3}{RGB}{215,228,189}

\definecolor{IFATred1}{RGB}{99,37,35}
\definecolor{IFATred2}{RGB}{149,55,53}
\definecolor{IFATred3}{RGB}{242,220,219}

\definecolor{IFATgray1}{RGB}{127,127,127}
\definecolor{IFATgray2}{RGB}{191,191,191}
\definecolor{IFATgray3}{RGB}{242,242,242}

\definecolor{darktangerine}{rgb}{1.0, 0.66, 0.07}
\definecolor{IFATorange1}{rgb}{0.72, 0.25, 0.05}
\definecolor{IFATorange2}{rgb}{0.82, 0.41, 0.12}
\definecolor{IFATorange3}{rgb}{1.0, 0.8, 0.6}
\definecolor{IFATorange4}{RGB}{236,233,212}

%% file: Sections/1_Introduction.tex
The chemical and biochemical industry is one of the largest and central elements in the manufacturing sector, covering processes for the production of batteries, biochemicals, chemicals, fertilizers, food and beverages, petrochemicals, pharmaceuticals, etc. \cite{MdNor2020, Panerati2019, severson2019data, Zendehboudi2018}.
The control of the involved physical, chemical, and biochemical processes is a key component to obtain high quality products, ensure safety, meet regulation requirements, and improve economic and sustainability aspects. 
However, controlling these processes is challenging and demands reliable monitoring of the process operation \cite{severson2016perspectives}. 
Moreover, most biochemical and chemical industrial processes are complex, multi-scale, high-dimensional systems that exhibit large nonlinearity and are affected by uncertainties \cite{Himmelblau2008, Zendehboudi2018, Paulson2018}. The design of the control strategy must take all these aspects into account. One way to consider these effects is to build mathematical models that describe the plant dynamics and use model-based control \cite{lucia2016predictive}. Traditionally, these models are built by using first principle knowledge, such as thermodynamics, energy, and mass conservation laws. Usually a relative small amount of measurements are required to fit the parameters of these models. Nevertheless, building first-principle models is often challenging or very expensive.

Recently, a renewed interest in machine learning has spurred many applications of machine learning in control problems. This was fueled by the ever increasing amount of data and by breakthroughs like deep learning that solved the problem of features selection, allowing the applications of machine learning models for very complex problems. Machine learning is successfully applied in many fields such as computer science, robotics, and data analytics, while the application to (bio)chemical processes is more challenging. Some of the most important bottlenecks for machine learning applications in the (bio)chemical field are safety requirements and lack of data.
Often the measurement techniques are available only offline or are expensive (especially in the biochemical industry). Historical data collected do not represent the current plant operation due to ,e.g., design changes or modifications, equipment degradation due to fouling or aging, changes in the operating regimes, etc. 
Hence, experiments must be carried out for new data acquisition, which often requires stopping production. Since the dynamics of these processes are often slower than, e.g., the dynamics of robots, these experiments require a considerable amount of time, which can in turn cause loss of production time. These losses might not be affordable, especially for the production of bulk chemicals, where plants run with tight economic margins. 
Due to these reasons, great opportunities can arise for the control of manufacturing processes when machine learning algorithms are developed that can deal with limited amounts of data. For instance, physics informed learning can be used to incorporate prior knowledge to enhance extrapolation capabilities of models in regions with few or no data.

When data is available, it almost always contains some uncertainty, which has to be taken into account.
Additionally, malfunctioning of a plant, during the experiment or during production, can have severe repercussions in terms of safety and product quality. To gain the trust of the operators, machine learning models must be equipped with some degree of robustness and safety guarantees. 
Hence, machine learning algorithms should be developed to support the control of (bio)chemical plants such that control performance can be enhanced while guaranteeing satisfaction of all safety regulations.
Often, various control strategies are applied on multiple hierarchical levels, spanning from classical PID controllers to optimization-based control such as model predictive control. Each of these levels interacts with the process at a different level of granularity, hence deals with different timescales and parts of the plant. Machine learning applications should be able to integrate with such hierarchical structure.  

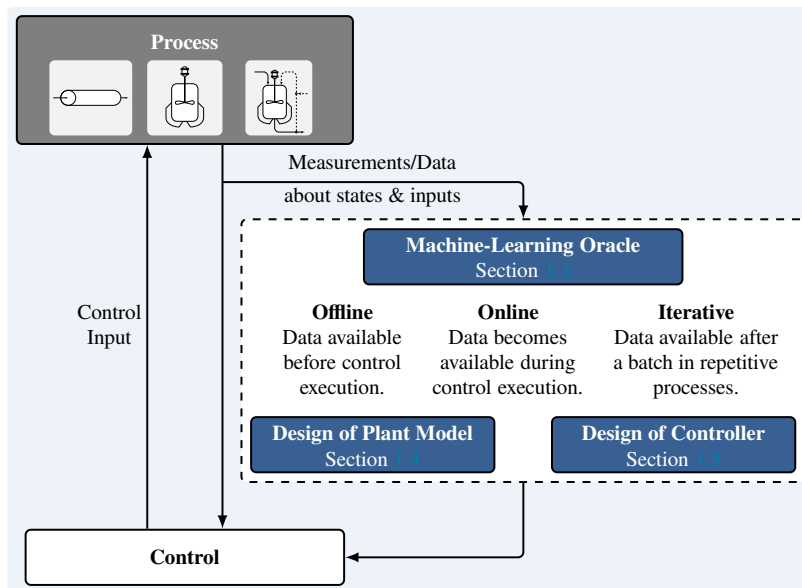
\begin{figure*}[htp]
    \centering
    \input{images_tot/Introduction}
    \caption{Outline of the paper structure how machine-learning can be used to support the control of (bio)chemical manufacturing processes.}
    \label{fig:Introduction}
\end{figure*}

In this paper, we want to outline how machine learning can address the mentioned challenges and support the control and controller design specifically for the (bio)chemical industry. Figure~\ref{fig:Introduction} depicts possibilities how machine learning can be used to do so. Basically, the machine-learning oracle (\MLterm) will use data from the plant to design or adapt models or parts of the control algorithm. The learned relations will be provided to the controller to  improve performance, robustness or safety. The generation of the machine-learning components can be performed offline (before control execution), online (during controller runtime) or iteratively (in between controller executions). 
In terms on machine learning algorithms, we focus especially on neural networks and Gaussian processes to represent commonly used machine learning techniques \cite{Ge2017a, MdNor2020, Panerati2019}. However, our aim is to outline a generic machine learning framework such that arbitrary machine learning methods can also be embedded in the presented setup. 
Mainly, the support of control technology by machine learning will be addressed in this paper from two perspectives (see also Figure~\ref{fig:Introduction}):
\begin{enumerate}
\item Machine learning can help to derive models of the (bio)chemical plant for analysis, simulation based controller design, as well as model-based estimation and control.
\item Machine learning can be used to learn or replace controllers and control laws directly from data.
\end{enumerate}
The first item is related to \textit{machine learning supported control} via data-based system identification, state and parameter estimation, and monitoring. 
Reviews on how machine learning can be used for modeling of (bio)chemical plants can be found in  \cite{Dobbelaere2021,Mowbray2021}. In contrast, we focus on modeling which is explicitly done to perform some notion of control or controller design, i.e. learning-supported simulation-, optimization-, and model-based controller design.
The second item relates to \textit{learning controllers}, i.e. controller parametrization, adaptive control, reinforcement learning, as well as imitation learning. 
Here, the controller is directly learned via machine learning techniques without a system identification step.

Our main contribution is to present a structured overview how different machine learning methods can be applied to support the control of (bio)chemical manufacturing processes.
To do so, a generic framework is presented which covers the most common tasks and algorithms of machine learning in control. Based on this framework, we present a vast literature review to describe the usage of machine learning in manufacturing processes.

The remainder of this paper is structured as follows: 
First, the general setup of chemical processes is presented in Section~\ref{sec:Process_setup} from which we  derive tasks for machine learning to support the control of these processes. 
These tasks span from modeling and estimation for monitoring and control, to a data-based controller design. 
Section~\ref{sec:Generator} presents the mathematical notation of the machine leaning components in an abstract/generic form and describes the training procedure of these components via a generic \MLterm. 
Based on this generic description, Sections~\ref{sec:ml4model} and \ref{sec:ML_Controller} give an overview of the applications of machine learning for the identified subtasks, see also Figure~\ref{fig:Introduction}. 
Section~\ref{sec:ml4model} focuses on model identification of (bio)process plants. It also describes how these models can be used for monitoring, simulation, and model based control. 
Section~\ref{sec:ML_Controller} outlines the direct design of controllers based on data. Here, the machine learning algorithms are directly used to represent the controller of the system instead of the dynamical system model. 
Section~\ref{sec:Summary} summarizes this overview article and presents future research directions. 
       
%% file: images_tot/Introduction.tex
\tikzset{every picture/.style={line width=0.75pt}} 

\begin{tikzpicture}[node distance=0.5cm]
    \node (T2) at (0,0) [rounded corners=2pt, draw=none, fill=IFATblue3!50]
        {\begin{tikzpicture}[>=latex]
            
            \node (P) at (2.5,-1.15) [rounded corners=2pt, draw, fill=CCPSgray1]
            {\begin{tikzpicture}[>=latex]
                \node (Name) at (0.0,0.0) [rounded corners=2pt, draw=none, fill=none]
                {\begin{minipage}[t][0.5cm]{4.0cm}\centering\footnotesize
                    \textbf{\textcolor{white}{Process}}
                \end{minipage}
                };
                
                \node (Pipe) [below of=Name, shift={(-1.25cm, -0.1cm)}, rounded corners=2pt, draw=none, inner sep=0pt, fill=CCPSgray3]
                {\resizebox{!}{1.0cm}
                {\begin{tikzpicture}[>=latex]
                    \path (-1.25,-1) rectangle ++(5.5,5.0);
                    
                    \draw [rounded corners=0.5cm, fill=white] (-0.5,1) rectangle ++(4.0,1.0);
                    \draw (0,1.5) circle (0.5cm);
                    \draw [->] (-1.0, 1.5) -- ++(1.0, 0);
                    \draw [->] (3.5, 1.5) -- ++(0.5, 0);
                \end{tikzpicture}}
                };
                
                \node (Batch) [below of=Name, yshift=-0.1cm, rounded corners=2pt, draw=none, inner sep=0pt, fill=CCPSgray3]
                {\resizebox{!}{1.0cm}
                {\begin{tikzpicture}[>=latex]
                    \path (-2.0,-1.75) rectangle ++(4.0,4.0);
                    
                    \draw[rounded corners =0pt,fill=CCPSgray3] (-0.7,0) -- ++(-0.3,-0.3) -- ++(0,-0.5) to[out=270,in=180] ++(0.7,-0.5) -- ++(0.2,0.3) -- ++(0.2,0.0) -- ++(0.2,-0.3) to[out=0,in=270] ++(0.7,0.5) -- ++(0,0.5) -- ++(-0.3,0.3) -- cycle; 
                    \draw [rounded corners=8pt, fill=white] (-0.7,-1) rectangle ++(1.4,2); 
                    \draw[rounded corners =0pt] (0,1.5) -- ++(0,-1.5) to[out=10,in=90] ++ (0.5,0.0) to[out=270,in=-10] ++ (-0.5,0.0) to[out=-10,in=90] ++ (-0.5,0.0) to[out=270,in=190] ++ (0.5,0.0) -- cycle; 
                    \draw [rounded corners=4pt, fill=white] (-0.15,1.4) rectangle ++(0.3,0.5); 
                    \draw [rounded corners=0pt, fill=CCPSgray1] (-0.2,1.5) rectangle ++(0.4,0.05); 
                    \draw [rounded corners=0pt, fill=CCPSgray1] (-0.2,1.8) rectangle ++(0.4,-0.05); 
                \end{tikzpicture}}
                };
                
                \node (CSTR) [below of=Name, shift={(1.25cm, -0.1cm)}, rounded corners=2pt, draw=none, inner sep=0pt, fill=CCPSgray3]
                {\resizebox{!}{1.0cm}
                {\begin{tikzpicture}[>=latex]
                    \path (-1.75,-2.0) rectangle ++(4.0,4.5);
                    
                    \draw[rounded corners =0pt,fill=CCPSgray3] (-0.7,0) -- ++(-0.3,-0.3) -- ++(0,-0.5) to[out=270,in=180] ++(0.7,-0.5) -- ++(0.2,0.3) -- ++(0.2,0.0) -- ++(0.2,-0.3) to[out=0,in=270] ++(0.7,0.5) -- ++(0,0.5) -- ++(-0.3,0.3) -- cycle; 
                    \draw [rounded corners=8pt, fill=white] (-0.7,-1) rectangle ++(1.4,2); 
                    \draw[rounded corners =0pt] (0,1.5) -- ++(0,-1.5) to[out=10,in=90] ++ (0.5,0.0) to[out=270,in=-10] ++ (-0.5,0.0) to[out=-10,in=90] ++ (-0.5,0.0) to[out=270,in=190] ++ (0.5,0.0) -- cycle; 
                    \draw [rounded corners=4pt, fill=white] (-0.15,1.4) rectangle ++(0.3,0.5); 
                    \draw [rounded corners=0pt, fill=CCPSgray1] (-0.2,1.5) rectangle ++(0.4,0.05); 
                    \draw [rounded corners=0pt, fill=CCPSgray1] (-0.2,1.8) rectangle ++(0.4,-0.05); 
                    
                    \draw[->,line width=1pt] (-1.2,1.8) -- ++(0.5,0) to [out=0,in=90] ++(0.3,-0.3)-- ++(0,-0.5); 
                    \draw[->,line width=1pt] (0,-1) -- ++(0.0,-0.5) to [out=270,in=180] ++(0.3,-0.3)-- ++(1.5,0); 
                    
                    \node (BN) at (1.3,-1.8) [circle,inner sep=1pt,fill=white,draw] {};
                    \draw[->,line width=1pt, dashed] (BN.north) -- (1.3,1.5) to [out=90,in=0] (1,1.8) -- ++(-0.3,0) to [out=180,in=90] ++(-0.3,-0.3)-- ++(0,-0.5); 
                    
                    \node (TN) at (1.3,0.5) [circle,inner sep=1pt,fill=white,draw] {};
                    \draw[->,line width=1pt, dashed] ([xshift=0.6cm]TN.east) -- (TN.east); 
                \end{tikzpicture}}
                };
            \end{tikzpicture}
            };

            \node (G) [below of=P, shift={(4.5cm, -3.10cm)}, rounded corners=2pt, draw, dashed, fill=white]
            {\begin{tikzpicture}[>=latex]
                \node (Name) at (0.0,0.0) [rounded corners=2pt, draw, solid, fill=CCPSblue2]
                {\begin{minipage}[c][0.5cm]{4.0cm}\centering\footnotesize
                    \textbf{\textcolor{white}{Machine-Learning Oracle}}\\
                    \textcolor{white}{Section~\ref{sec:Generator}}
                \end{minipage}
                };
                \node (Offline) [below of=Name, xshift=-2.4cm, rounded corners=2pt, draw=none, fill=none, anchor=north]
                {\begin{minipage}[t][1.2cm]{1.6cm}\centering\footnotesize
                    \textbf{Offline}\\
                    Data available before control execution.
                \end{minipage}
                };
                \node (Online) [below of=Name, xshift=-0.2cm, rounded corners=2pt, draw=none, fill=none, anchor=north]
                {\begin{minipage}[t][1.2cm]{2.1cm}\centering\footnotesize
                    \textbf{Online}\\
                    Data becomes available during control execution.
                \end{minipage}
                };
                \node (Iterative) [below of=Name, xshift=2.3cm, rounded corners=2pt, draw=none, fill=none, anchor=north]
                {\begin{minipage}[t][1.2cm]{2.2cm}\centering\footnotesize
                    \textbf{Iterative}\\
                    Data available after a batch in repetitive processes.
                \end{minipage}
                };
                \node (PM) [below of=Name, shift={(-2.0cm, -2.0cm)}, rounded corners=2pt, draw, solid, fill=CCPSblue2]
                {\begin{minipage}[t][0.5cm]{3.0cm}\centering\footnotesize
                    \textbf{\textcolor{white}{Design of Plant Model}}\\
                    \textcolor{white}{Section~\ref{sec:ml4model}}
                \end{minipage}
                };
                
                \node (C) [below of=Name, shift={(2.0cm, -2.0cm)}, rounded corners=2pt, draw, solid, fill=CCPSblue2]
                {\begin{minipage}[t][0.5cm]{3.0cm}\centering\footnotesize
                    \textbf{\textcolor{white}{Design of Controller}}\\
                    \textcolor{white}{Section~\ref{sec:ML_Controller}}
                \end{minipage}
                };
            \end{tikzpicture}
            };
            
            \node (C) [below of=P,  yshift=-5.85cm, rounded corners=2pt, draw, fill=white]
            {\begin{minipage}[c][0.5cm]{4.0cm}\centering\footnotesize
                \textbf{Control}
            \end{minipage}
            };
            
            \draw [->] ([xshift=-0.5cm] C.north) --
                node[anchor=south, shift={(-0.5cm, -0.25cm)}] {\begin{minipage}[c][0.5cm]{1.0cm}\centering\footnotesize
                    {Control\\Input}
                \end{minipage}}
            ([xshift=-0.5cm] P.south);
            \draw [->] ([xshift=0.5cm] P.south) -- ([xshift=0.5cm] C.north);
            
            \draw [->] ([shift={(0.5cm, -0.5cm)}] P.south) node[anchor=south, xshift=2cm, yshift=-0.45cm] 
            {\begin{minipage}{4cm}\footnotesize \centering
                Measurements/Data \\[3pt]
                about states \& inputs                
            \end{minipage}} -| (G.north);
            \draw [->] (G.south) |- (C.east);
        \end{tikzpicture}
    };
\end{tikzpicture}

%% file: Sections/2_Structure_of_manufacturing_process.tex
    
This section describes the structure of controlled (bio)chemical manufacturing processes.
We discuss the modeling of (bio)chemical plants, the continuous or repeated batch operation of chemical processes and hierarchical control structures.

\subsection{General System Structure of Manufacturing Processes}
Manufacturing processing plants uses various inputs (feeds) to produce a wide range of products.
These inputs can consist of material streams, i.e. chemical compounds, as well as various energy streams such as heat, electrical, or mechanical energy.
The plant is composed of several different processes and processing elements that are used to change the states of the process.
Reactors, separators, or mixers are mainly used to modify the composition of the material steams.
States such as the pressure, temperature, level, and flow rate are changed using compressors, heat exchangers, pumps, etc.
Storage processes, such as batteries, are also considered as a manufacturing process that takes electrical energy as input and store it as chemical energy. The stored chemical energy is later discharge as electrical energy that is produced by a chemical reaction.

The plant can be controlled through manipulation of the different processing elements.
Various control techniques on different control layers can be used to control the plant in the desired way. 
To do so, often a plant model is used. 
The plant model can be used for process monitoring, parameter and state estimation, and model-based controller design. 
In principle, each of these components of the process architecture, i.e. the process model and the controller, can be supported or replaced by machine learning algorithms to obtain an increased performance. 

\subsection{Plant Modeling} \label{subsec:Plant_Modelling} 
For analysis and control design, the plant dynamics are often described by complex mathematical models. 
These models allow for a better understanding of the process, enable improved plant design, and are crucial for process optimization and control \cite{Pantelides2013, Seborg2011, Subramanian2021}. 
In general, a mathematical model can be obtained from first principles based on fundamental engineering, physical, and chemical principles and/or experimental data \cite{Pantelides2013, Pirdashti2013}.

A general form to describe the plant model in an abstract, mathematical form is:
\begin{subequations}\label{eq:tot_equation}
\begin{alignat}{3}
\dfun{x}(t)  &=   f\bigl( \fun{x}(t), \fun{z}(t), \fun{u}(t), p \bigr) 
                + \mathantt{F}\bigl( \fun{x}(t), \fun{z}(t), p \bigr), \label{eq:ode}\\
    0           &= g\bigl( \fun{x}(t), \fun{z}(t), \fun{u}(t), p \bigr) 
                + \mathantt{G}\bigl( \fun{x}(t) , \fun{z}(t), p \bigr), \label{eq:ae}\\
    \fun{y}(t)  &= h \bigl( \fun{x}(t) , \fun{z}(t) \bigr),\label{eq:out}
\end{alignat}
\end{subequations}
where $\fun{x}(t)\in\mathcal{X}\subseteq\mathbb{R}^{n_\mathrm{x}}$ and $\fun{z}(t)\in\mathcal{Z}\subseteq\mathbb{R}^{n_\mathrm{z}}$ are dynamical and algebraic states, $\fun{u}(t)\in\mathcal{U}\subseteq\mathbb{R}^{n_\mathrm{u}}$ denotes the manipulating variables and $p\in\mathcal{P}\subseteq\mathbb{R}^{n_\mathrm{p}}$ represents model parameters.
The variable $\fun{y}(t)\in\mathcal{Y}\subseteq\mathbb{R}^{n_\mathrm{y}}$ denotes the outputs of the plant model where function $\mapS{h}{\mathcal{X} \times \mathcal{Z} }{\mathcal{Y}}$ describes which states can be measured.
The maps $\mapS{f}{\mathcal{X} \times \mathcal{Z} \times \mathcal{U} \times \mathcal{P}}{\mathbb{R}^{n_\mathrm{x}}}$ and $\mapS{\mathantt{F}}{\mathcal{X} \times \mathcal{Z} \times \mathcal{P}}{\mathbb{R}^{n_\mathrm{x}}}$ describe the change of the dynamical states represented by first principle equations and experimental data respectively. 
Similarly, the maps $\mapS{g}{\mathcal{X} \times \mathcal{Z} \times \mathcal{U} \times \mathcal{P}}{\mathbb{R}^{n_\mathrm{z}}}$ and $\mapS{\mathantt{G}}{\mathcal{X} \times \mathcal{Z} \times \mathcal{P}}{\mathbb{R}^{n_\mathrm{z}}}$ indicate the relationship between the algebraic states and other variables, where $g$ is generated by physical considerations and $\mathantt{G}$ by measurement data.
\begin{remark}
In many cases, \eqref{eq:ae} is given explicitly by 
\begin{alignat*}{3}
   \fun{z}(t) &= g_\mathrm{ex}\bigl( \fun{x}(t), \fun{u}(t), p\bigr)+\mathantt{G}_\mathrm{ex}\bigl( \fun{x}(t) ,p \bigr)
\end{alignat*}
which allows simplifying the model \eqref{eq:tot_equation} into an ordinary differential equation.
\end{remark}
In many applications it is favorable to describe the system behavior only at discrete times $t_k$.
This allows to examine the evolution of the states using a discrete-time model by defining a one-step integrator
\begin{alignat*}{3}
    I : \mathcal{X} \times \mathcal{U} \times \mathcal{P} \to \mathcal{X} \times \mathcal{Z},
\end{alignat*}
using the model equations~\eqref{eq:ode} and~\eqref{eq:ae}, 
a piece-wise constant input (i.e. $\fun{u}(t) =u_k, \, t\in\left[t_k,t_{k+1}\right)$) and the sampling time $T:=t_{k+1}-t_k$.
For simplicity, we assume an equidistant time grid, but all further explanations can also be extended to variable grid sizes. 
Using the one-step integrator, we can formulate a discrete-time model as follows 
\begin{subequations}\label{eq:tot_equation_D}
\begin{alignat}{3}
\bigl( \fun{x}(t_{k+1}), \fun{z}(t_{k+1}) \bigr) &= I\bigl( \fun{x}(t_k), \fun{u}(t_k), p \bigr), \label{eq:ode_D}\\
\fun{y}(t_k)  &= h \bigl( \fun{x}(t_k) , \fun{z}(t_k) \bigr),\label{eq:out_D}
\end{alignat}
\end{subequations}
Please note that the algebraic equation~\eqref{eq:ae} is directly included in the definition of the one-step integrator and must be integrated together with the ODE. 
Furthermore, all data-based functions (i.e. $\fun{F}$ and $\fun{G}$) have to be considered during the integration process and, thus, are part of $I$.
\begin{remark}
 In the following we will write $x_{k}:=\fun{x}(t_k)$, $z_{k}:=\fun{z}(t_k)$ and $u_{k}:=\fun{u}(t_k)$.
\end{remark}

\paragraph*{First principle models.}
Typical examples of first principle models ($f$ and/or $g$ in \eqref{eq:tot_equation}) are balance equations, thermodynamic relations, or kinetic information.
These equations are capable of describing the physical process with a high degree of spatial and temporal resolution \cite{Willard2003}.
Moreover, under suitable assumptions, they can extrapolate with a high degree of certainty and are applicable over a wide range of operating conditions while providing physical insight \cite{Bhutani2006, Seborg2011}.
However, some reasons stand in the way of the widespread use of first principle models in manufacturing processes.  
The performance of first principle models is limited by the modeling assumptions that are made \cite{Zendehboudi2018}.
Furthermore, deep process knowledge and understanding is required to develop first principle models capable of describing the underlying relations \cite{Bhutani2006}.
This renders the development of process models for large-scale systems with many integrated processes impractical \cite{Zhang2019}.
Also, the maintenance of complex first principle models is often not economically feasible and a trade-off between sustainability and rigor is required \cite{Pantelides2013, Zhang2019}.
First principle models can also suffer from computational complexity due to the high dimensionality and nonlinearities of chemical and biochemical process dynamics \cite{Basheer2000, Bhutani2006, Georgieva2007, Zendehboudi2018}.

\paragraph*{Data-based models.}
Data-based models (representing $\mathantt{F}$ and $\mathantt{G}$ in \eqref{eq:tot_equation}) can be used to address some of the mentioned drawbacks of first principle models.  
Machine learning algorithms can be used to describe relationships between input and output data based on general structures without knowing the process in advance \cite{Zendehboudi2018, Zhang2019}.
Thus, unknown phenomena and parameter effects can be approximated by machine learning techniques, reducing manual effort by experts, design time, and development costs \cite{Curteanu2006}.
Additionally, the flexibility in the production is increased since machine learning methods are adaptable and easy to retrain on new data when process changes occur \cite{Pirdashti2013}.
In particular, machine learning methods can also deal with noisy data, model uncertainties, large search spaces, nonlinearities, ill-defined subproblems, or tasks that require fast and real-time feasible solutions \cite{Bhutani2006, Himmelblau2008, Venkatasubramanian2019}.
Some techniques, such as Gaussian processes, can quantify approximation errors or model uncertainties allowing to obtain reliable models with corresponding error bounds \cite{Willard2003,Bradford2018,Mowbray2021}.
Furthermore, machine learning can help to develop process models for complex systems with reduced computational burden, enabling online applications \cite{MohdAli2015, Himmelblau2008, Zendehboudi2018}.
\paragraph*{Hybrid models. }
Machine learning models can be fused with first principle models, hence forming the so called hybrid models \cite{Oliveira2004,VonStosch2014}. 
Focus has recently been given on data-driven modeling which encloses or embeds certain properties of the underlying process principles to promote process understanding and reduce dataset size \cite{Shang2019,Karniadakis2021,Venkatasubramanian2019}.
Hybrid models have better extrapolation capabilities compared to pure data-based models as a result of incorporating first principles.  
They also provide physical significance to the modeling structure and parameters while bounding the model predictions to the physical constraints of the system \cite{Shang2019, Zendehboudi2018}.
Hybrid models have been used in reaction systems, thermodynamics, and fluid dynamics to estimate difficult to measure parameters and states using process data \cite{Shang2019, Zendehboudi2018}.
A recent review article \cite{Karniadakis2021} gives a more in-depth overview of this field.
In biotechnology, the interest on machine learning, particularly hybrid models, has increased in the last twenty years. 
A recent review \cite{Mowbray2021} covers machine learning applications in biotechnology of more than 200 papers published in the last 20 years. 
For example in metabolic engineering, in particular hybrid models have been used to model intracellular dynamics (e.g. metabolism, enzyme regulation), simplifying model identification, control, and optimization \cite{Teixeira2007, Oliveira2004}. 
For example, they have been used for modeling algae \cite{Garcia-Camacho2016,DelRio-Chanona2016}, bacteria \cite{VonStosch2014, Teixeira2006}, and for signaling pathways of animal cells \cite{Henriques2017,Lee2020,Lee2021}.

\subsection{Operating Modes of Plant} \label{subsec:Operating_modes_of_plant}

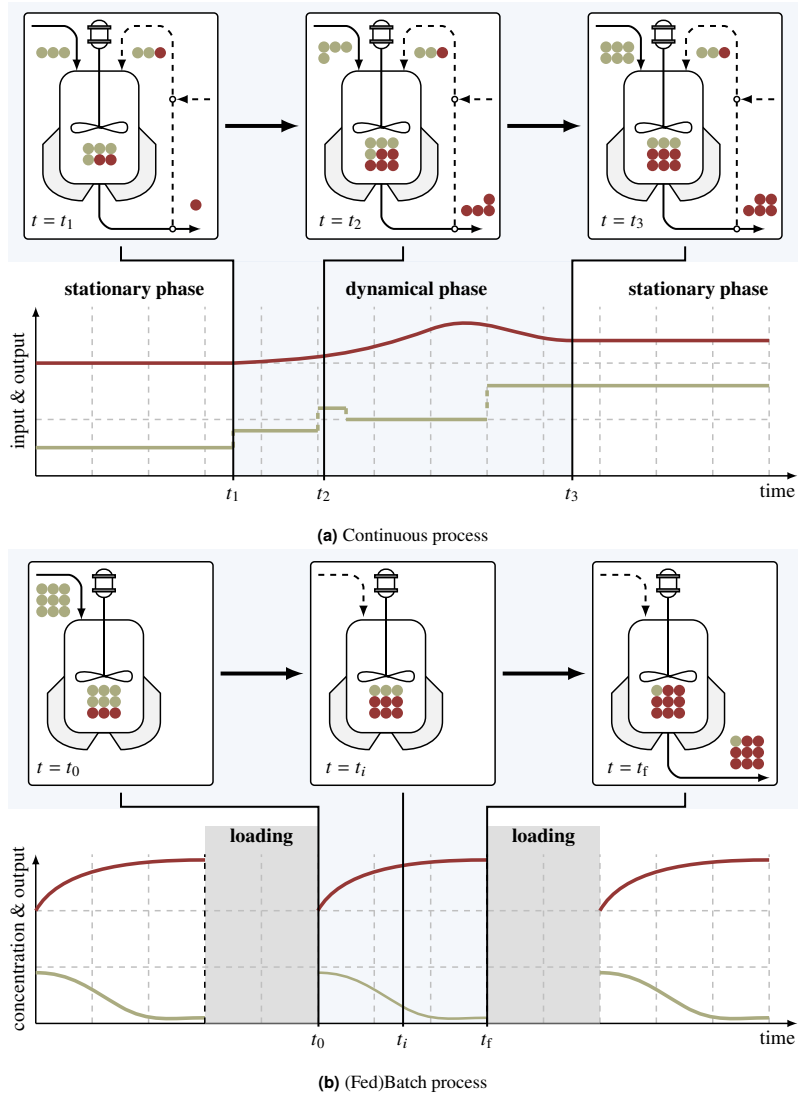
\begin{figure}[hbtp]
    \centering
    \begin{subfigure}[b]{\textwidth}
        \centering
        \scalebox{0.75}{\input{images_tot/process_structure_CSTR}}
        \caption{Continuous process}
        \label{fig:Continuous_process}
    \end{subfigure}

    \begin{subfigure}[b]{\textwidth}
        \centering
        \scalebox{0.75}{\input{images_tot/process_structure_batch}}
        \caption{(Fed)Batch process}
        \label{fig:Batch_process}
    \end{subfigure}
    \caption{Illustration of the two typical operating modes of manufacturing processes. The dashed lines represent an optional recycle (Figure a) or an optional feed (Figure b) to cover the fed-batch case.}
    \label{fig:BatchVContinuous}
\end{figure}

(Bio)chemical processes can be divided into continuous, batch, and fed-batch processes, cf. Figure~\ref{fig:BatchVContinuous}.
While dynamical system models of the form \eqref{eq:tot_equation} can be used to describe systems in these three operating modes in a similar way, the control goals for each of the cases are different. 
In the continuous case, the goal is to stabilize the system states at specified steady-state conditions or to follow a planned transition trajectory to reach the desired operating conditions, see also the time-graph in Figure~\ref{fig:Continuous_process} \cite{Mears2017}.
These desired setpoints and trajectories are dictated usually by a higher-level controller/optimizer, see also Section~\ref{subsec:controlarchitecture}.
The desired products are produced continuously without the interruption from loading stages in contrast to (fed)batch operations. 
Changes in the inputs (feed) to the process results in a dynamical or transitional phase in the output states (production).

However, the residence time of the reactors might not be large enough to convert all substances into the desired products. Hence, recycling is often utilized to improve the convergence rate, cf. the dashed lines in Figure~\ref{fig:Continuous_process}.
Recycling process streams can also come from different down stream units as indicated with the dashed arrow entering from the right in Figure~\ref{fig:Continuous_process}.
Plant models can be generated using the dynamical data, online or offline, obtained from the process during operation.

In contrast, batch and fed-batch processes usually start with the same initial conditions, terminate in a finite time, and are operated repeatedly, cf. Figure~\ref{fig:Batch_process} \cite{Mears2017, Seborg2011}.
Fed-batch processes have a continuous feed similar to the continuous operating mode, however, the product is only extracted after a finite time similar to the batch operating mode, cf. dashed line in Figure~\ref{fig:Batch_process}.
Fed-batch operation ensures a constant concentration of substrate, however, larger reactor volumes are required \cite{Seborg2011}.
Plant models can be generated using the dynamical data, online or offline, obtained from each batch.

Asymptotic properties such as stability are usually not of large importance for controller design for (fed)batch processes. Instead some of the states have to follow a specified time-varying profile or reach a given value at the end of the batch, i.e. in finite time.
Controlling and optimizing (fed)batch processes is especially hard due to the large state variation which requires the consideration of the nonlinear behavior of the controlled process. 
Linearizations of the nonlinear process behavior can often be used for control, estimation, and modeling of continuous processes that are operated in steady state. This is often not adequate for (fed)batch processes. 
However, the repetitive nature of (fed)batch processes can be utilized to tune and improve control performance over batches, cf. Figure~\ref{fig:Batch_process}. 
Iterative learning control (see Section \ref{subsec:Iterative_learning_control}) and run-to-run (or batch-to-batch) learning and optimization are suited to these operation modes but should be able to cope with changing batch initial conditions.

The different operating modes can also influence the way machine learning is integrated in the process. For example, for (fed)batch processes, some data might be available only at the end of the batch. In this case, the machine learning model can be trained once the batch is completed, resulting in a batch-to-batch learning approach, i.e., where the models are updated only in between batches. For continuous time processes, the models can instead be updated with different frequency depending on the particular case.
\subsection{Control Architecture}
\label{subsec:controlarchitecture}

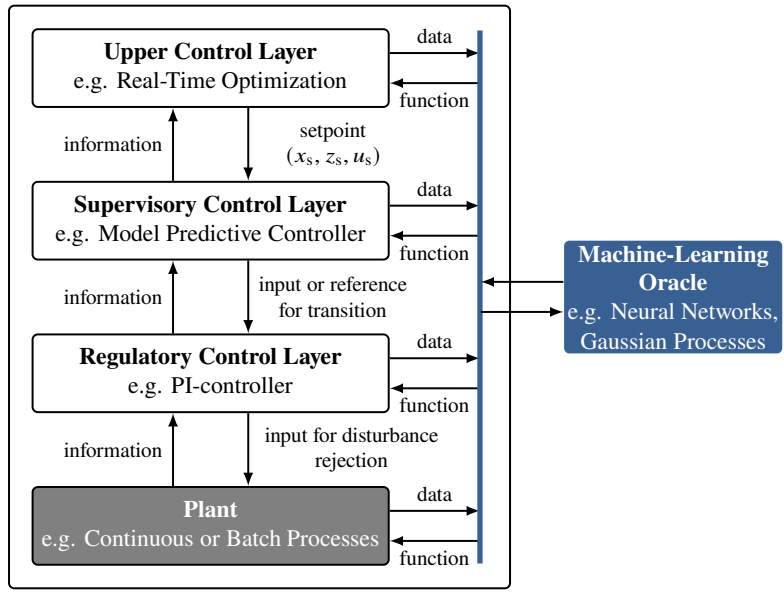
\begin{figure*}[tp]
    \centering
    \input{images_tot/manufacturing_structure}
    \caption{Manufacturing process information flow}
    \label{fig:manufacturing_structure}
\end{figure*}

Each manufacturing process requires a control architecture that implements the mentioned objectives for continuous, batch, or fed-batch processes. 
This control architecture is responsible for achieving specified operating points concerning the production quantity and quality. 
Another important aspect is the guarantee of disturbance rejection and noise compensation, thus ensuring the safe and stable operation of the plant. 
\\\\
Modern chemical manufacturing processes usually possess a hierarchical control structure to realize an economically optimal and safe operational process, cf. \cite{Darby2011,Qin2003,Skogestad2005}. 
Figure~\ref{fig:manufacturing_structure} shows a simplified representation of these different control layers.
The actions/responses of each layer is executed in different time horizons.
At the upper control level, optimal economical operation points are determined using the external specifications set at the planning and scheduling level for long periods and measurements coming from the plant. 
These economic decisions about setpoints of current production levels are executed at periods of hours or a few days, depending on the application.
For this purpose, a static optimization problem, using an economic objective function,
$J_\mathrm{eco}: \mathcal{X}\times\mathcal{Z}\times\mathcal{U} \times  \mathbb{R}^{n_\mathrm{eco}} \to \mathbb{R}$
(e.g., maximizing yield or profit, etc.), and the steady-state plant model, 
is computed.
The corresponding problem is referred to as real-time optimization (RTO).
For a given economic parameter, $p_\mathrm{eco} \in \mathbb{R}^{n_\mathrm{eco}}$ (e.g. describing the raw material or product prices), the general RTO control law reads 
\begin{alignat*}{3}
    \mathantt{K}_\mathrm{RTO} : \mathbb{R}^{n_\mathrm{eco}} \to \mathcal{X}\times\mathcal{Z}\times\mathcal{U}, \;
                                \left(p_\mathrm{eco}\right) \mapsto \mathantt{K}_\mathrm{RTO} \left(p_\mathrm{eco}\right) := \left(x_\mathrm{s},z_\mathrm{s},u_\mathrm{s} \right) 
\end{alignat*}
where the economically optimal setpoint $\left(x_\mathrm{s},z_\mathrm{s},u_\mathrm{s} \right)$ is obtained by solving
\begin{argmini!}
{\left(x,z,u\right)}{ J_\mathrm{eco} \left(x,z,u,p_\mathrm{eco} \right)}
{\label{eq:rto}}{\left(x_\mathrm{s},z_\mathrm{s},u_\mathrm{s} \right) :=}
\addConstraint{0}{=f\left(x,z,u,p\right)+\mathantt{F}\left(x,z,p\right)}
\addConstraint{0}{=g\left(x,z,u,p\right)+\mathantt{G}\left(x,z,p\right)}
\addConstraint{\left(x,z,u\right)}{\in\mathcal{X}\times\mathcal{Z}\times\mathcal{U}.}
\end{argmini!}
Subsequently, the setpoint is sent to the control layer below in order to implement it, see Figure~\ref{fig:manufacturing_structure}.
\\
The underlying supervisory control level implements this setpoint, usually utilizing a model predictive controller (MPC).
The main idea of an MPC is based on the repeated solution of an optimal control problem on a moving horizon while incorporating new measurements of the plant state.
In this way, the input of the system is continuously updated by comparing actual measurements with the predicted system behavior, considering various constraints. 
For general information on MPC, we refer to \cite{Mayne2000,Gruene2013,Rawlings2018}. 
Even though there are many specific ways to define an MPC, the following implementation is referred to as a general setpoint tracking problem.
The control law 
\begin{alignat*}{3}
    \mathantt{K}_\mathrm{MPC} : \mathcal{X} \to \mathcal{U}, \;
                                \left(x_k \right) \mapsto \mathantt{K}_\mathrm{MPC} \left( x_k \right) = u_k :=  \mathrm{pr}_1 \left(\mathbf{u}^\ast \right) 
\end{alignat*}
provides for a current state $x_k$ the input $u_k$ to be applied to the plant.
This input value is given by solving the nonlinear program 
\begin{argmini!}
{\substack{ u_0,\ldots,u_{N-1} }}
{ \sum_{l=1}^{N}        \left \| \hat{x}_l-x_\mathrm{s} \right\|_{Q} 
                    +  \left \| \hat{z}_l-z_\mathrm{s} \right\|_{R}
                    +   \left \| u_{l-1}-u_\mathrm{s} \right\|_{S}  }
{\label{eq:MPC}}{ \mathbf{u}^\ast :=}
\addConstraint{\left( \hat{x}_{l+1}, \hat{z}_{l+1}\right) }{=I \left(\hat{x}_l,u_l,p\right)}{\quad l=0,\ldots,N-1}
\addConstraint{0}{=\hat{x}_0-x_k \label{eq:MPCiniital}}{}
\addConstraint{\left(\hat{x}_l,\hat{z}_l,u_l\right)}{\in\mathcal{X}\times\mathcal{Z}\times\mathcal{U} \label{eq:MPCpath} }{\quad l=0,\ldots,N-1}
\addConstraint{\left(\hat{x}_N,\hat{z}_N\right)}{\in\mathcal{X}_\mathrm{f}\times\mathcal{Z}_\mathrm{f}. \label{eq:MPCterminal}}
\end{argmini!}
Here, the dynamical plant model is replaced by a discrete-time model, where the map
$I$ 
denotes a one-step integrator using~\eqref{eq:tot_equation} and the sampling time $T\in\mathbb{R}_+$ \cite{Biegler1998}.
In addition to the path constraints~\eqref{eq:MPCpath}, we can also define additional terminal constraints~\eqref{eq:MPCterminal} that must be satisfied for the states at the end of the prediction horizon $N$.
The idea of feedback comes from the fact that the current state value $x_k$ of the system is used as the initial value for prediction in \eqref{eq:MPCiniital}.
The current input value is then transmitted to the regulatory layer where plant information is processed over a short time horizon.
\\
The regulatory layer ensures directly on the plant side that the decisions of the supervisory layer are implemented within seconds on the process elements.
Furthermore, this layer supplies the rejection of high-frequency disturbances ensuring a stable process operation.
Typical control laws used here are PID controllers or linear quadratic regulators.

Each of these control layers can be supported by machine learning algorithms. The blue box on the right side of Figure~\ref{fig:manufacturing_structure} visualizes the machine learning generation block which supplies mappings or functions to each of the control layers while using data. These functions can represent system models, such as static or dynamic and linear or nonlinear state space models. 
Moreover, the \MLterm can also provide control laws to replace components of the hierarchical control structure, cf. \ref{sec:ML_Controller}. For example it can provide setpoints or feedback laws to replace the real-time optimization or a model predictive controller, respectively, using machine learning.

\subsection{Monitoring}
To enable closed-loop control, the current state information must be provided to the control units. 
It can be estimated or extracted from online measurements where noise should be filtered and outliers and faults should be detected.
Parameter estimation can be necessary to obtain reliable models for these tasks. 
A monitoring unit tackles these tasks. 
It often relies on using the model \eqref{eq:tot_equation} of the plant. However, also data-based algorithms can be used for monitoring purposes without the need of a plant model. In both cases, i.e., model-based and data-based monitoring, machine learning can be helpful to increase the monitoring or estimation quality. In the first case, machine learning can be used to model the system while the (partially) learned model can be used in model based estimation such as moving horizon estimators \cite{Santos2020}. In the second case, state, parameter and fault estimation can be performed directly on the data using machine learning algorithms. In both cases, the \MLterm block builds the basis to design and train the machine learning components, which is outlined in Subsections~\ref{subsec: generator} and \ref{subsec:MLapplication_monitoring}.

%% file: images_tot/process_structure_CSTR.tex
\tikzset{every picture/.style={line width=0.75pt}} 
\begin{tikzpicture}[>=latex]
    
    \draw[fill=CCPSblue3,draw=none,fill opacity=0.3] (-7,2.2) rectangle ++ (14,-4.6);
    
    \node (CSTR1) at (-5,0) [rounded corners=2pt,draw=none]
    {\begin{tikzpicture}[>=latex]
        \node (T1) at (0,0) [rounded corners=2pt,draw=,,fill=white]
        {\begin{tikzpicture}[>=latex]
            \draw[rounded corners =0pt,fill=CCPSgray3] (-0.7,0) -- ++(-0.3,-0.3) -- ++(0,-0.5) to[out=270,in=180] ++(0.7,-0.5) -- ++(0.2,0.3) -- ++(0.2,0.0) -- ++(0.2,-0.3) to[out=0,in=270] ++(0.7,0.5) -- ++(0,0.5) -- ++(-0.3,0.3) -- cycle; 
            \draw [rounded corners=8pt, fill=white] (-0.7,-1) rectangle ++(1.4,2); 
            \draw[rounded corners =0pt] (0,1.5) -- ++(0,-1.5) to[out=10,in=90] ++ (0.5,0.0) to[out=270,in=-10] ++ (-0.5,0.0) to[out=-10,in=90] ++ (-0.5,0.0) to[out=270,in=190] ++ (0.5,0.0) -- cycle; 
            \draw [rounded corners=4pt, fill=white] (-0.15,1.4) rectangle ++(0.3,0.5); 
            \draw [rounded corners=0pt, fill=CCPSgray1] (-0.2,1.5) rectangle ++(0.4,0.05); 
            \draw [rounded corners=0pt, fill=CCPSgray1] (-0.2,1.8) rectangle ++(0.4,-0.05); 
            
            \draw[->,line width=1pt] (-1.2,1.8) -- ++(0.5,0) to [out=0,in=90] ++(0.3,-0.3)-- ++(0,-0.5); 
            \draw[->,line width=1pt] (0,-1) -- ++(0.0,-0.5) to [out=270,in=180] ++(0.3,-0.3)-- ++(1.5,0); 
            
            \node (BN) at (1.3,-1.8) [circle,inner sep=1pt,fill=white,draw] {};
            \draw[->,line width=1pt, dashed] (BN.north) -- (1.3,1.5) to [out=90,in=0] (1,1.8) -- ++(-0.3,0) to [out=180,in=90] ++(-0.3,-0.3)-- ++(0,-0.5); 
            
            \node (TN) at (1.3,0.5) [circle,inner sep=1pt,fill=white,draw] {};
            \draw[->,line width=1pt, dashed] ([xshift=0.6cm]TN.east) -- (TN.east); 
        \end{tikzpicture}};
        
        \node (D1) at (-1.2,1.3) [rounded corners=0pt,draw=none,inner sep=0pt]
        {\begin{tikzpicture}[>=latex]
            \draw[fill=CCPSgreen2,draw=none] (0,0) circle (0.1cm);
            \draw[fill=CCPSgreen2,draw=none] (0.2,0) circle (0.1cm);
            \draw[fill=CCPSgreen2,draw=none] (0.4,0) circle (0.1cm);
            
        \end{tikzpicture}};
    
        \node (D2) at (-0.375,-0.5) [rounded corners=0pt,draw=none,inner sep=0pt]
        {\begin{tikzpicture}[>=latex]
            \draw[fill=CCPSgreen2,draw=none] (0,0) circle (0.1cm);
            \draw[fill=CCPSgreen2,draw=none] (0.2,0) circle (0.1cm);
            \draw[fill=CCPSgreen2,draw=none] (0.4,0) circle (0.1cm);
            
            \draw[fill=CCPSgreen2,draw=none] (0,-0.2) circle (0.1cm);
            \draw[fill=CCPSred2,draw=none] (0.2,-0.2) circle (0.1cm);
            \draw[fill=CCPSred2,draw=none] (0.4,-0.2) circle (0.1cm);
        \end{tikzpicture}};
    
        \node (D3) at (0.5,1.3) [rounded corners=0pt,draw=none,inner sep=0pt]
        {\begin{tikzpicture}[>=latex]
            \draw[fill=CCPSgreen2,draw=none] (0,-0.4) circle (0.1cm);
            \draw[fill=CCPSgreen2,draw=none] (0.2,-0.4) circle (0.1cm);
            \draw[fill=CCPSred2,draw=none] (0.4,-0.4) circle (0.1cm);
        \end{tikzpicture}};

        \node (D3) at (1.33,-1.4) [rounded corners=0pt,draw=none,inner sep=0pt]
        {\begin{tikzpicture}[>=latex]
            \draw[fill=CCPSred2,draw=none] (0,-0.4) circle (0.1cm);
        \end{tikzpicture}};
        
        \node at (-1.2,-1.7) [] {$t=t_1$};
    \end{tikzpicture}};
    \node (CSTR2) at (0,0) [rounded corners=2pt,draw=none]
    {\begin{tikzpicture}[>=latex]
        \node (T1) at (0,0) [rounded corners=2pt,draw=,fill=white]
        {\begin{tikzpicture}[>=latex]
            \draw[rounded corners =0pt,fill=CCPSgray3] 
            (-0.7,0) -- ++(-0.3,-0.3) -- ++(0,-0.5) to[out=270,in=180] ++(0.7,-0.5) -- ++(0.2,0.3) -- ++(0.2,0.0) -- ++(0.2,-0.3) to[out=0,in=270] ++(0.7,0.5) -- ++(0,0.5) -- ++(-0.3,0.3) -- cycle; 
            \draw [rounded corners=8pt, fill=white] (-0.7,-1) rectangle ++(1.4,2); 
            \draw[rounded corners =0pt] (0,1.5) -- ++(0,-1.5) to[out=10,in=90] ++ (0.5,0.0) to[out=270,in=-10] ++ (-0.5,0.0) to[out=-10,in=90] ++ (-0.5,0.0) to[out=270,in=190] ++ (0.5,0.0) -- cycle; 
            \draw [rounded corners=4pt, fill=white] (-0.15,1.4) rectangle ++(0.3,0.5); 
            \draw [rounded corners=0pt, fill=CCPSgray1] (-0.2,1.5) rectangle ++(0.4,0.05); 
            \draw [rounded corners=0pt, fill=CCPSgray1] (-0.2,1.8) rectangle ++(0.4,-0.05); 
            
            \draw[->,line width=1pt] (-1.2,1.8) -- ++(0.5,0) to [out=0,in=90] ++(0.3,-0.3)-- ++(0,-0.5); 
            \draw[->,line width=1pt] (0,-1) -- ++(0.0,-0.5) to [out=270,in=180] ++(0.3,-0.3)-- ++(1.5,0); 
            
            \node (BN) at (1.3,-1.8) [circle,inner sep=1pt,fill=white,draw] {};
            \draw[->,line width=1pt, dashed] (BN.north) -- (1.3,1.5) to [out=90,in=0] (1,1.8) -- ++(-0.3,0) to [out=180,in=90] ++(-0.3,-0.3)-- ++(0,-0.5); 
            
            \node (TN) at (1.3,0.5) [circle,inner sep=1pt,fill=white,draw] {};
            \draw[->,line width=1pt, dashed] ([xshift=0.6cm]TN.east) -- (TN.east); 
        \end{tikzpicture}};
    
        \node (D1) at (-1.2,1.3) [rounded corners=0pt,draw=none,inner sep=0pt]
        {\begin{tikzpicture}[>=latex]
            \draw[fill=CCPSgreen2,draw=none] (0,0) circle (0.1cm);
            \draw[fill=CCPSgreen2,draw=none] (0.2,0) circle (0.1cm);
            \draw[fill=CCPSgreen2,draw=none] (0.4,0) circle (0.1cm);
            
            \draw[fill=CCPSgreen2,draw=none] (0,-0.2) circle (0.1cm);
        \end{tikzpicture}};
    
        \node (D2) at (-0.375,-0.5) [rounded corners=0pt,draw=none,inner sep=0pt]
        {\begin{tikzpicture}[>=latex]
            \draw[fill=CCPSgreen2,draw=none] (0,0) circle (0.1cm);
            \draw[fill=CCPSgreen2,draw=none] (0.2,0) circle (0.1cm);
            \draw[fill=CCPSgreen2,draw=none] (0.4,0) circle (0.1cm);
            
            \draw[fill=CCPSgreen2,draw=none] (0,-0.2) circle (0.1cm);
            \draw[fill=CCPSred2,draw=none] (0.2,-0.2) circle (0.1cm);
            \draw[fill=CCPSred2,draw=none] (0.4,-0.2) circle (0.1cm);
            
            \draw[fill=CCPSred2,draw=none] (0,-0.4) circle (0.1cm);
            \draw[fill=CCPSred2,draw=none] (0.2,-0.4) circle (0.1cm);
            \draw[fill=CCPSred2,draw=none] (0.4,-0.4) circle (0.1cm);
        \end{tikzpicture}};
    
        \node (D3) at (0.5,1.3) [rounded corners=0pt,draw=none,inner sep=0pt]
        {\begin{tikzpicture}[>=latex]
            \draw[fill=CCPSgreen2,draw=none] (0,0) circle (0.1cm);
            \draw[fill=CCPSgreen2,draw=none] (0.2,0) circle (0.1cm);
            \draw[fill=CCPSred2,draw=none] (0.4,0) circle (0.1cm);
        \end{tikzpicture}};
        
        \node (D3) at (1.33,-1.4) [rounded corners=0pt,draw=none,inner sep=0pt]
        {\begin{tikzpicture}[>=latex]
            \draw[fill=CCPSred2,draw=none] (0.4,-0.2) circle (0.1cm);            
            \draw[fill=CCPSred2,draw=none] (0,-0.4) circle (0.1cm);
            \draw[fill=CCPSred2,draw=none] (0.2,-0.4) circle (0.1cm);
            \draw[fill=CCPSred2,draw=none] (0.4,-0.4) circle (0.1cm);
        \end{tikzpicture}};        
    
        \node at (-1.1,-1.7) [] {$t=t_2$};
    \end{tikzpicture}};
    
    \draw [->,line width=2pt] (CSTR1.east) -- (CSTR2.west);
    \node (CSTR3) at (5,0) [rounded corners=2pt,draw=none]
    {\begin{tikzpicture}[>=latex]
        \node (T1) at (0,0) [rounded corners=2pt,draw=,fill=white]
        {\begin{tikzpicture}[>=latex]
            \draw[rounded corners =0pt,fill=CCPSgray3] 
            (-0.7,0) -- ++(-0.3,-0.3) -- ++(0,-0.5) to[out=270,in=180] ++(0.7,-0.5) -- ++(0.2,0.3) -- ++(0.2,0.0) -- ++(0.2,-0.3) to[out=0,in=270] ++(0.7,0.5) -- ++(0,0.5) -- ++(-0.3,0.3) -- cycle; 
            \draw [rounded corners=8pt, fill=white] (-0.7,-1) rectangle ++(1.4,2); 
            \draw[rounded corners =0pt] (0,1.5) -- ++(0,-1.5) to[out=10,in=90] ++ (0.5,0.0) to[out=270,in=-10] ++ (-0.5,0.0) to[out=-10,in=90] ++ (-0.5,0.0) to[out=270,in=190] ++ (0.5,0.0) -- cycle; 
            \draw [rounded corners=4pt, fill=white] (-0.15,1.4) rectangle ++(0.3,0.5); 
            \draw [rounded corners=0pt, fill=CCPSgray1] (-0.2,1.5) rectangle ++(0.4,0.05); 
            \draw [rounded corners=0pt, fill=CCPSgray1] (-0.2,1.8) rectangle ++(0.4,-0.05); 
            
            \draw[->,line width=1pt] (-1.2,1.8) -- ++(0.5,0) to [out=0,in=90] ++(0.3,-0.3)-- ++(0,-0.5); 
            \draw[->,line width=1pt] (0,-1) -- ++(0.0,-0.5) to [out=270,in=180] ++(0.3,-0.3)-- ++(1.5,0); 
            
            \node (BN) at (1.3,-1.8) [circle,inner sep=1pt,fill=white,draw] {};
            \draw[->,line width=1pt, dashed] (BN.north) -- (1.3,1.5) to [out=90,in=0] (1,1.8) -- ++(-0.3,0) to [out=180,in=90] ++(-0.3,-0.3)-- ++(0,-0.5); 
            
            \node (TN) at (1.3,0.5) [circle,inner sep=1pt,fill=white,draw] {};
            \draw[->,line width=1pt, dashed] ([xshift=0.6cm]TN.east) -- (TN.east); 
        \end{tikzpicture}};
    
        \node (D1) at (-1.2,1.3) [rounded corners=0pt,draw=none,inner sep=0pt]
        {\begin{tikzpicture}[>=latex]
            \draw[fill=CCPSgreen2,draw=none] (0,0) circle (0.1cm);
            \draw[fill=CCPSgreen2,draw=none] (0.2,0) circle (0.1cm);
            \draw[fill=CCPSgreen2,draw=none] (0.4,0) circle (0.1cm);
            
            \draw[fill=CCPSgreen2,draw=none] (0,-0.2) circle (0.1cm);
            \draw[fill=CCPSgreen2,draw=none] (0.2,-0.2) circle (0.1cm);
            \draw[fill=CCPSgreen2,draw=none] (0.4,-0.2) circle (0.1cm);
        \end{tikzpicture}};
    
        \node (D2) at (-0.375,-0.5) [rounded corners=0pt,draw=none,inner sep=0pt]
        {\begin{tikzpicture}[>=latex]
            \draw[fill=CCPSgreen2,draw=none] (0,0) circle (0.1cm);
            \draw[fill=CCPSgreen2,draw=none] (0.2,0) circle (0.1cm);
            \draw[fill=CCPSgreen2,draw=none] (0.4,0) circle (0.1cm);
            
            \draw[fill=CCPSred2,draw=none] (0,-0.2) circle (0.1cm);
            \draw[fill=CCPSred2,draw=none] (0.2,-0.2) circle (0.1cm);
            \draw[fill=CCPSred2,draw=none] (0.4,-0.2) circle (0.1cm);
            
            \draw[fill=CCPSred2,draw=none] (0,-0.4) circle (0.1cm);
            \draw[fill=CCPSred2,draw=none] (0.2,-0.4) circle (0.1cm);
            \draw[fill=CCPSred2,draw=none] (0.4,-0.4) circle (0.1cm);
        \end{tikzpicture}};
    
        \node (D3) at (0.5,1.3) [rounded corners=0pt,draw=none,inner sep=0pt]
        {\begin{tikzpicture}[>=latex]
            \draw[fill=CCPSgreen2,draw=none] (0,0) circle (0.1cm);
            \draw[fill=CCPSgreen2,draw=none] (0.2,0) circle (0.1cm);
            \draw[fill=CCPSred2,draw=none] (0.4,0) circle (0.1cm);
        \end{tikzpicture}};

        \node (D3) at (1.33,-1.4) [rounded corners=0pt,draw=none,inner sep=0pt]
        {\begin{tikzpicture}[>=latex]
            \draw[fill=CCPSred2,draw=none] (0.2,-0.2) circle (0.1cm);
            \draw[fill=CCPSred2,draw=none] (0.4,-0.2) circle (0.1cm);            
            \draw[fill=CCPSred2,draw=none] (0,-0.4) circle (0.1cm);
            \draw[fill=CCPSred2,draw=none] (0.2,-0.4) circle (0.1cm);
            \draw[fill=CCPSred2,draw=none] (0.4,-0.4) circle (0.1cm);
        \end{tikzpicture}};     
        \node at (-1.1,-1.7) [] {$t=t_3$};
    \end{tikzpicture}};
    
    \draw [->,line width=2pt] (CSTR2.east) -- (CSTR3.west);
    
    \node (Graph) at (0,-4.55) [rounded corners=0pt,draw=none]
    {\begin{tikzpicture}[>=latex]
    
    \draw[fill=CCPSblue3,draw=none,fill opacity=0.3] (3.5,0) rectangle ++ (6.0,3.8);
    
    \draw[dashed,CCPSgray2] (0,1) -- ++ (13,0); 
    \draw[dashed,CCPSgray2] (0,2) -- ++ (13,0); 
    \draw[->] (0,0) -- node [rotate=90,above] {input \& output} ++ (0,3);
    \foreach \x in {1,...,13}
    \draw[dashed,CCPSgray2] (\x,0) -- ++ (0,3.0);
    \draw[->] (0,0) --  ++ (13.5,0) node [xshift=-10pt,below] {time};
    
    \def\Yoffset{-0.5}
    \draw [line width=1.8pt, CCPSgreen2]          (0.0,1.0+\Yoffset) -- ++(3.5,0.0);
    \draw [line width=1.8pt, CCPSgreen2,dashed]   (3.5,1.0+\Yoffset) -- ++(0.0,0.3);
    \draw [line width=1.8pt, CCPSgreen2]          (3.5,1.3+\Yoffset) -- ++(1.5,0.0);
    \draw [line width=1.8pt, CCPSgreen2,dashed]   (5.0,1.3+\Yoffset) -- ++(0.0,0.4);
    \draw [line width=1.8pt, CCPSgreen2]          (5.0,1.7+\Yoffset) -- ++(0.5,0.0);
    \draw [line width=1.8pt, CCPSgreen2,dashed]   (5.5,1.7+\Yoffset) -- ++(0.0,-0.2);
    \draw [line width=1.8pt, CCPSgreen2]          (5.5,1.5+\Yoffset) -- ++(2.5,0.0);
    \draw [line width=1.8pt, CCPSgreen2,dashed]   (8.0,1.7+\Yoffset) -- ++(0.0,0.4);
    \draw [line width=1.8pt, CCPSgreen2]          (8.0,2.1+\Yoffset) -- ++(5.0,0.0);
    
    
    
    \draw [line width=1.8pt, CCPSred2] (0,2) -- ++(3.5,0) to[out=3,in=200] ++ (3.5,0.6) to[out=200+180,in=180] ++ (2.5,-0.2) -- ++ (3.5,0.0);
    
    \node at (1.75,3.3) [] {\bfseries stationary phase}; 
    \node at (6.75,3.3) [] {\bfseries dynamical phase};
    \node at (11.75,3.3) [] {\bfseries stationary phase}; 
    \end{tikzpicture}};
    
    \draw[line width=1pt] (CSTR1.south) -- ++ (0.0,-0.25) -- ++ (1.99,0) -- ++(0,-3.9) node [below] {$t_1$};
    \draw[line width=1pt] (CSTR2.south) -- ++ (0.0,-0.25) -- ++ (-1.4, 0.0) -- ++(0,-3.9) node [below] {$t_2$};
    \draw[line width=1pt] (CSTR3.south) -- ++ (0.0,-0.25) -- ++ (-2.0,0) -- ++(0,-3.9) node [below] {$t_3$};  
    
\end{tikzpicture}

%% file: images_tot/process_structure_batch.tex
\tikzset{every picture/.style={line width=0.75pt}} 
\begin{tikzpicture}[>=latex]
    
    \draw[fill=CCPSblue3,draw=none,fill opacity=0.3] (-7,2.2) rectangle ++ (14,-4.6);
    \node (C1) at (-5,0) [rounded corners=2pt,draw=none]
    {\begin{tikzpicture}[>=latex]
        \node (T1) at (0,0) [rounded corners=2pt,draw=,fill=white]
        {\begin{tikzpicture}[>=latex]
            \draw[rounded corners =0pt,fill=CCPSgray3] (-0.7,0) -- ++(-0.3,-0.3) -- ++(0,-0.5) to[out=270,in=180] ++(0.7,-0.5) -- ++(0.2,0.3) -- ++(0.2,0.0) -- ++(0.2,-0.3) to[out=0,in=270] ++(0.7,0.5) -- ++(0,0.5) -- ++(-0.3,0.3) -- cycle; 
            \draw [rounded corners=8pt, fill=white] (-0.7,-1) rectangle ++(1.4,2); 
            \draw[rounded corners =0pt] (0,1.5) -- ++(0,-1.5) to[out=10,in=90] ++ (0.5,0.0) to[out=270,in=-10] ++ (-0.5,0.0) to[out=-10,in=90] ++ (-0.5,0.0) to[out=270,in=190] ++ (0.5,0.0) -- cycle; 
            \draw [rounded corners=4pt, fill=white] (-0.15,1.4) rectangle ++(0.3,0.5); 
            \draw [rounded corners=0pt, fill=CCPSgray1] (-0.2,1.5) rectangle ++(0.4,0.05); 
            \draw [rounded corners=0pt, fill=CCPSgray1] (-0.2,1.8) rectangle ++(0.4,-0.05); 
            
            \draw[->,line width=1pt] (-1.2,1.8) -- ++(0.5,0) to [out=0,in=90] ++(0.3,-0.3)-- ++(0,-0.5); 
            \draw[->,line width=1pt,white] (0,-1.1) -- ++(0.0,-0.4) to [out=270,in=180] ++(0.3,-0.3)-- ++(1.5,0); 
        \end{tikzpicture}};
    
        \node (D1) at (-1.2,1.3) [rounded corners=0pt,draw=none,inner sep=0pt]
        {\begin{tikzpicture}[>=latex]
            \draw[fill=CCPSgreen2,draw=none] (0,0) circle (0.1cm);
            \draw[fill=CCPSgreen2,draw=none] (0.2,0) circle (0.1cm);
            \draw[fill=CCPSgreen2,draw=none] (0.4,0) circle (0.1cm);
            
            \draw[fill=CCPSgreen2,draw=none] (0,-0.2) circle (0.1cm);
            \draw[fill=CCPSgreen2,draw=none] (0.2,-0.2) circle (0.1cm);
            \draw[fill=CCPSgreen2,draw=none] (0.4,-0.2) circle (0.1cm);
            
            \draw[fill=CCPSgreen2,draw=none] (0,-0.4) circle (0.1cm);
            \draw[fill=CCPSgreen2,draw=none] (0.2,-0.4) circle (0.1cm);
            \draw[fill=CCPSgreen2,draw=none] (0.4,-0.4) circle (0.1cm);
        
        \end{tikzpicture}};
    
        \node (D2) at (-0.3,-0.5) [rounded corners=0pt,draw=none,inner sep=0pt]
        {\begin{tikzpicture}[>=latex]
            \draw[fill=CCPSgreen2,draw=none] (0,0) circle (0.1cm);
            \draw[fill=CCPSgreen2,draw=none] (0.2,0) circle (0.1cm);
            \draw[fill=CCPSgreen2,draw=none] (0.4,0) circle (0.1cm);
            
            \draw[fill=CCPSgreen2,draw=none] (0,-0.2) circle (0.1cm);
            \draw[fill=CCPSgreen2,draw=none] (0.2,-0.2) circle (0.1cm);
            \draw[fill=CCPSgreen2,draw=none] (0.4,-0.2) circle (0.1cm);
            
            \draw[fill=CCPSred2,draw=none] (0,-0.4) circle (0.1cm);
            \draw[fill=CCPSred2,draw=none] (0.2,-0.4) circle (0.1cm);
            \draw[fill=CCPSred2,draw=none] (0.4,-0.4) circle (0.1cm);
        \end{tikzpicture}};
        
        \node at (-1.1,-1.7) [] {$t=t_0$};
    \end{tikzpicture}};
    \node (C2) at (0,0) [rounded corners=2pt,draw=none]
    {\begin{tikzpicture}[>=latex]
        \node (T1) at (0,0) [rounded corners=2pt,draw=,fill=white]
        {\begin{tikzpicture}[>=latex]
            \draw[rounded corners =0pt,fill=CCPSgray3] (-0.7,0) -- ++(-0.3,-0.3) -- ++(0,-0.5) to[out=270,in=180] ++(0.7,-0.5) -- ++(0.2,0.3) -- ++(0.2,0.0) -- ++(0.2,-0.3) to[out=0,in=270] ++(0.7,0.5) -- ++(0,0.5) -- ++(-0.3,0.3) -- cycle; 
            \draw [rounded corners=8pt, fill=white] (-0.7,-1) rectangle ++(1.4,2); 
            \draw[rounded corners =0pt] (0,1.5) -- ++(0,-1.5) to[out=10,in=90] ++ (0.5,0.0) to[out=270,in=-10] ++ (-0.5,0.0) to[out=-10,in=90] ++ (-0.5,0.0) to[out=270,in=190] ++ (0.5,0.0) -- cycle; 
            \draw [rounded corners=4pt, fill=white] (-0.15,1.4) rectangle ++(0.3,0.5); 
            \draw [rounded corners=0pt, fill=CCPSgray1] (-0.2,1.5) rectangle ++(0.4,0.05); 
            \draw [rounded corners=0pt, fill=CCPSgray1] (-0.2,1.8) rectangle ++(0.4,-0.05); 
            
            \draw[->,line width=1pt, dashed] (-1.2,1.8) -- ++(0.5,0) to [out=0,in=90] ++(0.3,-0.3)-- ++(0,-0.45); 
            \draw[->,line width=1pt,white] (0,-1.1) -- ++(0.0,-0.4) to [out=270,in=180] ++(0.3,-0.3)-- ++(1.5,0); 
        \end{tikzpicture}};

        \node (D2) at (-0.3,-0.5) [rounded corners=0pt,draw=none,inner sep=0pt]
        {\begin{tikzpicture}[>=latex]
            \draw[fill=CCPSgreen2,draw=none] (0,0) circle (0.1cm);
            \draw[fill=CCPSgreen2,draw=none] (0.2,0) circle (0.1cm);
            \draw[fill=CCPSgreen2,draw=none] (0.4,0) circle (0.1cm);
            
            \draw[fill=CCPSred2,draw=none] (0,-0.2) circle (0.1cm);
            \draw[fill=CCPSred2,draw=none] (0.2,-0.2) circle (0.1cm);
            \draw[fill=CCPSred2,draw=none] (0.4,-0.2) circle (0.1cm);
            
            \draw[fill=CCPSred2,draw=none] (0,-0.4) circle (0.1cm);
            \draw[fill=CCPSred2,draw=none] (0.2,-0.4) circle (0.1cm);
            \draw[fill=CCPSred2,draw=none] (0.4,-0.4) circle (0.1cm);
        \end{tikzpicture}};
        
        \node at (-1.0,-1.7) [] {$t=t_i$};
    
    \end{tikzpicture}};
    
    \draw [->,line width=2pt] (C1.east) -- (C2.west);
    \node (C3) at (5,0) [rounded corners=2pt,draw=none]
    {\begin{tikzpicture}[>=latex]
        \node (T1) at (0,0) [rounded corners=2pt,draw=,fill=white]
        {\begin{tikzpicture}[>=latex]
            \draw[rounded corners =0pt,fill=CCPSgray3] (-0.7,0) -- ++(-0.3,-0.3) -- ++(0,-0.5) to[out=270,in=180] ++(0.7,-0.5) -- ++(0.2,0.3) -- ++(0.2,0.0) -- ++(0.2,-0.3) to[out=0,in=270] ++(0.7,0.5) -- ++(0,0.5) -- ++(-0.3,0.3) -- cycle; 
            \draw [rounded corners=8pt, fill=white] (-0.7,-1) rectangle ++(1.4,2); 
            \draw[rounded corners =0pt] (0,1.5) -- ++(0,-1.5) to[out=10,in=90] ++ (0.5,0.0) to[out=270,in=-10] ++ (-0.5,0.0) to[out=-10,in=90] ++ (-0.5,0.0) to[out=270,in=190] ++ (0.5,0.0) -- cycle; 
            \draw [rounded corners=4pt, fill=white] (-0.15,1.4) rectangle ++(0.3,0.5); 
            \draw [rounded corners=0pt, fill=CCPSgray1] (-0.2,1.5) rectangle ++(0.4,0.05); 
            \draw [rounded corners=0pt, fill=CCPSgray1] (-0.2,1.8) rectangle ++(0.4,-0.05); 
            
            \draw[->,line width=1pt, dashed] (-1.2,1.8) -- ++(0.5,0) to [out=0,in=90] ++(0.3,-0.3)-- ++(0,-0.45); 
            \draw[->,line width=1pt] (0,-1) -- ++(0.0,-0.5) to [out=270,in=180] ++(0.3,-0.3)-- ++(1.5,0); 
        \end{tikzpicture}};

        \node (D2) at (-0.3,-0.5) [rounded corners=0pt,draw=none,inner sep=0pt]
        {\begin{tikzpicture}[>=latex]
            \draw[fill=CCPSgreen2,draw=none] (0,0) circle (0.1cm);
            \draw[fill=CCPSred2,draw=none] (0.2,0) circle (0.1cm);
            \draw[fill=CCPSred2,draw=none] (0.4,0) circle (0.1cm);
            
            \draw[fill=CCPSred2,draw=none] (0,-0.2) circle (0.1cm);
            \draw[fill=CCPSred2,draw=none] (0.2,-0.2) circle (0.1cm);
            \draw[fill=CCPSred2,draw=none] (0.4,-0.2) circle (0.1cm);
            
            \draw[fill=CCPSred2,draw=none] (0,-0.4) circle (0.1cm);
            \draw[fill=CCPSred2,draw=none] (0.2,-0.4) circle (0.1cm);
            \draw[fill=CCPSred2,draw=none] (0.4,-0.4) circle (0.1cm);
        \end{tikzpicture}};
        
        \node (D3) at (1.1,-1.4) [rounded corners=0pt,draw=none,inner sep=0pt]
        {\begin{tikzpicture}[>=latex]
            \draw[fill=CCPSgreen2,draw=none] (0,0) circle (0.1cm);
            \draw[fill=CCPSred2,draw=none] (0.2,0) circle (0.1cm);
            \draw[fill=CCPSred2,draw=none] (0.4,0) circle (0.1cm);
            
            \draw[fill=CCPSred2,draw=none] (0,-0.2) circle (0.1cm);
            \draw[fill=CCPSred2,draw=none] (0.2,-0.2) circle (0.1cm);
            \draw[fill=CCPSred2,draw=none] (0.4,-0.2) circle (0.1cm);
            
            \draw[fill=CCPSred2,draw=none] (0,-0.4) circle (0.1cm);
            \draw[fill=CCPSred2,draw=none] (0.2,-0.4) circle (0.1cm);
            \draw[fill=CCPSred2,draw=none] (0.4,-0.4) circle (0.1cm);
        \end{tikzpicture}};          
        
        \node at (-1.0,-1.7) [] {$t=t_\mathrm{f}$};
    \end{tikzpicture}};
    
    \draw [->,line width=2pt] (C2.east) -- (C3.west);
    
    \node (Graph) at (0,-4.55) [rounded corners=0pt,draw=none]
    {\begin{tikzpicture}[>=latex]
    
    \draw[fill=CCPSblue3,draw=none,fill opacity=0.3] (5.0,0) rectangle ++(3.0,3.8);

    \draw[dashed,CCPSgray2] (0,1) -- ++ (13,0); 
    \draw[dashed,CCPSgray2] (0,2) -- ++ (13,0); 
    \draw[->] (0,0) -- node [rotate=90,above] {concentration \& output} ++ (0,3); 
    
    \foreach \x in {1,...,13}
    \draw[dashed,CCPSgray2] (\x,0) -- ++ (0,3); 
    \draw[->] (0,0) --  ++ (13.5,0) node [xshift=-10pt,below] {time}; 
    
    \draw [line width=1.8pt, CCPSgreen2] (0.0, 0.9) to[out=0,in=150] ++(1.5, -0.6) to[out=-30,in=180] ++(1.5, -0.2);
    \draw [line width=1.8pt, CCPSred2]   (0.0,2.0) .. controls (0.4,2.7) and (1.5,2.9) .. (3.0,2.9);
    \draw [line width=1.0pt, black,dashed](3.0, 0.0) -- (3.0, 3.0);
    
    \draw[fill=CCPSgray2,draw=none,fill opacity=0.5] (3.0,0) rectangle (5.0,3.5);
    \node at (4.0,3.3) [] {\bfseries loading}; 
    
    \draw [line width=1.3pt, CCPSgreen2] (5.0, 0.9) to[out=0,in=150] ++(1.5, -0.6) to[out=-30,in=180] ++(1.5, -0.2);
    \draw [line width=1.8pt, CCPSred2]   (5.0,2.0) .. controls (5.4,2.7) and (6.5,2.9) .. (8.0,2.9);
    \draw [line width=1.0pt, black,dashed]           (8.0, 0.0) -- (8.0, 3.0);
    
    \draw[fill=CCPSgray2,draw=none,fill opacity=0.5] (8.0,0) rectangle (10.0,3.5);
    \node at (9.0,3.3) [] {\bfseries loading}; 
    
    \draw [line width=1.8pt, CCPSgreen2] (10.0, 0.9) to[out=0,in=150] ++(1.5, -0.6) to[out=-30,in=180] ++(1.5, -0.2);
    \draw [line width=1.8pt, CCPSred2]   (10.0,2.0) .. controls (10.4,2.7) and (11.5,2.9) .. (13.0,2.9);
    
    \end{tikzpicture}};
    
    \draw[line width=1pt] (C1.south) -- ++ (0.0,-0.25) -- ++ (3.50,0) -- ++ (0,-3.9) node [below] {$t_0$};
    \draw[line width=1pt] (C2.south) -- ++ (0.0,-0.25) -- ++ (0,-3.9) node [below] {$t_i$};
    \draw[line width=1pt] (C3.south) -- ++ (0.0,-0.25) -- ++ (-3.51,0) -- ++ (0,-3.9) node [below] {$t_\mathrm{f}$}; 
    
\end{tikzpicture}

%% file: images_tot/manufacturing_structure.tex
\tikzset{every picture/.style={line width=0.75pt}} 

\begin{tikzpicture}[>=latex]

\node (T1) at (0,0) [rounded corners=2pt,draw=none]
{\begin{tikzpicture}[>=latex]

    \node (A0) at (0,0) [rounded corners=2pt,draw,fill=black!50]
    {\begin{minipage}[c][0.8cm]{4.5cm}\centering\small\color{white}
    \textbf{Plant}\\
    e.g. Continuous or Batch Processes
    \end{minipage}};
    
    \node (A1) at ([yshift=1.5cm]A0.north) [rounded corners=2pt,draw,fill=white]
    {\begin{minipage}[c][0.8cm]{4.5cm}\centering\small
    \textbf{Regulatory Control Layer}\\
    e.g. PI-controller
    \end{minipage}};
    
    \node (A2) at ([yshift=1.5cm]A1.north) [rounded corners=2pt,draw,fill=white]
    {\begin{minipage}[c][0.8cm]{4.5cm}\centering\small
    \textbf{Supervisory Control Layer}\\
    e.g. Model Predictive Controller
    \end{minipage}};
    
    \node (A3) at ([yshift=1.5cm]A2.north) [rounded corners=2pt,draw,fill=white]
    {\begin{minipage}[c][0.8cm]{4.5cm}\centering\small
    \textbf{Upper Control Layer}\\
    e.g. Real-Time Optimization
    \end{minipage}};
    \draw [->] ([xshift=-0.5cm]A0.north) --
    node[left] {\footnotesize information}
    ([xshift=-0.5cm]A1.south);
    \draw [->] ([xshift=-0.5cm]A1.north) -- 
    node[left] {\footnotesize information}
    ([xshift=-0.5cm]A2.south);
    \draw [->] ([xshift=-0.5cm]A2.north) -- 
    node[left] {\footnotesize information}
    ([xshift=-0.5cm]A3.south);
    \draw [<-] ([xshift=+0.5cm]A0.north) -- 
    node[right] {
    \begin{minipage}{2.5cm}\footnotesize\centering 
    input for disturbance rejection
    \end{minipage}} ([xshift=+0.5cm]A1.south);
    \draw [<-] ([xshift=+0.5cm]A1.north) -- 
    node[right] {\begin{minipage}{2cm}\footnotesize\centering  
    input or reference for transition
    \end{minipage}}
    ([xshift=+0.5cm]A2.south);
    \draw [<-] ([xshift=+0.5cm]A2.north) -- 
    node[right] {\begin{minipage}{2cm}\footnotesize\centering
    setpoint\\ $\left(x_\mathrm{s},z_\mathrm{s},u_\mathrm{s} \right)$
    \end{minipage}}
    ([xshift=+0.5cm]A3.south);
    \draw [->] ([yshift=0.2cm]A0.east) -- 
    node[above] {\footnotesize data} 
    ([xshift=+1.2cm,yshift=0.2cm]A0.east);
    \draw [<-] ([yshift=-0.2cm]A0.east) -- 
    node[below] {\footnotesize function} 
    ([xshift=+1.2cm,yshift=-0.2cm]A0.east);
    \draw [->] ([yshift=0.2cm]A1.east) -- 
    node[above] {\footnotesize data} 
    ([xshift=+1.2cm,yshift=0.2cm]A1.east);
    \draw [<-] ([yshift=-0.2cm]A1.east) -- 
    node[below] {\footnotesize function} 
    ([xshift=+1.2cm,yshift=-0.2cm]A1.east);
    \draw [->] ([yshift=0.2cm]A2.east) -- 
    node[above] {\footnotesize data} 
    ([xshift=+1.2cm,yshift=0.2cm]A2.east);
    \draw [<-] ([yshift=-0.2cm]A2.east) -- 
    node[below] {\footnotesize function} 
    ([xshift=+1.2cm,yshift=-0.2cm]A2.east);
    \draw [->] ([yshift=0.2cm]A3.east) -- 
    node[above] {\footnotesize data} 
    ([xshift=+1.2cm,yshift=0.2cm]A3.east);
    \draw [<-] ([yshift=-0.2cm]A3.east) -- 
    node[below] {\footnotesize function} 
    ([xshift=+1.2cm,yshift=-0.2cm]A3.east);
    \draw [line width=2pt,CCPSblue2] ([xshift=+1.2cm,yshift=-0.5cm]A0.east) -- ([xshift=+1.2cm,yshift=0.5cm]A3.east);
    
    \draw ([xshift=-0.3cm,yshift=-0.3cm]A0.south west) rectangle 
    ([xshift=+1.6cm,yshift=+0.3cm]A3.north east);
\end{tikzpicture}};

\node (T2) at (5.5,0) [rounded corners=2pt,draw=none, inner sep=0pt]
{\begin{tikzpicture}[>=latex]

    \node (A0) at ([yshift=1cm]A2.north) [rounded corners=2pt,draw=none,fill=CCPSblue2]
    {\begin{minipage}[c][1.5cm]{2.9cm}\centering\small\color{white}
    \textbf{Machine-Learning Oracle}\\
    e.g. Neural Networks, Gaussian Processes
    \end{minipage}};

\end{tikzpicture}};

\draw [->] ([yshift=0.2cm]T2.west) -- ([xshift=-1.1cm,yshift=0.2cm]T2.west);
\draw [<-] ([yshift=-0.2cm]T2.west) -- ([xshift=-1.1cm,yshift=-0.2cm]T2.west);

\end{tikzpicture}

%% file: Sections/3_Generator.tex
This section introduces a  machine-learning oracle to support the control of chemical processes. 
Here, the \MLterm can provide system models to different control layers, or replace some of the controller parts. 
 The intention of this section is to introduce an abstract formulation of machine learning algorithms to obtain a generic framework how machine learning can be embedded into the control task. 
We introduce the general principles of modern machine learning algorithms such as feed-forward and recurrent neural networks, Gaussian processes, and physics informed learning and outline how these can be embedded in the generic framework.

\subsection{The machine-learning Oracle}\label{subsec: generator}
The task of the machine-learning oracle is to generate surrogate models of the plant, the  controller, or parts of them. 
Figure~\ref{fig:manufacturing_structure} illustrates the relationship between this entity and the manufacturing process.
In general, a \MLterm $\GE$ is an abstract map
\begin{alignat}{2}\label{eq:generator}
   \mathcal{D}_\mathrm{tot} \overset{\GE}{\longmapsto}  \mathantt{S}  ,
\end{alignat}
that constructs, from a data set $D_\mathrm{tot}$, a continuous function $\mapS{\mathantt{S}}{\mathcal{F}}{\mathcal{L}}$.
This function provides a correlation between the feature set $\mathcal{F}$ and the label set $\mathcal{L}$.
Hence, the \MLterm is responsible for providing machine learning algorithms to train data-based functions that perform either regression or classification tasks (either for modeling the plant or a control law). 
The set $D_\mathrm{tot}$ contains various information depending on the application, such as measured data points, current parameter configurations of the plant, or measurement signals that support the \MLterm to train a function $\mathantt{S}$.  
\\
According to the specific application and design of the \MLterm, $\mathantt{S}$ can represent (parts of the)  dynamic plant model or a control unit. 
In the first case, the plant model is created or supported by the \MLterm using data-based techniques. 
The function $\mathantt{S}$ can be representative of the terms $\mathantt{F}$ and/or $\mathantt{G}$ in~\eqref{eq:tot_equation}. 
In the second case, the \MLterm and thus the function $\mathantt{S}$ represent parts of the control law.
In this way, classical controller design procedures can be extended by using continuously updated data from the plant. 
On the one hand, this allows to consider effects that are difficult to model in advance. 
On the other hand, classical control laws can be approximated by a map $\mathantt{S}$, whose online evaluation on embedded hardware is computationally expensive.

In order to design such a function $\mathantt{S}$, the \MLterm needs data containing measured or observed information from the plant or the controllers. 
Moreover, the \MLterm can incorporate additional information and requirements if available. 
For instance, interpretability,  differentiability and real-time capabilities of a generated plant model might be desired to increase acceptance but also to enable gradient-based optimizations with this model.  
Moreover, a quantification of the model uncertainty to estimate the feasible operating area of the  models might be necessary.
If $\mathantt{S}$ is used as control law, it often is intended to maintain closed-loop stability or ensure robustness in the presence of unpredictable disturbances or model uncertainties.
These requirements should be considered in the \MLterm by using appropriate priors, approaches or constraints in the design of the desired functions. 
In the following, we will outline a mathematical framework to describe arbitrary machine learning techniques including this prior information and elaborate on two specific algorithms, namely neural networks and Gaussian processes. 

\subsection{Mathematical description of  the machine-learning Oracle}

\begin{figure}[!ht]
    \centering
    \input{images_tot/Generator_Structure3}
    \caption{General structure of the \MLterm block to design data-based functions. a) Structure of the \MLterm. b) Subtypes of constraints used in the \MLterm.  }
    \label{fig:Generator_Structure}
\end{figure}
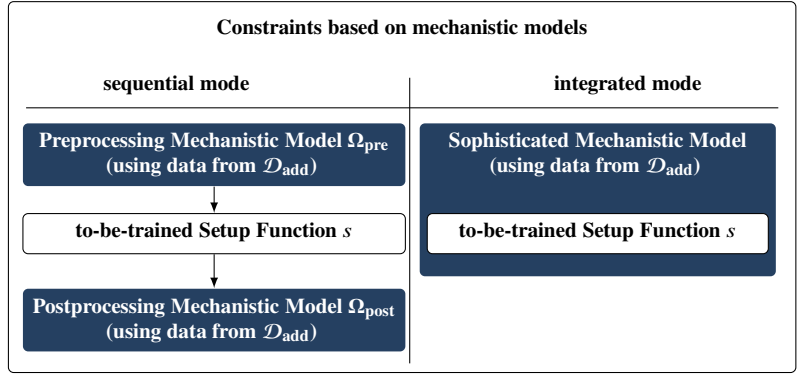

In the previous subsection, we introduced the \MLterm as a map that uses a total data set $\mathcal{D}_\mathrm{tot}$ to design a function 
\begin{alignat*}{3}
    \mathantt{S} \in \mathcal{M} := \left\{ \xi: \mathcal{F} \to \mathcal{L} \right\}
\end{alignat*}
between the elements of a feature space $\mathcal{F}$ and a space $\mathcal{L}$ of the labeled data.
Figure~\ref{fig:Generator_Structure}~a) illustrates this \MLterm and describes its basic structure.
Based on the data set $\mathcal{D}_\mathrm{tot}$, the \MLterm produces the map $\mathantt{S}$ using a coordination/planning step (lower blue box) and a training step (upper blue box).

\subsubsection{Information in the data set}

We assume that the total data set $\mathcal{D}_\mathrm{tot}= \left\{ \mathcal{D}, \mathcal{D}_\mathrm{add} \right\}$ used to generate $\mathantt{S}$ contains the two subsets $\mathcal{D}$ and $\mathcal{D}_\mathrm{add}$ encoding different information about the plant. 
The data set 
$$\mathcal{D} := \bigl\{ \left(f_{\mathrm{pre},1}, \ell_{\mathrm{post},1} \right), \ldots
\left(f_{\mathrm{pre},n}, \ell_{\mathrm{post},n} \right) \; \mid \; 
f_{\mathrm{pre},i} \in \mathcal{F}_\mathrm{pre},
\ell_{\mathrm{post},i} \in \mathcal{L}_\mathrm{post}
\bigr\} $$
is a collection of measurements that  indirectly represent the features and the labels to learn $\mathantt{S}$.
To obtain the features $f_i\in \mathcal F$ and labels $\ell_i \in \mathcal L$ needed to generate $\mathantt{S}$, processing of the data $\mathcal{D}$ might be required.  
This processing is performed by the training block, cf. Figure~\ref{fig:Generator_Structure} and Section~\ref{sec:Trainingblock}.

\begin{remark}
The data set $\mathcal{D}$ is assumed to consist of labeled features, indicating supervised learning. 
\end{remark}
The labels in the data set $\mathcal D$ can be either discrete or a continuous values. 
If a continuous set is considered, the learning task is called regression, otherwise it is called classification.

Additionally to the measured input-output data $\mathcal{D}$, the \MLterm might need to consider additional parameters. 
For example, in physics informed learning, the training of a machine learning algorithm takes into account a first principles model that contains certain parameters or should satisfy certain constraints. 
The additional information about physical assumptions, knowledge, or restrictions which should be considered during the training process is stored in $\mathcal{D}_\mathrm{add}$. 
The total data set $\mathcal{D}_\mathrm{tot} =\left\{ \mathcal{D}, \mathcal{D}_\mathrm{add} \right\}$ is passed to the coordination/planning block responsible for executing the training block and evaluating the trained model.

\subsubsection{Coordinator Block.}
Commonly, the data set $\mathcal D$ is divided into a training, test, and validation dataset.
The coordinator/planner block is responsible for selecting these divisions and evaluating the performance of the trained algorithms.
The training data is used for learning, the test data set is used for evaluating the performance on unseen examples, and the validation dataset is used to compare different machine learning models, to select the best performing model. To speed up the training, the training dataset can be further divided into \emph{batches}. These are a subset of the training dataset that are passed to the machine learning model before its parameters or weights are updated. 
The number of time the training dataset has been used for training is called the \emph{epoch}.
 
The coordinator uses the data set $\mathcal D_\mathrm{tot}$ and performs three duties:
    \begin{itemize}
    \item[1.] Selecting and forwarding of training data and additional parameters to the training block 
    \item[2.] Executing the training block 
    \item[3.] Evaluating the quality of trained function
    \end{itemize} 
    Algorithm~\ref{algo:coordinator} formalizes this task, where $\mathcal{P}\left( \mathcal{D} \right)$ denotes the power set of the data set $\mathcal{D}$. 
    
    \begin{algorithm}[H]
    \label{algo:coordinator}
        \SetAlgoLined
        \caption{Coordination of the \MLterm designing the data-based function.}
        \KwIn{Data set $\mathcal{D}_\mathrm{tot}$}
        \KwResult{function $S\in \mathcal{M}$ trained on the data set $\mathcal{D}_\mathrm{tot}$}
        \While{true}{
            selecting subsets $D_\mathrm{train}$ and $D_\mathrm{test}$ from the data set $\mathcal{P}\left( \mathcal{D} \right)$\\
            running the training block using $D_\mathrm{train}$ and $\mathcal D_\mathrm{add}$ \\
            evaluating the function $S$ using $D_\mathrm{test}$ in terms of e.g. inter- and extrapolation quality or constraint satisfaction\\
            \If{Function $S$ satisfies all conditions}{
            break
            }
            }
    \end{algorithm} 

\subsubsection{Setup functions}
In order to design the function $\fun{S}$, the \MLterm uses a to-be-trained setup function \begin{alignat*}{3}
   \mapS{ \fun{s} }{\mathcal{F} \times \mathcal{W} \times \mathcal{P}\left( \mathcal{D} \right) }{ \mathcal{L} },
\end{alignat*}
where $\mathcal{W}$ is the set of parameters and $\mathcal{P}\left( \mathcal{D} \right)$ denotes the power set of the incoming data $\mathcal{D}$. 
The to-be-trained setup function $\fun{s}$ is chosen externally in accordance to the particular application.
For a neural network, for instance, one chooses the number of layers and the way these layers are connected, as well as the activation function. 
The weights of the network and possibly the parameters of the activation function are not yet determined.
Whereas, in a Gaussian process, one chooses the prior mean as well as the kernel function, which is used to build the covariance matrix from a subset of the data. 
The hyperparameters of the kernel function are not yet specified and must still be trained.
Once  the parameters in $\fun{s}$ have been determined, $\mathantt{S}$ is obtained by fixing the parameters and training data such that $\mathantt{S}(\cdot)= \fun{s} (\cdot, w,D)$, i.e.  $\mathantt{S}$ is the evaluation of $ \fun{s} $ for a given set of parameters and data.
\\
In general, two types of to-be-trained setup functions can be distinguished:     
\begin{description}[labelwidth=*]
    \item[Directly data-based setup functions.]
    The function $s$ depends directly on a subset $D$ of the original set $\mathcal{D}$ of data points, i.e., $\mathcal P (\mathcal D)\ni D \neq \emptyset$.
    Machine learning algorithms that use these type of setup functions do not only use data during training, i.e., to determine the parameters in $\mathcal W$, but they also use data for evaluation or prediction during training.
    Typical examples are  Gaussian process or kernel-based interpolation.
    \item[Indirectly data-based setup functions.]
    The function $\fun{s}$ is not directly dependent on the data points. Hence, the selected element $D$ from the power set of the data is empty, i.e., $D = \emptyset$. 
    Instead of using data directly, the elements of $\mathcal{D}$ are used  to determine parameters in $\mathcal W$. The prediction or evaluation of the trained algorithm then only relies on $\mathcal F$ and $\mathcal W$. 
    Typical examples of machine-learning algorithms that use this type of setup functions are neural networks or linear regression techniques.
\end{description}

The chosen to-be-trained setup function is used inside the training block to parameterize the machine learning algorithm, which is described in the following section. 

\subsubsection{Training Block.}
\label{sec:Trainingblock}
In most machine learning approaches, the training of a function is formulated in terms of an optimization problem in order to find the best model. 
Depending on the particular application, the objective function can be classified into two main categories.
\begin{description}[labelwidth=*]
    \item[Error-based objectives.]
        Here, the aim is to minimize the deviation between the measured label data and the predicted label data based on $\mathantt{S}$.
        The objective function
        \begin{alignat}{3}\label{eq:obj_error}    
         \mapS{J_\mathrm{er}}{
                 \mathcal{L}_\mathrm{post} \times \mathcal{L}_\mathrm{post} }{
         \mathbb{R}} 
        \end{alignat}
        maps from a preselected set of data points $D \in \mathcal{P}\left( \mathcal{D} \right) $
        and the corresponding predicted information using the to-be-trained setup function on the real numbers.
        Typical examples are functions for mean absolute error (MAE), mean squared error (MSE), binary cross entropy (BCE), or cross entropy loss, etc.
    \item[Evidence-based objectives.]    
        This type of objective aims to describe the capability of the distribution of the parameter $w\in\mathcal{W}$ to explain the observed data.
        The objective function 
        \begin{alignat}{3}\label{eq:obj_evidence}    
          \mapS{J_\mathrm{ev}}{ \mathcal{P} \left( \mathcal{D} \right) \times \mathcal{W} }{\mathbb{R}}, 
        \end{alignat}
        maps from a preselected set of data points $D \in \mathcal{P}\left( \mathcal{D} \right) $ and the parameters required to define a to-be-trained setup function to the real numbers.
        These types of objectives relate to directly data-based setup functions. 
        An example for this type of objective functions is the likelihood function.
    \end{description}
In addition to these two types of objective functions, regularization functions are often used to regulate the influence of the parameters and to avoid overfitting.
In this way it is possible to modify the structure of the to-be-trained setup function by weighting different terms or components within such a function.

The goal of the training block is to identify the optimal parameter $w^\ast\in\mathcal{W}$. 
Once the parameter is found, $\fun{S}$ can be constructed based on $\fun{s}$ evaluated at $w^\ast$. 
In the case of indirectly data-based setup functions $\fun{S}$ maps only from features (and not parameters) for further applications.
In case of directly data-based setup functions, the training data $D\in\mathcal{P}\left(\mathcal{D}\right)$ set in $\fun{s}$ must also be fixed to obtain  $\fun{S}$.

\subsubsection{Physics informed machine learning} \label{subsec:PIML}
For many applications, besides the main task of minimizing the objectives~\eqref{eq:obj_error} or~\eqref{eq:obj_evidence} (in conjunction with regularization terms), it must also be ensured that certain constraints are satisfied.
In these cases, it is common to use the term physically informed machine learning.

One type of constraints is provided by the additional conditions of the function space where the setup function lives, such as continous differentiability, monotonicity, or periodicity. 
To account for this, we define the \MLterm outcome as something that has to be an element of the restricted function space
\begin{alignat*}{3}
     \mathcal{M}_\mathrm{r} := \left\{ \xi: \mathcal{F} \to \mathcal{L} \; \mid \; \xi \text{ has to satisfy additional conditions} \right\}.
\end{alignat*}
For instance, in regression problems, one is often only interested in continuously differentiable functions defined by the set
\begin{alignat*}{3}
    C^k\left(\mathcal{F},\mathcal{L} \right) := \left\{ \xi: \mathcal{F} \to \mathcal{L} \; \mid \; \xi \text{ is $k$-times continuously differentiable} \right\}.
\end{alignat*}
Other restrictions could be to permit only monotonous functions, positive/negative definite functions, or those that are bounded. 
We include these constraints in the training block of the \MLterm, see also Figure~\ref{fig:Generator_Structure}~a).

A second type of constraint arises from mechanistic models as illustrated in Figure~\ref{fig:Generator_Structure}~b). 
Mechanistic models are always applied when the sets $\mathcal{F}_\mathrm{pre}$ and $\mathcal{L}_\mathrm{post}$ associated with the data points in $\mathcal{D}$ do not directly correspond to the feature and label space, i.e., $\mathcal{F}$ and $\mathcal{L}$.
In other words, the learning of the desired map from $\mathcal{F}$ to $\mathcal{L}$ can only use data from $\mathcal{F}_\mathrm{pre} \neq \mathcal{F}$ and $\mathcal{L}_\mathrm{post} \neq \mathcal{L}$.  
In such cases it is necessary to translate between these sets, which can be done by means of additional models.
These models usually have the form of implicitly or explicitly given algebraic or dynamic systems of equations. 
For instance, it is necessary to convert between different physical variables (e.g. temperature, pressure, concentrations, chemical potentials, etc.) when not all of them can be measured directly.
Depending on the way these physics informed mechanistic models are applied, one can distinguish between sequential and integrated methods. 
These methods use additional parameters $d\in\mathcal{D}_\mathrm{add}$ that must be provided to the \MLterm in addition to the data points. 
\begin{description}[labelwidth=*]
 \item[Sequential Methods.]
    Sequential methods use physics informed models
    \begin{alignat*}{2}
       0 &=\tilde{\Omega}_\mathrm{pre} \left(  f_{\mathrm{pre}}, f \right) , \quad\text{and}\quad
       0 =\tilde{\Omega}_\mathrm{post} \left(  \ell, \ell_{\mathrm{post}} \right),
    \end{alignat*}
    to convert the measured data so that the setup function can process it.
    Typical examples might include thermodynamic equations that describe the relationship between different state variables (e.g. temperature, pressure, volume, etc.), thermodynamic potentials (e.g. enthalpy, entropy, etc.) or material properties (e.g. densities, capacities, etc.).
    If these models are (locally) given explicitly, i.e. in the form of mappings 
    \begin{alignat*}{2}
       \Omega_\mathrm{pre} :  \mathcal{F}_\mathrm{pre} \to \mathcal{F}, \quad\text{and}\quad
       \Omega_\mathrm{post} :  \mathcal{L} \to \mathcal{L}_\mathrm{post},
    \end{alignat*}
    it is possible to integrate these mappings directly into the setup function.
    The result is a hybrid model. 
    In general, sequential methods of physics informed machine learning allow to outsource the transformation of the data in $\mathcal{D}$. 
    More precisely, the conversion from $f_\mathrm{pre}$ to $f$ and from $\ell_\mathrm{post}$ to $\ell$ can be performed offline and outside of the training block.
    
    Table~\ref{tab:exampl_sm_framework} shows two examples of a training procedure where the data is converted sequentially.
    It provides the mathematical formulation of the training of a given setup function $s$ to obtain optimal parameters $w^*$. In both cases presented in Table~\ref{tab:exampl_sm_framework} additional constraints enter the optimization, which reflect the mechanistic model. While in Case~1, this mechanistic model is given explicitly, Case~2 uses implicit mechanistic constraints. Even though these constraints are stated in Table~\ref{tab:exampl_sm_framework} as a part of the training optimization, the related transformations could be performed outside of the optimizations, since ${\Omega}_\alpha$ or $\tilde{\Omega}_\alpha$, $\alpha\in\left\{\mathrm{pre}, \mathrm{post} \right\}$ do not depend on the parameters $w$.
   \end{description}

    \renewcommand{\arraystretch}{1.4}
    \begin{table}[!ht]
        \centering
        \caption{Example for training a map $\fun{G}_\mathrm{ex}$ using sequential methods.
                 The system in both cases is given as an algebraic model.}
        \label{tab:exampl_sm_framework}
        \begin{tabular}{|rl|}\hline
            \cellcolor{CCPSgray3}  \textbf{Setup function:} &
            \cellcolor{CCPSgray3} indirectly data-based setup function to be trained 
            \\
            \cellcolor{CCPSgray3} \textbf{Objective function:} &
            \cellcolor{CCPSgray3} error-based to minimize deviation between simulated \& measured data
            \\ \hline
            { \cellcolor{CCPSblue1} \color{white}\bfseries Case 1: } &
            { \cellcolor{CCPSblue1} \color{white}\bfseries 
            $\boldsymbol{z_1 = g_\mathrm{ex} \left( u \right)}$, 
            $\boldsymbol{ z_2 = \fun{G}_\mathrm{ex}\left( z_1 \right), y = h\left( z_2 \right)} $
            with $\boldsymbol{z=\left(z_1, z_2 \right)}$}          
            \\ \hline
            \textbf{Data set:} & 
            $\mathcal{D} := \left\{ \left( u_k, y_k \right) \;\mid\;
                u_k \in \mathcal{F}_\mathrm{pre}, y_k \in \mathcal{L}_\mathrm{post}, 
                k=1,\ldots,N  \right\}$ 
            \\
            \textbf{Physical model:} & 
            define $\Omega_\mathrm{pre}:= g_\mathrm{ex} $, $\Omega_\mathrm{post}:=h $ 
            \\
            \textbf{Training:} & 
            \begin{minipage}{8cm}\centering
                \begin{argmini}|l| 
                    {w}{\sum_{k}\left\| y_k - \hat{y}_k \right\|_Q}{\notag}{w^\ast :=}
                    \addConstraint{\hat{y}_k}{=h \circ \fun{s}\bigl( g_\mathrm{ex} \left( u_k \right),w \bigr) }{\quad k=1,\ldots,N}
                \end{argmini}
            \end{minipage}
            \\
            \textbf{\MLterm~output:} &
            $ \fun{G}_\mathrm{ex}(z_1) \approx	\fun{S}(z_1):= \fun{s}(z_1,w^\ast)$ 
            \\\hline
            { \cellcolor{CCPSblue1} \color{white}\bfseries Case 2: } &
            { \cellcolor{CCPSblue1} \color{white}\bfseries 
            $\boldsymbol{0 = g \left(z_1, u \right)}$, 
            $\boldsymbol{ z_2 = \fun{G}_\mathrm{ex}\left( z_1 \right), y = h\left( z_2 \right)} $
            with $\boldsymbol{z=\left(z_1, z_2 \right)}$    }       
            \\ \hline
            \textbf{Data set:} & 
            $\mathcal{D} := \left\{ \left( u_k, y_k \right) \;\mid\;
                u_k \in \mathcal{F}_\mathrm{pre}, y_k \in \mathcal{L}_\mathrm{post}, 
                k=1,\ldots,N  \right\}$ 
            \\
            \textbf{Physical model:} & 
            define $\tilde{\Omega}_\mathrm{pre}:= g $, $\Omega_\mathrm{post}:=h $
            \\
            \textbf{Training:} & 
            \begin{minipage}{8cm}\centering
                \begin{argmini}|l| 
                    {w}{\sum_{k}\left\| y_k - \hat{y}_k \right\|_Q}{\notag}{w^\ast :=}
                    \addConstraint{0}{=g \left(z_{1,k}, u_k \right) }{\quad k=1,\ldots,N}
                    \addConstraint{\hat{y}_k}{=h \circ \fun{s}\bigl( z_{1,k},w \bigr) }{\quad k=1,\ldots,N}
                \end{argmini}
            \end{minipage}
            \\[5pt]
            \textbf{\MLterm~output:} &
            $ \fun{G}_\mathrm{ex}(z_1) \approx \fun{S}(z_1):= \fun{s}(z_1,w^\ast)$ 
            \\\hline
        \end{tabular}
    \end{table}

\begin{description}[labelwidth=*]
  \item[Integrated Methods.]
    Integrated methods deal with physics informed models that are mutually coupled with the setup function.
    In this case, it is not possible to calculate the features $f$ and labels $\ell$  from $\mathcal{D}$ without evaluating the model using parameters $w$ which are still unknown and need to be trained.
    One common example of such models are dynamical systems, where the setup function takes the role of algebraic equations within the system description. 
    For instance, the setup function can represent the kinetic equation in a dynamic model of a reactor. 
    However, often the reaction rates are not directly measurable and the measurement data is composed of the concentration profiles. 
    Learning the reaction rate from this data includes the solution of a dynamical system equation, which must be performed in each training step. 
    Table~\ref{tab:exampl_im_framework} illustrates the training procedure, where the data-based model is part of an algebraic model (Case 1) and a dynamical model (Case 2). 
    Here, the system must be solved during training to obtain data that can be compared with the measured information.
    The constraints in the optimization describe the system equations and directly depend on the parameters $w$.
    Hence, there is no possibility to precalculate features and labels detached from the optimization to simplify training. 
    
\end{description}

    \renewcommand{\arraystretch}{1.4}
    \begin{table}[!ht]
        \centering
        \caption{Example for training a map $\fun{G}_\mathrm{ex}$ using integrated methods.
                 The systems are either given by an algebraic or a dynamical model.}
        \label{tab:exampl_im_framework}
        \begin{tabular}{|rl|}\hline
            \cellcolor{CCPSgray3}  \textbf{Setup function:} &
            \cellcolor{CCPSgray3} indirectly data-based setup function to be trained 
            \\
            \cellcolor{CCPSgray3} \textbf{Objective function:} &
            \cellcolor{CCPSgray3} error-based to minimize deviation between simulated \& measured data
            \\ \hline
            { \cellcolor{CCPSblue1} \color{white}\bfseries Case 1: } &
            { \cellcolor{CCPSblue1} \color{white}\bfseries 
            $\boldsymbol{0=g\left(z_1, z_2 , u \right)}$,
            $\boldsymbol{z_2 = \fun{G}_\mathrm{ex}\left( z_1 \right)}$,
             $\boldsymbol{y = h\left(z_1, z_2 \right) }$
            with $\boldsymbol{z=\left(z_1, z_2 \right)}$  }       
            \\ \hline
            \textbf{Data set:} & 
            $\mathcal{D} := \left\{ \left( u_k, y_k \right) \;\mid\;
            u_k \in \mathcal{F}_\mathrm{pre}, y_k \in \mathcal{L}_\mathrm{post}, 
            k=1,\ldots,N  \right\}$ 
            \\
            \textbf{Physical model:} & 
            $0 =  g \bigl(z_1, \fun{s} \left( z_1, w\right), u\bigr) =: \tilde{g} \left(z_1, w , u\right)  $
            \\
            &
            and define $\tilde{h}  \left(z_1, w\right) := h\bigl(z_1, \fun{s} \left( z_1, w\right)\bigr)$ using $\fun{s}$ 
            \\
            \textbf{Training:} & 
            \begin{minipage}{8cm}\centering
                \begin{argmini}|l| 
                    {w}{\sum_{k}\left\| y_k - \hat{y}_k \right\|_Q}{\notag}{w^\ast :=}
                    \addConstraint{0}{=\tilde{g} \left(z_{1,k}, w , u_k\right)  }{\quad k=1,\ldots,N}
                    \addConstraint{\hat{y}_k}{= \tilde{h}\left(z_{1,k}, w\right)            }{\quad k=1,\ldots,N}
                \end{argmini}
            \end{minipage}
            \\
            \textbf{\MLterm~output:} &
            $ \fun{G}_\mathrm{ex}(z_1) \approx \fun{S}(z_1):= \fun{s}(z_1,w^\ast)$ 
            \\\hline
            { \cellcolor{CCPSblue1} \color{white}\bfseries Case 2: } &
            { \cellcolor{CCPSblue1} \color{white}\bfseries 
            $\boldsymbol{\dot{x} = f\left(x,z,u\right)}$, 
            $\boldsymbol{z=\fun{G}_\mathrm{ex}\left( x \right)}$,
            $\boldsymbol{y = h\left(x\right) }$   }       
            \\ \hline
             \textbf{Data set:} & 
            $\mathcal{D} := \left\{ \left( t_k, y_k \right) \;\mid\;
                t_k \in \mathcal{F}_\mathrm{pre}, y_k \in \mathcal{L}_\mathrm{post}, 
                k=1,\ldots,N  \right\}$ 
            \\
            \textbf{Additional data:} & 
            $\mathcal{D}_\mathrm{add} := \left\{ x_0, \left(u_k\right)_{k\in\mathbb{T}}  \right\}$
            with $\mathbb{T}:=\left\{ 0,1,\ldots, N-1 \right\}$
            \\ 
            \textbf{Physical model:} & 
            one-step integrator $\left(x_{k+1},z_{k+1} \right) = I \left(x_k,u_k,w\right) $ using $\fun{s}$
            \\
            \textbf{Training:} & 
            \begin{minipage}{8cm}\centering
                \begin{argmini}|l| 
                    {w}{\sum_{k}\left\| y_k - \hat{y}_k \right\|_Q}{\notag}{w^\ast :=}
                    \addConstraint{\left(\hat{x}_{k+1},z_{k+1} \right)}{=I \left( \hat{x}_k,u_k,w\right) }{\quad k=0,\ldots,N-1}
                    \addConstraint{\hat{x}_0}{=x_0 }{}
                    \addConstraint{\hat{y}_k}{=h\left( \hat{x}_k \right) }{\quad k=1,\ldots,N}
                \end{argmini}
            \end{minipage}
            \\
            \textbf{\MLterm~output:} &
            $ \fun{G}_\mathrm{ex}(x) \approx \fun{S}(x):= \fun{s}(x,w^\ast)$ 
            \\\hline
        \end{tabular}
    \end{table}

\bigskip
\noindent
In summary, physics informed knowledge can be integrated into the design process of the setup function and also into the training of the parameters in various ways. 
In this way, physical models can be used to convert between measured data and variables needed for training. 
Also the choice of a setup function to be trained can be motivated by physical and technical aspects. 
For instance, the initial network topology of a neural network (e.g., number of hidden layers and choice of the activation function) or the kernel function of a Gaussian process can be crucial in approximating the data.  
In particular, the choice of setup function determines how efficiently additional constraints on the function space $\mathcal{M}_\mathrm{r}$ can be incorporated into the optimization. 
Indeed, some setup functions satisfy certain properties (e.g. periodicity) directly by their structure. 
Also, the choice of initial parameters or boundaries results from prior knowledge of experienced engineers or from technical restrictions, see \cite{Willard2003}.

\subsection{Specific machine-learning oracle examples} 
We have introduced the \MLterm in the previous section in a generic form, such that arbitrary machine learning algorithms can be naturally embedded in the framework. 
In this subsection, we want to outline specific examples for \MLterm{}s, namely neural networks and Gaussian processes. 
For this, we assume that the features and labels are measured directly and no conversion between $\mathcal{F}_\mathrm{pre}/\mathcal{F}$ or $\mathcal{L}_\mathrm{post}/\mathcal{L}$ is required via physical models. 
Nevertheless, the principles of physics informed machine learning can be extended to these setup functions as well. 

\subsubsection{Feed-forward Artificial Neural Networks}
\label{sec:FNN}
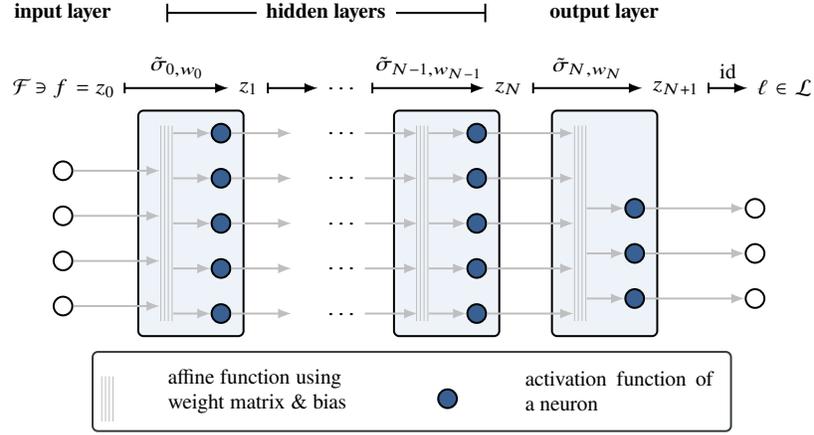
\begin{figure}[tp]
    \centering
    \input{images_tot/ANN_new2}
    \caption{Representation of a feed-forward neural network.}
    \label{fig:ann}
\end{figure}

Artificial neural networks are one of the most flexible and wide-spread machine learning models. 
They are inspired by the neural networks of the brain \cite{McCulloch1943}. 
Artificial neural networks are formed by a set of neurons that can be organized in layers (see Figure~\ref{fig:ann}) or in more complex structures (\cite{Kusiak20201594,Zhou20171229}) where each neuron is connected to other neurons in the network. 
Here, we consider feed-forward neural networks (FNNs) (cf. Figure~\ref{fig:ann}), i.e. each neuron in a layer is connected to the neurons in the next layer without self connections or loops. 
Each of these connections is weighted with a parameter that indicates the importance of the signal to the next neuron. 
Information is entered in the input layer of the FNN and transported and transformed through the network to obtain the corresponding output. 
The transformation is performed in the neurons by applying nonlinear activation functions to the weighted sum of all inputs and a bias to each neuron. 
The feature $f=:z_0$ is the input to the overall network. 
Training a network aims at finding parameters, such as the weights, to produce the network output $z_{N+1}\in \mathcal L $ that mimics the label data.

Neural networks belong to the group of machine learning techniques that use indirectly data-based setup functions.
This means that the measured data is only used to train the parameters and not during inference.
The to-be-trained setup function $s_\mathrm{FNN} $ is composed of elementary functions 
\begin{equation}
\label{eq:activation}
      z_{l+1}=\tilde \sigma_{l, w_l} \left(z_{l}\right) := \sigma_l\left(W_l z_{l}+b_l\right),
\end{equation}
which describe the processing in each layer.
Here, $\mapS{ \sigma_l }{\mathbb{R}^{n_{l+1}}}{\mathbb{R}^{n_{l+1}}}$ is the activation function, and 
$w_l = \left( W_l, b_l \right) \in \mathcal{W}_l:=\mathbb{R}^{n_{l+1}, n_{l}} \times \mathbb{R}^{n_{l+1}} $ is the parameter of the layer $l$ with the weight matrix $W_l$ and bias $b_l$. 
The output information $z_{l} \in \mathbb{R}^{n_{l}}$ of layer $l$ builds the input to layer $l+1$, where $n_l$ indicates the number of neurons in the layer $l$. 
The numbers $n_0=n_\mathrm{f}$ and $n_{N+1}=n_\mathrm{\ell}$ are identical to the dimension of the feature space and the label space.
The parameters of a neural network, i.e. the elements of $\mathcal W$, consist of the entries in the weighting matrices and the bias terms. 

By repeating this operation for all the $N$ layers, we obtain the to-be-trained setup function 
\begin{alignat*}{3}
     \fun{s}_\mathrm{FNN} \colon &\mathcal{F} \times \mathcal{W} \to \mathcal{L},\;
                           \left( f, w \right) \mapsto \fun{s}_\mathrm{FNN} \left( f, w \right) 
         :=\tilde \sigma_{N,w_N} \circ  \ldots \circ \tilde \sigma_{0,w_0} \left(f\right). 
\end{alignat*} 
The individual parameters $w_l$ form the entire parameter $w = \left( w_0, \ldots, w_N \right) \in \mathcal{W}:= \mathcal{W}_0 \times \ldots \mathcal{W}_N$. 
The weights $W_i$ and bias $b_i$ of every layer are optimized during the training procedure using error-based objectives~\eqref{eq:obj_error}. 
In other words, the error between the predicted and measured labels is minimized \cite{Psichogios1992}.  
In practice, gradient-based optimization algorithms are often used to tackle the training of artificial neural networks.  
Based on the specific structure of deep neural networks (networks with many hidden layers), special subtypes of gradient-based techniques are used for the training procedure. 
One of them is backpropagation, which allows iterative updating of parameters while preventing the network from getting stuck in local minima due to vanishing gradients.
For more detailed discussion on vanishing gradient and backpropagation, readers are referred to \cite{Goodfellow2016} and \cite{Sandro2018}.

\subsubsection{Recurrent Neural Network}
\label{sec:RNN}
\begin{figure}[!ht]
    \centering
    \input{images_tot/RNN_new2}
    \caption{Representation of a RNN. Note that the we used the subscript $l$ to highlight the memory effect of the network due to the feedback.}
    \label{fig:rnn2}
\end{figure}
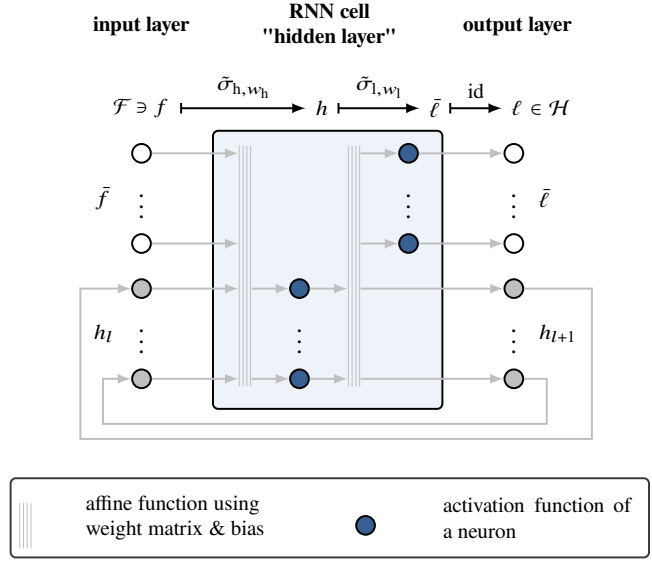

Feed-forward neural networks are static maps between inputs and outputs. 
In some cases it is useful to express the current output not only as a function of the current input, but also of the previous ones. 
This is the case where the order of the measured features is important to predict the labels, e.g., in time series data, text generation, or text prediction. 
Recurrent neural networks (RNNs) are neural networks that serve this purpose by providing the network with a \emph{memory effect} which is often useful when modeling dynamical systems. 
To integrate this memory effect into the network, one uses so-called \emph{hidden states}, which allow to transfer the information from previous evaluations into further ones via information feedback.
For the sake of simplicity, we assume only one processing layer to describe the principle of an RNN. 
More precisely, we consider a network with only one other layer besides the input and output layer.  
In general, there are different options within such a layer and thus different architectures to incorporate the memory effect into the network.
In the following, we focus only on a single RNN cell, further structures are e.g. gated recurrent unit (GRU) \cite{Chung2014empirical} or long short-term memory (LSTM) \cite{Sherstinsky2020}. 
\\
An RNN cell is a kind of hidden layer formed by two  sublayers of neurons, see Figure~\ref{fig:rnn2}. 
The first layer is used to calculate the hidden states, while the second layer defines how the labels are derived from these states. 
To bring the RNN cell architecture into our framework, the following two assumptions are used.
\begin{itemize}
    \item[(i)]  The feature and label space are of the form
                    \begin{alignat}{2}\label{eq:FandLstructureRNN}
                       \mathcal{F} = \bar{\mathcal{F}} \times \mathcal{H} 
                       \quad \text{and} \quad 
                       \mathcal{L} = \bar{\mathcal{L}} \times \mathcal{H}.
                    \end{alignat}
                Here, the set $\mathcal{H}$ contains the so-called hidden states allowing to describe a parameter-dependent mapping.
                The sets $\bar{\mathcal{F}}$ and $\bar{\mathcal{L}}$ represent the original domain and the codomain if one would use an FNN to describe the setup function.  
    \item[(ii)] The part of the labels containing the internal state $h\in\mathcal{H}$ is fed back to the input of the neuron,             see Figure~\ref{fig:rnn2}.
\end{itemize}
The feature and label space is artificially extended in Assumption (i) by the hidden states to allow to feeding them back as a part of the cell output and input.
The second assumption gives us the ability to store and reuse information once it has been computed. 
In other words, the neuron is equipped with a memory. 
The elementary functions of the individual sublayers within an RNN cell are defined as follows
\begin{alignat*}{3}
\begin{array}{rl}
    \text{1\textsuperscript{st} layer:} & 
        h_{l+1}  = \tilde{\sigma}_{\mathrm{h}, w_\mathrm{h}} \left( f, h_l \right) 
                := \sigma_\mathrm{h}\left( W_\mathrm{f} f + W_\mathrm{h} h_{l} + b_{\mathrm{h}} \right),\\
    \text{2\textsuperscript{nd} layer:} & 
        \quad\,
        \ell     = \tilde{\sigma}_{\mathrm{l}, w_\mathrm{l}} \left( h_{l+1} \right)  \;\;
                := \sigma_\mathrm{l}\left( W_\mathrm{l} h_{l+1} + b_\mathrm{l} \right).
\end{array}
\end{alignat*}
Here, the maps $\mapS{ \sigma_\mathrm{h} }{ \mathbb{R}^{n_\mathrm{h}} }{ \mathbb{R}^{n_\mathrm{h}} }$ and 
$\mapS{ \sigma_\mathrm{l} }{ \mathbb{R}^{n_\mathrm{h}} }{ \mathbb{R}^{n_\mathrm{\ell}} }$
are the activation functions. 
The parameters 
$w_\mathrm{h} = \left( W_\mathrm{f}, W_\mathrm{h}, b_{\mathrm{h}} \right) \in \mathcal{W}_\mathrm{h}
:=\mathbb{R}^{n_\mathrm{h},n_\mathrm{f} } \times \mathbb{R}^{n_\mathrm{h},n_\mathrm{h} } \times \mathbb{R}^{n_\mathrm{h}} $ 
and
$w_\mathrm{l} = \left( W_\mathrm{l}, b_{\mathrm{l}} \right) \in \mathcal{W}_\mathrm{l}
:=\mathbb{R}^{n_\mathrm{h},n_\mathrm{h} } \times \mathbb{R}^{n_\mathrm{h}} $ 
consist of the weight matrices ($ W_\mathrm{f}$,  $W_\mathrm{h}$ and $W_\mathrm{l}$) and bias values ($b_{\mathrm{h}}$ and $b_{\mathrm{l}}$) that have to be trained.
Using~\eqref{eq:FandLstructureRNN}, the setup function of a single RNN cell can be defined as
\begin{alignat*}{3}
    \fun{s}_\mathrm{RNN} \colon & \bar{\mathcal{F}} \times \mathcal{H} \times \mathcal{W} \to \bar{\mathcal{L}} \times \mathcal{H},\;
                           \left( f, h_l , w \right) \mapsto \fun{s}_\mathrm{RNN} \left( f, h_l, w \right) 
                           :=\left( \ell, h_{l+1} \right) ,
\end{alignat*} 
where $w:=\left(w_\mathrm{h},w_\mathrm{l} \right) \in\mathcal{W}:= \mathcal{W}_\mathrm{h} \times \mathcal{W}_\mathrm{l}$.

\subsubsection{Gaussian Process} 
\label{sec:gaussian_process}
\def\hyp{\mathantt{h}}
The approach of Gaussian processes (GPs) differs from that of neural networks in terms of the to-be-trained setup function, the objective during training, and the type of data processing.
The basic idea is that the labels are interpreted as normally distributed random variables, which depend on the features. 
To describe the similarity of data points, one uses a kernel function 
$ \mapS{\sigma}{\mathcal{F} \times \mathcal{F} \times \mathcal{W} }{\mathbb{R}^+_0}$, 
assigning a positive real number to two features.
The kernel specific parameter $\hyp\in\mathcal{H}$ is either assumed to be given or  trained using data and an evidence-based objective function.
The kernel function can be used to evaluate neighborhoods or relationships between two points in the feature space. 
If the value of $\sigma$ is greater for two points than for two other points, it can be concluded that there is a stronger coupling between them. 
This in turn affects the evaluation of the labels among each other.
In general, the kernel function must satisfy the following two properties:
\begin{description}[labelwidth=*]
  \item[Symmetric]
      For a fixed parameter $\hyp\in\mathcal{H}$, the kernel function is symmetric, 
      i.e., $\sigma \left(f_1, f_2, \hyp \right) = \sigma \left(f_2, f_1, \hyp \right)$.
  \item[Upper bounded]
      For a fixed parameter $\hyp\in\mathcal{H}$ and a fixed feature $f\in\mathcal{F}$,
      the function 
      $\tilde{\sigma}: \mathcal{F} \to \mathbb{R}^+_0$, 
      $ \tilde{\sigma} \left( \tilde{f} \right) := \sigma \left(f, \tilde{f} , \hyp \right)$,
     is upper bounded by $\tilde{\sigma} \left( f \right)$.
\end{description}
The symmetry is important because two arbitrary points in the feature space $\mathcal{F}$ are in the same relationship no matter which point is used for which argument of $\sigma$.
The second property ensures that the similarity between an arbitrary point $\tilde{f}$ in the feature space $\mathcal{F}$ and a base point $f$ is not greater than the similarity of $f$ to itself.
A common example of a kernel function used in the literature is the squared exponential function 
\begin{alignat*}{3}
  \sigma_\mathrm{se}\left( f_1, f_2, \hyp \right) 
  := \hyp_1 \exp\left( \frac{ \sum_{\alpha}  \left( {f_1}^\alpha-{f_2}^\alpha \right)^2 }{\hyp_2} \right),
\end{alignat*} 
where ${f_i}^\alpha$ denotes the $\alpha$\textsuperscript{th} component of $f_i$ and $\hyp_1,\hyp_2\in\mathbb{R}_+$ are the parameters.

\bigskip
\noindent
By means of the kernel function, a directly data-based setup function 
\begin{alignat*}{3}
  \fun{s}_\mathrm{GP} \colon & \mathcal{F} \times \mathcal{W} \times \mathcal{P}\left( \mathcal{D} \right)  \to \mathcal{L},\\
                    &\left( \tilde{f}, w, D\right) \mapsto 
                    \fun{s}_\mathrm{GP} \left( \tilde{f}, w, D\right) 
                    := \sum_{\alpha,\beta=1}^{n} \left(K(w,D)^{-1}\right)^{\alpha \beta} \ell_\beta \,\sigma \left(f_\alpha, \tilde{f},\hyp\right)   
\end{alignat*} 
can be designed using the incoming data $D=\left\{ \left(f_1,\ell_1 \right), \ldots, \left(f_n,\ell_n \right) \right\} \in \mathcal{P}\left( \mathcal{D} \right)$ and the hyperparameter $w:=\left( \hyp, \nu \right)\in \mathcal{W}:=\mathcal{H}\times\mathbb{R}_+$.
Here, $\hyp$ denotes the kernel specific parameter and $\nu\in\mathbb{R}_+$ the variance.
The matrix-valued function  
\begin{alignat*}{3}
K \colon & \mathcal{W} \times \mathcal{P}\left( \mathcal{D} \right)  \to \mathbb{R}^{n,n},\\
         & \left(  w, D \right) \mapsto K\left(  w, D \right) 
            := \begin{pmatrix}
                        \sigma \left(f_1,f_1,\hyp\right) & \hdots & \sigma \left(f_1,f_n,\hyp\right)  \\
                         \vdots & \ddots & \vdots \\ 
                        \sigma \left(f_n,f_1,\hyp\right) & \hdots & \sigma \left(f_n,f_n,\hyp\right)  \\
                 \end{pmatrix}
                + \nu \mathbb{I},
\end{alignat*}
yields the covariance matrix and thus allows to describe the relationship of the features to each other.
Due to the symmetry property of the kernel function, it follows that $K\left( w, D \right) $ is a symmetric matrix.
In the literature, the function $ \fun{s}_\mathrm{GP}$ is referred to as a posterior mean with a prior mean of zero, see \cite{Rasmussen2006}.

The benefit of Gaussian processes arises from the stochastic nature of this approach. 
In addition to the $\fun{s}_\mathrm{GP}$ function, another function $\kappa$ can be generated that describes a covariance around $\fun{s}_\mathrm{GP}$ and thus defines a confidence interval.  
This posterior covariance function reads 
\begin{alignat*}{3}
  \kappa \colon &  \mathcal{F} \times \mathcal{W} \times \mathcal{P}\left( \mathcal{D} \right)  \to \mathbb{R}_+,
                 \left( \tilde{f}, w, D\right) \mapsto \kappa \left( \tilde{f}, w, D\right) 
\intertext{where}                 
                 \kappa \left( \tilde{f}, w, D\right) 
                &:= \sigma \left(\tilde{f} ,\tilde{f} ,\hyp\right) - 
                \sum_{\alpha,\beta=1}^{n} \left(K(w,D)^{-1}\right)^{\alpha \beta} \,\sigma \left(f_\alpha, \tilde{f},\hyp\right) \,\sigma \left(f_\beta, \tilde{f},\hyp\right)
\end{alignat*}
In order to train the hyperparameters $w\in\mathcal{W}$ one chooses an evidence-based objective function to maximize the reliability of the data.

\begin{table}[!ht]
    \centering
    \caption{Some references for the three different machine learning models considered. This list is not comprehensive since the literature is very vast. In this list we focused of application of machine learning models for Model Predictive Control.}
    \label{tab:my_label}
    \begin{tabularx}{300pt}{lX}
    \toprule
        Network type & References \\ \midrule
         Feed-forward NNs& \cite{Oliveira2004,Teixeira2006,Morabito2021,Samek2005335,Embaby2020109,Kheirabadi20213077,Chen20213273,Shao2014717,Sarali2019,VanDenBoom2005639,Varshney2009543,Mjalli2006539,Ou2002195,Dahunsi2010,Mohanty2009991,BETHGE202014356, Georgieva2007, Kumar2021, Zhang2019RTO} \\
         Recurrent NNs& \cite{Yan20164377,Zhang2008322,Temeng199519,Zarkogianni20112467,Patan20151147,Dalamagkidis2011818,Thuruthel2019127,Yan2012746,AlSeyab2008568,Kittisupakorn2009579, Lu20081366,Yan2012717,Pan20123089,Zhang2019524,Huang2015256,Pereira20213213,Yang2021,Pan20101597,Atuonwu20101418,Wang201929,Núñez20202859,Wu2020,Wu20202275,Wu202074,Pan2009683,Chen20198461,Pan20081685,Yan20164377,Zhang2008322,Temeng199519,Zarkogianni20112467,Patan20151147,Dalamagkidis2011818,Thuruthel2019127,Yan2012746,AlSeyab2008568,Kittisupakorn2009579,Lu20081366,Yan2012717,Pan20123089}  \\
         Gaussian process & \cite{BETHGE2018517,Kocijan2003, Kocijan2004, Murray2003, Likar2007, Grancharova2007, Grancharova2008,Cao2016, Cao2017b,Nghiem2017,Maiworm2018b,maiworm2019online, Hewing2017,Yang2015b,Soloperto2018,bradford2021hybrid,caldwell2021towards, li2021adaptive}\\
     \bottomrule
    \end{tabularx}

\end{table}

%% file: images_tot/Generator_Structure3.tex
\begin{tikzpicture}[>=latex]





\node (BASIC) at (0,0) [rounded corners=2pt,draw]
{\begin{tikzpicture}[>=latex]

\node (A0) at (-0.0,-3.2) [rounded corners=2pt,draw,fill=IFATblue3!50]
{\begin{minipage}[c][0.3cm]{10cm}\centering\bfseries\footnotesize
Coordinator/Planner Block
\end{minipage}};

\node (A1) at (-0.0,-0.9) [rounded corners=2pt,draw,fill=IFATblue3!50]
{\begin{minipage}[t][3cm]{10cm}\bfseries\centering\footnotesize
Training Block
\begin{argmini*}
{w}{J(w)}{}{w^\ast:=}
\addConstraint{\hspace*{6cm}}{}{}
\end{argmini*}
\end{minipage}};

\node (Acenter) at (-0,-1) [anchor=north,rounded corners=2pt,draw,fill=IFATblue3!00]
{\begin{minipage}[t][0.3cm]{9.65cm}\centering\bfseries\footnotesize
Constraints based on mechanistic models  
\end{minipage}};

\node (Acenter) at (-0,-1.8) [anchor=north,rounded corners=2pt,draw,fill=IFATblue3!00]
{\begin{minipage}[t][0.3cm]{9.65cm}\centering\bfseries\footnotesize
Constraints obtained from the function space of $\mathantt{S}$
\end{minipage}};

\draw [->,line width=1pt] ([xshift=+2cm,yshift=+0.0cm]A0.north) -- ([xshift=+2cm,yshift=+0.0cm]A1.south);
\draw [<-,line width=1pt] ([xshift=-2cm,yshift=+0.0cm]A0.north) -- ([xshift=-2cm,yshift=+0.0cm]A1.south);

\end{tikzpicture}};

\draw [->,line width=1pt] ([xshift=+1.5cm,yshift=-1cm]BASIC.south) -- node[right] {\footnotesize
data set $\mathcal{D}_\mathrm{tot}$} ([xshift=+1.5cm,yshift=+0.0cm]BASIC.south);

\draw [->,line width=1pt] ([xshift=-1.5cm,yshift=0cm]BASIC.south) -- node[left] {\footnotesize
map $\mathantt{S}\in \mathcal{M}$} ([xshift=-1.5cm,yshift=-1cm]BASIC.south);

\node (DETAIL) at (0,-7) [rounded corners=2pt,draw]
{\begin{tikzpicture}[>=latex]

\node (Acenter) at (-0,1.5) [anchor=north]
{\begin{minipage}{9.65cm}\centering\bfseries\footnotesize
Constraints based on mechanistic models  
\end{minipage}};

\node (AcenterL) at ([yshift=-0.3cm,xshift=-3cm]Acenter.south) [anchor=north]
{\begin{minipage}{4cm}\centering\bfseries\footnotesize
sequential mode
\end{minipage}};

\node (AcenterR) at ([yshift=-0.3cm,xshift=3cm]Acenter.south) [anchor=north]
{\begin{minipage}{4cm}\centering\bfseries\footnotesize
integrated mode
\end{minipage}};

\node (A3s) at (2.6,-0) [anchor=north,rounded corners=2pt,draw=none,fill=CCPSblue1]
{\begin{minipage}[t][1.8cm]{4.5cm}\centering\bfseries\footnotesize\textcolor{white}{
Sophisticated Mechanistic Model \\
(\footnotesize using data from $\boldsymbol{\mathcal{D}_{\mathrm{add}}}$)
}
\end{minipage}};

\node (A2l) at (-2.5,-0.0) [anchor=north,rounded corners=2pt,draw=none,fill=CCPSblue1]
{\begin{minipage}[t][0.6cm]{4.85cm}\centering\bfseries\footnotesize\textcolor{white}{
Preprocessing Mechanistic Model $\boldsymbol{\Omega_\mathrm{pre}}$\\
(\footnotesize using data from $\boldsymbol{\mathcal{D}_{\mathrm{add}}}$)
}
\end{minipage}};

\node (A3l) at (-2.5,-1.2) [anchor=north,rounded corners=2pt,draw=,fill=white]
{\begin{minipage}[t][0.3cm]{4.85cm}\centering\bfseries\footnotesize
to-be-trained Setup Function $s$
\end{minipage}};

\node (A3r) at (2.6,-1.2) [anchor=north,rounded corners=2pt,draw=,fill=white]
{\begin{minipage}[t][0.3cm]{4.3cm}\centering\bfseries\footnotesize
to-be-trained Setup Function $s$
\end{minipage}};

\node (A4l) at (-2.5,-2.2) [anchor=north,rounded corners=2pt,draw,draw=none,fill=CCPSblue1]
{\begin{minipage}[c][0.6cm]{4.85cm}\centering\bfseries\footnotesize\textcolor{white}{
Postprocessing Mechanistic Model $\boldsymbol{\Omega_\mathrm{post}}$\\
(\footnotesize using data from $\boldsymbol{\mathcal{D}_{\mathrm{add}}}$)
}
\end{minipage}};

\draw[] (0.1,0.6) --++(0,-3.8);
\draw[] (-5,0.2) --++(10,0);

\draw [->] ([xshift=0cm,yshift=+0.0cm]A2l.south) -- ([xshift=0cm,yshift=+0.0cm]A3l.north);
\draw [->] ([xshift=0cm,yshift=+0.0cm]A3l.south) -- ([xshift=0cm,yshift=+0.0cm]A4l.north);
\end{tikzpicture}};

\node (F1) at (-5,-3.1) [anchor=north] {a)};
\node (F2) at (-5,-9.7) [anchor=north] {b)};
\end{tikzpicture}

%% file: images_tot/ANN_new2.tex
\tikzset{every picture/.style={line width=0.75pt}} 

\begin{tikzpicture}[node distance=0.6cm]
    \node (T2) at (0,0) [rounded corners=2pt, draw=none] 
    {\begin{tikzpicture}[>=latex]
        
            
        \node at (0.5,1) [] {\footnotesize\bfseries input layer};     
        \node (HL) at (4,1) [] {\footnotesize\bfseries hidden layers}; 
        \draw [|-] ([xshift=-1.2cm]HL.west) -- (HL.west);
        \draw [|-] ([xshift=1.2cm]HL.east) -- (HL.east);

        \node at (7.7,1) [] {\footnotesize\bfseries output layer}; 
        
        \node (E0) at (0.5,0) {\footnotesize $\mathcal{F} \ni f=z_0$};
        \node (z01) [below of=E0, yshift=-0.5cm, circle, inner sep=2.5pt, fill=white,draw] {};
        \node (z02) [below of=z01, circle, inner sep=2.5pt, fill=white,draw] {};
        \node (z03) [below of=z02, circle, inner sep=2.5pt, fill=white,draw] {};
        \node (z04) [below of=z03, circle, inner sep=2.5pt, fill=white,draw] {};
        
        \node (E1) [right of=E0, xshift=1.5cm] {};
        \node (E1t) [right of=E1, xshift=-0.5cm,anchor=west] {\footnotesize $z_1$};
        \draw[rounded corners=2pt, draw, inner sep=2pt, fill=IFATblue3!50] (1.5, -0.3) rectangle ++(1.4,-3);
        \node (z11) [below of=E1, circle, inner sep=2.5pt, fill=IFATblue2,draw]  {};
        \node (z12) [below of=z11, circle, inner sep=2.5pt, fill=IFATblue2,draw] {};
        \node (z13) [below of=z12, circle, inner sep=2.5pt, fill=IFATblue2,draw] {};
        \node (z14) [below of=z13, circle, inner sep=2.5pt, fill=IFATblue2,draw] {};
        \node (z15) [below of=z14, circle, inner sep=2.5pt, fill=IFATblue2,draw] {};
        
        \node (E2) [right of=E1, xshift=1cm] {\footnotesize $\ldots$};
        
        \node (z21) [below of=E2, circle, inner sep=0.5pt] {\footnotesize $\ldots$};
        \node (z22) [below of=z21, circle, inner sep=0.5pt] {\footnotesize $\ldots$};
        \node (z23) [below of=z22, circle, inner sep=0.5pt] {\footnotesize $\ldots$};
        \node (z24) [below of=z23, circle, inner sep=0.5pt] {\footnotesize $\ldots$};
        \node (z25) [below of=z24, circle, inner sep=0.5pt] {\footnotesize $\ldots$};
        
        \node (En) [right of=E2, xshift=1.2cm] {};  
        \node (Ent) [right of=En, xshift=-0.5cm,anchor=west] {\footnotesize $z_{N}$};
        \draw[rounded corners=2pt, draw, inner sep=2pt, fill=IFATblue3!50] (4.9, -0.3) rectangle ++(1.4,-3);
        \node (zn0) [below of=En, circle, inner sep=2.5pt, fill=IFATblue2,draw] {};
        \node (zn1) [below of=zn0, circle, inner sep=2.5pt, fill=IFATblue2,draw] {};
        \node (zn2) [below of=zn1, circle, inner sep=2.5pt, fill=IFATblue2,draw] {};
        \node (zn3) [below of=zn2, circle, inner sep=2.5pt, fill=IFATblue2,draw] {};
        \node (zn4) [below of=zn3, circle, inner sep=2.5pt, fill=IFATblue2,draw] {};
        
        \node (EN) [right of=En, xshift=1.5cm] {};
        \node (ENt) [right of=EN, xshift=-0.5cm,anchor=west] {\footnotesize $z_{N+1}$};
        \draw[rounded corners=2pt, draw, inner sep=2pt, fill=IFATblue3!50] (7.0, -0.3) rectangle ++(1.4,-3);
        \node (zN1) [below of=EN, yshift=-1.0cm, circle, inner sep=2.5pt, fill=IFATblue2,draw] {};
        \node (zN2) [below of=zN1, circle, inner sep=2.5pt, fill=IFATblue2,draw] {};
        \node (zN3) [below of=zN2, circle, inner sep=2.5pt, fill=IFATblue2,draw] {};

        \node (ENN) [right of=EN, xshift=1cm] {\footnotesize };
        \node (ENNt) [right of=ENN, xshift=-0.7cm,anchor=west] {\footnotesize $\ell\in\mathcal{L}$};
        \node (zNN1) [below of=ENN, yshift=-1.0cm, circle, inner sep=2.5pt, fill=white,draw] {};
        \node (zNN2) [below of=zNN1, circle, inner sep=2.5pt, fill=white,draw] {};
        \node (zNN3) [below of=zNN2, circle, inner sep=2.5pt, fill=white,draw] {};

        \draw[->] (E0.east)-- node[above] {\footnotesize $\tilde{\sigma}_{0,w_0}$} (E1t.west);
        \draw ([yshift=-2pt]E0.east) -- ([yshift=2pt]E0.east);
        
        \draw[->] (E1t.east)--  (E2.west);  
        \draw ([yshift=-2pt]E1t.east) -- ([yshift=2pt]E1t.east);
        
        \draw[->] ([xshift=0.1cm]E2.east)-- node[above] {\footnotesize $\tilde{\sigma}_{N-1,w_{N-1}}$} (Ent.west);
        \draw ([yshift=-2pt,xshift=0.1cm]E2.east) -- ([yshift=2pt,xshift=0.1cm]E2.east);
        
        \draw[->] (Ent.east)-- node[above] {\footnotesize $\tilde{\sigma}_{N,w_N}$} (ENt.west);
        \draw ([yshift=-2pt]Ent.east) -- ([yshift=2pt]Ent.east);
        
        \draw[->] (ENt.east)-- node[above] {\footnotesize $\mathrm{id}$} (ENNt.west);
        \draw ([yshift=-2pt]ENt.east) -- ([yshift=2pt]ENt.east);
                
         \draw[->,IFATgray2] (z01.east) -- ++(1.15,0);
         \draw[->,IFATgray2] (z02.east) -- ++(1.15,0);
         \draw[->,IFATgray2] (z03.east) -- ++(1.15,0);
         \draw[->,IFATgray2] (z04.east) -- ++(1.15,0);
        
            \draw[IFATgray2,line width=0.3pt] (1.8,-0.5) -- ++(0,-2.6);
            \draw[IFATgray2,line width=0.3pt] (1.8+1*0.05,-0.5) -- ++(0,-2.6);
            \draw[IFATgray2,line width=0.3pt] (1.8+2*0.05,-0.5) -- ++(0,-2.6);
            \draw[IFATgray2,line width=0.3pt] (1.8+3*0.05,-0.5) -- ++(0,-2.6);
            
         \draw[<-,IFATgray2]  (z11.west) --++(-0.5,0);
         \draw[<-,IFATgray2]  (z12.west) --++(-0.5,0);
         \draw[<-,IFATgray2]  (z13.west) --++(-0.5,0);
         \draw[<-,IFATgray2]  (z14.west) --++(-0.5,0);
         \draw[<-,IFATgray2]  (z15.west) --++(-0.5,0);

        \draw[->,IFATgray2]  (z11.east) --++(0.8,0);
        \draw[->,IFATgray2]  (z12.east) --++(0.8,0);
        \draw[->,IFATgray2]  (z13.east) --++(0.8,0);
        \draw[->,IFATgray2]  (z14.east) --++(0.8,0);
        \draw[->,IFATgray2]  (z15.east) --++(0.8,0);

         \draw[->,IFATgray2]  ([xshift=-1.35cm]zn0.west) --++(0.7,0);
         \draw[->,IFATgray2]  ([xshift=-1.35cm]zn1.west) --++(0.7,0);
         \draw[->,IFATgray2]  ([xshift=-1.35cm]zn2.west) --++(0.7,0);
         \draw[->,IFATgray2]  ([xshift=-1.35cm]zn3.west) --++(0.7,0);
         \draw[->,IFATgray2]  ([xshift=-1.35cm]zn4.west) --++(0.7,0);
        
            \draw[IFATgray2,line width=0.3pt] (5.2,-0.5) -- ++(0,-2.6);
            \draw[IFATgray2,line width=0.3pt] (5.2+1*0.05,-0.5) -- ++(0,-2.6);
            \draw[IFATgray2,line width=0.3pt] (5.2+2*0.05,-0.5) -- ++(0,-2.6);
            \draw[IFATgray2,line width=0.3pt] (5.2+3*0.05,-0.5) -- ++(0,-2.6);
            
         \draw[<-,IFATgray2]  (zn0.west) --++(-0.5,0);
         \draw[<-,IFATgray2]  (zn1.west) --++(-0.5,0);
         \draw[<-,IFATgray2]  (zn2.west) --++(-0.5,0);
         \draw[<-,IFATgray2]  (zn3.west) --++(-0.5,0);
         \draw[<-,IFATgray2]  (zn4.west) --++(-0.5,0);

         \draw[->,IFATgray2]  (zn0.east) --++(1.15,0);
         \draw[->,IFATgray2]  (zn1.east) --++(1.15,0);
         \draw[->,IFATgray2]  (zn2.east) --++(1.15,0);
         \draw[->,IFATgray2]  (zn3.east) --++(1.15,0);
         \draw[->,IFATgray2]  (zn4.east) --++(1.15,0);
        
            \draw[IFATgray2,line width=0.3pt] (7.3,-0.5) -- ++(0,-2.6);
            \draw[IFATgray2,line width=0.3pt] (7.3+1*0.05,-0.5) -- ++(0,-2.6);
            \draw[IFATgray2,line width=0.3pt] (7.3+2*0.05,-0.5) -- ++(0,-2.6);
            \draw[IFATgray2,line width=0.3pt] (7.3+3*0.05,-0.5) -- ++(0,-2.6);
            
         \draw[<-,IFATgray2]  (zN1.west) --++(-0.5,0);
         \draw[<-,IFATgray2]  (zN2.west) --++(-0.5,0);
         \draw[<-,IFATgray2]  (zN3.west) --++(-0.5,0);

         \draw[->,IFATgray2]  (zN1.east) -- (zNN1.west);
         \draw[->,IFATgray2]  (zN2.east) -- (zNN2.west);
         \draw[->,IFATgray2]  (zN3.east) -- (zNN3.west);

    \end{tikzpicture}};

    \node (T3) at (0,-3) [rounded corners=2pt, draw=MPItext] 
    {\begin{tikzpicture}[>=latex]
    
            \draw[IFATgray2,line width=0.3pt] (0,0.2) -- ++(0,-0.6);
            \draw[IFATgray2,line width=0.3pt] (0+1*0.05,0.2) -- ++(0,-0.6);
            \draw[IFATgray2,line width=0.3pt] (0+2*0.05,0.2) -- ++(0,-0.6);
            \draw[IFATgray2,line width=0.3pt] (0+3*0.05,0.2) -- ++(0,-0.6);
            
            \node (leg1) at (0.75,0) [anchor=west] 
            {\begin{minipage}{3.5cm}\footnotesize  affine function using \\ weight matrix \& bias \end{minipage}};
            
            \node (n) at (4.5,0) [anchor=north west,circle, inner sep=2.5pt, fill=IFATblue2,draw] {};
            
            \node (leg2) at (5.5,0) [anchor=west,] 
            {\begin{minipage}{2.5cm}\footnotesize  activation function of a neuron \end{minipage}};
            
    \end{tikzpicture}};
    
\end{tikzpicture}

%% file: images_tot/RNN_new2.tex
\tikzset{every picture/.style={line width=0.75pt}} 

\begin{tikzpicture}[node distance=0.6cm]

    \node (T1) at (0,5) [draw=none] 
    {\begin{tikzpicture}[>=latex]

        \node (E0) at (0.5,0) {};  
        
        \node at (0.5,1.1) [] {
        \begin{minipage}{2cm}\footnotesize\bfseries\centering 
        input layer\\
        \end{minipage}};     
        \node at (3,1.1) [] {
        \begin{minipage}{2cm}\footnotesize\bfseries\centering 
        RNN cell\\
        "hidden layer"
        \end{minipage}};
        \node at (5.5,1.1) [] {
        \begin{minipage}{2cm}\footnotesize\bfseries\centering 
        output layer\\
        \end{minipage}};
        
        \node (ED1) at (0.5,0.0) [] {\footnotesize $\mathcal{F} \ni f$ };
        \node (ED2) at (2.9,0.0) [] {\footnotesize $h$ };
        \node (ED22) at (4.4,0) [] {\footnotesize $\bar{\ell}$ };
        \node (ED3) at (5.8,0.0) [] {\footnotesize $\ell \in\mathcal{H}$ };
        
        \draw [->] (ED1.east) -- node[above] {\footnotesize $\tilde{\sigma}_{\mathrm{h}, w_\mathrm{h}}$ }  (ED2.west);
        \draw ([yshift=-2pt]ED1.east) -- ([yshift=2pt]ED1.east);
        
        \draw [->] (ED2.east) -- node[above] {\footnotesize $ \tilde{\sigma}_{\mathrm{l}, w_\mathrm{l}}$ }  (ED22.west);
        \draw ([yshift=-2pt]ED2.east) -- ([yshift=2pt]ED2.east);
        
        \draw [->] (ED22.east) -- node[above] {\footnotesize $ \mathrm{id}$ }  (ED3.west);
        \draw ([yshift=-2pt]ED22.east) -- ([yshift=2pt]ED22.east);
        
        %

        \node (z01) [below of=E0, yshift=-0.0cm, circle, inner sep=2.5pt, fill=white,draw] {};
        \node (z02) [below of=z01] {$\vdots$};
        \node (z02Z) [left of=z02,xshift=0.1cm] {\footnotesize $\bar{f}$};
        \node (z03) [below of=z02, circle, inner sep=2.5pt, fill=white,draw] {};
        \node (z04) [below of=z03, circle, inner sep=2.5pt, fill=IFATgray2,draw] {};
        \node (z05) [below of=z04] {$\vdots$};
        \node (z05Z) [left of=z05,xshift=0.1cm] {\footnotesize $h_l$};
        \node (z06) [below of=z05, circle, inner sep=2.5pt, fill=IFATgray2,draw] {};
        
        \node (E1) [right of=E0, xshift=1.5cm] {};  
        \draw[rounded corners=2pt, draw, inner sep=2pt, fill=IFATblue3!50] (1.45, -0.3) rectangle ++(3.05, -3.7);
        \node (z11) [below of=E1, yshift=-1.8cm, circle, inner sep=2.5pt, fill=IFATblue2,draw] {};
        \node (z12) [below of=z11] {$\vdots$}; 
        \node (z13) [below of=z12, circle, inner sep=2.5pt, fill=IFATblue2,draw] {}; 

        \node (E2) [right of=E1, xshift=0.85cm] {};  
        \node (z21) [below of=E2, circle, inner sep=2.5pt, fill=IFATblue2,draw] {}; %
        \node (z22) [below of=z21] {$\vdots$}; 
        \node (z23) [below of=z22, circle, inner sep=2.5pt, fill=IFATblue2,draw] {}; 

        \node (E3) [right of=E2, xshift=0.8cm] {};  
        \node (z31) [below of=E3, yshift=-0.0cm, circle, inner sep=2.5pt, fill=white,draw] {};
        \node (z32) [below of=z31] {$\vdots$};
        \node (z32Z) [right of=z32,xshift=-0.4cm,anchor=west] {\footnotesize $\bar{\ell}$};
        \node (z33) [below of=z32, circle, inner sep=2.5pt, fill=white,draw] {};
        \node (z34) [below of=z33, circle, inner sep=2.5pt, fill=IFATgray2,draw] {};
        \node (z35) [below of=z34] {$\vdots$};
        \node (z35Z) [right of=z35,xshift=-0.4cm,anchor=west] {\footnotesize $h_{l+1}$};
        \node (z36) [below of=z35, circle, inner sep=2.5pt, fill=IFATgray2,draw] {};
        
         \draw[->,IFATgray2] (z01.east) -- ++(1.15,0);
         \draw[->,IFATgray2] (z03.east) -- ++(1.15,0);
         \draw[->,IFATgray2] (z04.east) -- ++(1.15,0);
         \draw[->,IFATgray2] (z06.east) -- ++(1.15,0);
            \draw[IFATgray2,line width=0.3pt] (1.8,-0.5) -- ++(0,-3.2);
            \draw[IFATgray2,line width=0.3pt] (1.8+1*0.05,-0.5) -- ++(0,-3.2);
            \draw[IFATgray2,line width=0.3pt] (1.8+2*0.05,-0.5) -- ++(0,-3.2);
            \draw[IFATgray2,line width=0.3pt] (1.8+3*0.05,-0.5) -- ++(0,-3.2);
            
         \draw[<-,IFATgray2]  (z11.west) --++(-0.5,0);
         \draw[<-,IFATgray2]  (z13.west) --++(-0.5,0);

         \draw[->,IFATgray2] (z11.east) -- ++(0.5,0);
         \draw[->,IFATgray2] (z13.east) -- ++(0.5,0);
         \draw[<-,IFATgray2] (z21.west) -- ++(-0.5,0);
         \draw[<-,IFATgray2] (z23.west) -- ++(-0.5,0);
            \draw[IFATgray2,line width=0.3pt] (3.25,-0.5) -- ++(0,-3.2);
            \draw[IFATgray2,line width=0.3pt] (3.25+1*0.05,-0.5) -- ++(0,-3.2);
            \draw[IFATgray2,line width=0.3pt] (3.25+2*0.05,-0.5) -- ++(0,-3.2);
            \draw[IFATgray2,line width=0.3pt] (3.25+3*0.05,-0.5) -- ++(0,-3.2);

        \draw[->,IFATgray2] (z21.east) -- (z31.west);
        \draw[->,IFATgray2] (z23.east) -- (z33.west);
        \draw[<-,IFATgray2] (z34.west) -- ++(-1.9,0);
        \draw[<-,IFATgray2] (z36.west) -- ++(-1.9,0);

        \draw[->,IFATgray2] (z34.east) 
        -- ++(0.9,0) -- ++ (0,-2) -- ++ (-6.8,0) -- ++ (0,2) -- (z04.west);
        \draw[->,IFATgray2] (z36.east) 
        -- ++(0.3,0) -- ++ (0,-0.6) -- ++ (-6.5+0.6,0) -- ++ (0,0.6) -- (z06.west);


    \end{tikzpicture}};

    \node (T3) at (0,1) [rounded corners=2pt, draw=MPItext] 
    {\begin{tikzpicture}[>=latex]
    
            \draw[IFATgray2,line width=0.3pt] (0,0.2) -- ++(0,-0.6);
            \draw[IFATgray2,line width=0.3pt] (0+1*0.05,0.2) -- ++(0,-0.6);
            \draw[IFATgray2,line width=0.3pt] (0+2*0.05,0.2) -- ++(0,-0.6);
            \draw[IFATgray2,line width=0.3pt] (0+3*0.05,0.2) -- ++(0,-0.6);
            
            \node (leg1) at (0.75,0) [anchor=west] 
            {\begin{minipage}{3.5cm}\footnotesize  affine function using \\ weight matrix \& bias \end{minipage}};
            
            \node (n) at (4.5,0) [anchor=north west,circle, inner sep=2.5pt, fill=IFATblue2,draw] {};
            
            \node (leg2) at (5.5,0)  [anchor=west,] 
            {\begin{minipage}{2.5cm}\footnotesize  activation function of a neuron \end{minipage}};
            
    \end{tikzpicture}};


        
        
        
        

    

\end{tikzpicture}

%% file: Sections/4_ML_within_the_model.tex
The increased availability of plant data as a result of cheaper data collection and storage combined with faster model development has increased the application of machine learning in the (bio)chemical industries \cite{MdNor2020}.
Machine learning has been applied to various fields including energy, chemicals, petrochemicals, oil and gas, polymer, pharmaceuticals, food, beverage, biotechnology, mineral industry, and water \cite{Panerati2019, Pirdashti2013}.

In this section, we outline how machine learning can be used to enable or enhance process monitoring and control in these fields via providing models of the system to the estimators or controllers. 
Hence, data is used in the \MLterm to develop a plant model \eqref{eq:tot_equation} using data-based and hybrid modeling techniques.
We outline how machine learning models of a system can be used for monitoring in Subsection~\ref{subsec:MLapplication_monitoring} and for control in Subsection~\ref{subsec:ML_process_control}.

\subsection{Learning System Models for Process Monitoring}
\label{subsec:MLapplication_monitoring} 
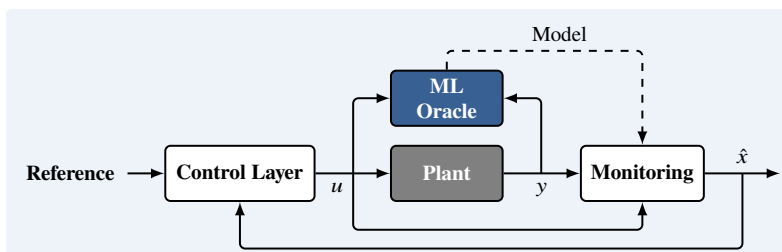
\begin{figure}[hbtp]
            \centering
            \input{images_tot/Section1_4_1}
            \caption{Learning System Models for Process Monitoring.}
            \label{fig:Learning_System_Models_for_Process_Monitoring}
        \end{figure}

    This subsection provides applications of data-based plant models to support plant monitoring (cf.~Figure~\ref{fig:Learning_System_Models_for_Process_Monitoring}) and is composed of offline applications, covering dimensionality reduction, and parameter estimation, as well as online applications, covering state estimation and fault detection. 
    
    \subsubsection{Dimensionality reduction with machine learning}
        High fidelity plant models are valuable for engineers as they improve the process understanding, enable plant design, monitoring, and control \cite{Pantelides2013, Seborg2011, Subramanian2021}.
        To obtain accurate models, first principle models have been developed for complex processes that capture the dynamics, hydraulics, physicochemical properties, or thermodynamics \cite{Pantelides2013, Sansana2021, Subramanian2021}.
        This first principle modeling involves solving nonlinear differential equations, numerical integration, parameter fitting, and root finding methods \cite{Sansana2021, Subramanian2021}.
        Since these operations can be computationally challenging, overly complex models might not be suited for real time applications, optimal process design, or control due to their complexity \cite{Pantelides2013}.
        Machine learning algorithms have been used to tackle this challenge by learning reduced order models for decreased computational offline and online demands.
        For example, governing mechanisms have been identified using machine learning techniques to obtain low dimensional models of the essential system dynamics \cite{Subramanian2021}.
        
        However, the use of machine learning models, instead of first principles models, does not necessarily lead to increased computational speed or model order reduction.
        Training these machine learning models on large data sets can require significant computational resources.
        Hence, sampling design methods used for reduced order model development have been developed, see for instance \cite{Nentwich2019} for more information.
        In \cite{Nentwich2019} Gaussian process and neural network surrogate models were developed using data generated by a complex thermodynamic model for liquid-liquid equilibrium phase prediction.  Both reduced order models provided accurate predictions 36 times faster than the complex thermodynamic model.
        In \cite{Raissi2018} a hybrid reduced order model for fluid dynamics was developed to speed up the prediction of both velocity and pressure fields for external and internal flow problems.  It was noted in \cite{Raissi2018} that the approach has the benefit of providing accurate predictions regardless of the initial and boundary conditions.  This implies that the model is flexible and can be used on the domain of interest, decreasing the amount of calculation and data required.  The integration of first principles into the reduced order model enabled its application to model blood flow in an intracranial aneurysm, where the model was able to predict the velocity and pressure fields accurately.
        
        \begin{remark}
            In general, direct data-based machine learning methods suffer more from computational complexity for big data sets compared to indirect data-based algorithms such as neural networks which come with reduced computational effort required during prediction.
            For instance, the computational complexity of Gaussian processes scales cubically with the size of the data set which makes online applications difficult \cite{Fezai2020}.
            However, many algorithms such as sparse Gaussian processes and data update schemes have been developed to circumvent this drawback. For example, \cite{Fezai2020} developed a Gaussian regression process method that uses only a subset of the data set for offline training. Online training is implemented on new data to reduce the computational complexity and ensure sufficient accuracy.
        \end{remark}
        
    \subsubsection{Parameter and state estimation with machine learning} \label{subsec:Parameter_State_Estimation_ML}
        Certain phenomena, mechanisms, and physicochemical properties in (bio)chemical processes are sometimes unknown or only partly known \cite{Bhutani2006, Chai2014, Pirdashti2013, Zendehboudi2018}.
        Some chemical properties cannot be measured online and require laboratory analysis, which suffers from large time-delays \cite{Shang2019}.
        The time delays for online analyzers, such as gas chromatography, poses significant challenges for accurate online monitoring and control \cite{Kano2009, MohdAli2015}. 
        Moreover, industrial analyzers are expensive to acquire and maintain as some might not be able to be calibrated online \cite{Ge2011, Himmelblau2008, Shang2019}.
        
        These challenges can be addressed using the plant model as an estimator, also referred to as a soft sensor \cite{Ge2011}.
        Machine learning has gained popularity in the application as estimators due to their robustness, simplicity, and flexibility.
        Different artificial neural network structures can be used for estimation, each with their own advantages and limitations, as summarized in \cite{MohdAli2015}.
        A guideline for estimator development and examples of data-based and hybrid estimators based on previous literature work, are also summarized in \cite{MohdAli2015}.
        The objective is to estimate the unmeasurable variables using the measurable variables for online monitoring, control, and process optimization \cite{Ge2011, Shang2019}.
        The estimator can be used to estimate the unknown parameters that are required in the plant model, referred to as parameter estimation.  If the parameters remain unchanged throughout the operating window, they can be estimated offline.  
        For example, a Gaussian process regression model with the radial basis function kernel was used in \cite{Bishnoi2021} for offline parameter estimation of nine different properties for glass compositions using 37 different chemical components that can be used to form the glass composition.
        Online parameter estimation is required when the parameters are dependent on the current state of the process.
        In \cite{Gharagheizi2011} the surface tension parameter for pure compounds, using a single layer feed-forward neural network, was estimated.  
        The model was capable of estimating the surface tension for 752 different pure compounds at various temperatures using the information about the chemical functional groups present in the pure compound.  
        
        Machine learned estimators can also be used to estimate the unmeasured states such as concentrations, heat flux, reaction rates, growth kinetics, battery life cycle \cite{attia2020closed, MohdAli2015, severson2019data}.
        For state estimation, the estimator is trained offline and applied online due to their dependency on the current plant state.
        The estimator's internal parameters can be updated by applying additional training using online data to improve the accuracy \cite{MohdAli2015}.
        For instance, a recurrent neural network was used in \cite{Nikolaou1993} to model a nonisothermal continuously stirred tank reactor (CSTR) that acted as an open-loop observer where the process states were fed back to a linear controller.  Local stability of the observer and the controller were obtained.  The proposed nonlinear controller had a better performance compared to an optimally tuned linear controller \cite{Nikolaou1993}.
        Another example is \cite{Psichogios1992}, where hybrid modeling was used to estimate the states of a fedbatch bioreactor.  A neural network was used to model the growth rate kinetics that was used in the first principle model to estimate the states of the bioreactor at the next sampling time.  Training of the machine learning model's internal parameters required the use of sensitivity equations that were derived from the objective function.  
        The same method was used in \cite{Galvanauskas2006, Georgieva2007} on a crystallization process where the machine learning model was used to model the kinetic parameters required in the first principle models to estimate the online process states.  A model predictive controller was further applied in \cite{Georgieva2007} to control the process using the hybrid state estimator, which is discussed in  \ref{subsec:MPC_ML_Plant_Model}.
        
    \subsubsection{Fault detection with machine learning}
        Industrial sensors can become faulty due to sensor drift, blockages, or damage to the sensor resulting in problems for accurate process control \cite{Kano2009}.
        A small malfunction can have a large effect on a manufacturing processes that can be prevented by early detection \cite{MdNor2020, Shang2019}.
        Hence, fault detection and diagnosis are important components in plant monitoring.
        They can ensure safe process operation, reliability, product quality, and decrease the costs from production loss \cite{Fezai2020, MdNor2020, Shang2019}.
        Often, fault detection models are used to classify an unmeasured fault based on the measured variables of the process \cite{MdNor2020}.
        These fault classification models can be based purely on first principles, data, or a hybrid combination of the two \cite{MdNor2020}.
        
        In hybrid and data based fault detection, the training dataset requires classification  and labeling of the fault which can be challenging as some faults might be undiagnosed \cite{Shohei2020}.
        Labeling the data requires manual analysis, which can increase model cost and development time \cite{Willard2003}.
        Data can be generated using an accurate plant model or plant simulation for normal as well as faulty conditions \cite{Fezai2020, Shohei2020}.
        The internal parameters of the machine learning model can be updated using online plant data, if required \cite{Fezai2020}.
        A general guide to fault detector development for chemical processes as well as various applications are presented in \cite{MdNor2020}, where issues regarding high-dimensionality, nonlinearity, complexity, and fault types are discussed.
        
        In \cite{Shohei2020} a convolutional neural network was used to predict faults in a heating, ventilation, and air conditioning (HVAC) system by transforming the data into an image space.  The convolutional neural network was trained using simulated data and evaluated on plant data.  The model was able to accurately detect the faults on the plant data.  This training approach can save model development and data cost while ensuring a high model accuracy.
        Gaussian process regression has been applied to fault detection problems while providing the advantages of only requiring a relatively small dataset, approximating the system uncertainty, and providing confidence intervals for its predictions \cite{Fezai2020}.
        A Gaussian process regression model was developed in \cite{Fezai2020} for the benchmark Tennessee Eastman chemical process with the objective to reduce the false alarm rate, missed detection rate, and model computational time.  
        
    \subsubsection{Outlook for the use of machine learned plant models for process monitoring}
        Even though the discussed approaches for data-based monitoring of manufacturing systems are promising, some open challenges remain to be solved.
        The increased scale and complexity of industrial chemical processes result in high dimensional datasets that contain some irrelevant and redundant features which adds complexity and reduces the performance \cite{MdNor2020}.
        Machine learning which is capable of feature extraction without any additional knowledge should be favored \cite{MdNor2020}.
        Machine learning based estimators can also be complex to implement, maintain, and to troubleshoot.  
        Hence, simpler designs are required along with more tests on pilot and industrial plants to persuade industry to implement these techniques \cite{MohdAli2015}. 
        Also, soft sensor performance can degrade over time due to the time-varying characteristics of industrial processes \cite{Shang2019}.
        Adaptive training methods that are capable of online learning and calibration to tackle this issue require further investigation \cite{Shang2019, MdNor2020, MohdAli2015}.
        Furthermore, inaccuracies in laboratory data caused by uncertain time delays, varying sampling intervals, and sampling habits should be taken into account in order to improve the quality of the training data \cite{Shang2019}. 
        Hence, robust approaches should be developed. 
        At the same time, focus should be given to provide better explainable and interpretable models by incorporating physics and domain knowledge \cite{MohdAli2015, Venkatasubramanian2019}.
    
        In particular for manufacturing systems, 
        machine learning methods capable of detecting multiple faults occurring in the same time window should be researched to account for process interactions with other systems \cite{MdNor2020}.
        More research into modeling transitional processes with machine learning, such as start-ups or shutdowns, where faults are most likely to occur is required \cite{MdNor2020}.
        Finally, the incorporation of estimators with process control strategies should also be investigated to improve the overall estimation and control performance in real time \cite{MohdAli2015}.

\subsection{Learning System Models for Process Control} \label{subsec:ML_process_control} 
    Process control is used to drive the process to a reference state in a safe and efficient way by manipulating the inputs to the process \cite{Seborg2011}.
    Traditionally, the control structure can be organized in a hierarchy, where  the process control activities are grouped into different layers operating  in different time scales and with different goals (cf.~Figure~\ref{fig:manufacturing_structure}). 
    
\begin{figure}[hbtp]
        \centering
        \input{images_tot/Section1_4_2}
        \caption{Learning System Models for Process Control.}
        \label{fig:Learning_System_Models_for_Process_Control}
    \end{figure}
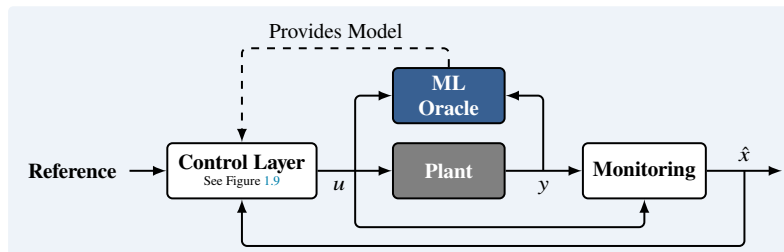

    Model-based control methods can be used in the lower control layer (e.g. linear quadratic regulators), in the supervisory control layer (e.g. model predictive control), and in the upper control layer (e.g. real-time optimization). These layers uses plant models \eqref{eq:tot_equation} of different granularity  (e.g. static models in RTO and dynamic models in MPC) directly in the design of the controller.
    The model accuracy is important since it greatly influences the controller performance. 
    Obtaining an accurate plant model can be difficult for industrial processes due to nonlinearities and changes in the process characteristics \cite{Kano2009}.
    Machine learning can be used to increase the accuracy of plant models using process data, as discussed in Subsection~\ref{subsec:Plant_Modelling}. 
    This subsection discusses the applications of using machine learned plant models in model-based control methods such as adaptive control, model predictive control, and inverse model-based control \cite{Hussain1999, Lee2009}.
    Hence, we do not train a machine learning algorithm to replace or approximate an existing controller (see Section~\ref{sec:ML_Controller}), but utilize machine learning for process modeling and supply the gathered information to the model-based control techniques (cf.~Figure~\ref{fig:Learning_System_Models_for_Process_Control}).
    
    \subsubsection{Adaptive control using machine learning plant models}
        Controller parameters can require adaptation throughout the plant operation to compensate for changes in the process dynamics, process specifications, or in the environment while achieving and maintaining the desired performance \cite{Landau2003, Seborg2011}. These changes are, for example, heat exchanger fouling or catalyst deactivation, frequent or large disturbances in feed quality or composition, changes in product specifications such as quality changes, and inherent nonlinear behavior \cite{Mears2017, Seborg2011}.
        Adaptive control can be used to directly adapt the control parameters when the process deviate from the desired behavior.  A model-based control design can also be used to adapt the model parameters online which in turn are used to adapt the control law (indirect adaptive control).

        \begin{figure}[btp]
            \centering
            \input{images_tot/Section1_4_2_Control_Layer}
            \caption{Application of the System Model for Process Control.}
            \label{fig:System_Model_used_for_Process_Control}
        \end{figure}
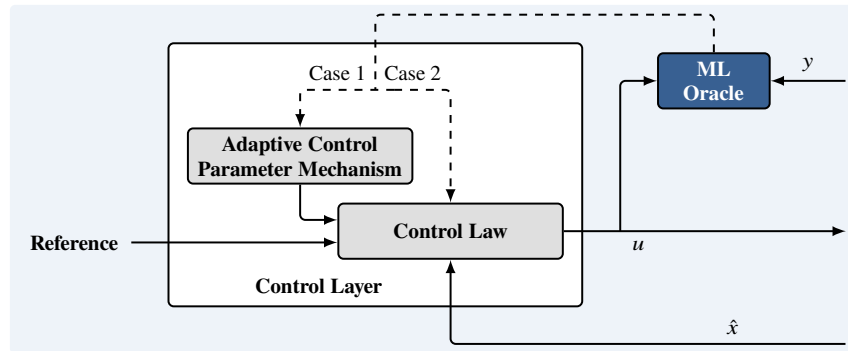
        
        Indirect adaptive control is also referred to as self-tuning control. 
        The controller parameters are indirectly adapted online using updated plant models \cite{Hussain1999, Landau2003, Seborg2011}.
        Here, the \MLterm is used to design and adapt the plant model online. 
        Based on these online learned/adapted models different adjustments of controllers can be performed. 
        For example, the design/adaptation of PID control gains is possible based on machine learning models (cf.~Figure~\ref{fig:System_Model_used_for_Process_Control} Case 1). Also, the updated model can be used in model-based controller such as LQR or MPC to obtain optimal inputs (cf.~Figure~\ref{fig:System_Model_used_for_Process_Control} Case 2). 
        In principle, also the parameters of an LQR or MPC (such as the weights in the cost function) could be adapted based on machine learning models regardless of the update of the prediction model. 
        
        In \cite{Parlos2001} an indirect adaptive control structure was used to control a steam \MLterm. A recurrent neural network was used to generate the plant model.  A weight adaption law was used to adjusted the weights of the recurrent neural network, using new observations as they became available, in order to enhance the predictive capabilities of the plant model.  A gain adaption law was used to adjust the PID controller gains online to ensure sufficient controller performance over the entire operating region.  The advantage of this method is that it can be implemented in the industry using the existing control hardware, thereby requiring little capital investment.
        Applications of neural networks in adaptive control for (bio)chemical processes are summarized in \cite{Hussain1999, Mears2017}.
        
        
    \subsubsection{Model predictive control using machine learning plant models} \label{subsec:MPC_ML_Plant_Model} 
        Model predictive control (MPC) is a model-based optimal control method that repeatedly solves an optimal control problem. It uses the model of the plant to find an optimal input trajectory by predicting the system into the future. Two main advantages are: It can be used for multi-inputs multi-outputs systems and it can guarantee satisfaction of input and state constraints \cite{Kano2009, Seborg2011, Zhang2019}.
        For instance, MPC is widely used in distillation and reaction processes in (bio)chemical industries as it allows to optimize performance while operating close to the process limits \cite{Hussain1999, Kano2009, Mears2017, Mowbray2021}.
        However, the performance of the MPC control method depends on the accuracy of the plant model \cite{Seborg2011, Zhang2019}.
        Hence, modeling and model maintenance are critical to prevent deterioration and to keep MPC controllers operational \cite{Saltık2018}.
        Solving the optimization problem using complex plant models can be computationally demanding and has been a challenge for industrial implementation.  The plant model should be able to describe the true process unknowns to prevent unrealistic state predictions but should also be simple enough to find the solution of the control problem online \cite{Saltık2018}.

        For this reason, often linear plant models are used in MPC in industrial applications as this simplifies the model development as well as online optimization \cite{Pantelides2013}.
        However, nonlinear model predictive control would allow to operate the system in wider operating ranges, as it takes the nonlinearities of the underlying processes into account \cite{Hussain1999}. 
        This is especially important in batch processes and dynamically operated plants with changing process conditions.
        Still, \cite{Kano2009} found that nonlinear MPC has not been widely implemented due to difficulties in developing nonlinear process models.
        
        The development of such nonlinear models can be simplified using machine learning. Machine learning has been successfully used to build the plant model required by MPC, see Table~\ref{tab:my_label} and \cite{Seborg2011}.
        Here, the \MLterm is used to obtain the plant model \eqref{eq:tot_equation} for MPC.
        This plant model, which is in general nonlinear, is used inside the MPC for prediction (cf.~Figure~\ref{fig:System_Model_used_for_Process_Control} Case 2) \cite{Hussain1999}.
        Thus, the machine learning model enters the optimization problem as an equality constraint similar to classical optimal control approaches \cite{Hussain1999}.
        In contrast to that, Section~\ref{sec:ML_Controller} tackles the case when the MPC feedback law is approximated using machine leaning techniques, e.g. to improve computational speed especially for large dimensional problems \cite{Kumar2021}.
        
        The performance of a machine learned plant model in MPC was studied by \cite{Macmurray1995} who compared the MPC performance using a simplified first principle model and a recurrent neural network model on a simulated packed distillation column. The recurrent neural network plant model was trained offline using data generated from a complex first principle dynamic model.  Similar performances between the two methods have been obtained by \cite{Macmurray1995}, concluding that machine learning techniques are adequate to develop dynamic plant models for complex or unknown processes with sufficient performances.
        Another example is crystallization processes, which have complex nucleation, particle growth, and agglomeration mechanisms \cite{Georgieva2007}.  These mechanisms can be modeled using machine learning methods combined with first principle models to obtain accurate hybrid plant models, as discussed in Subsection~\ref{subsec:PIML}.
        These hybrid crystallization plant models can then be further used in model-based control such as MPC, as demonstrated by \cite{Georgieva2007} on a sugar crystallization process.
        The difference between using a first principle model and a hybrid plant model in a Lyapunov-based MPC for a simulated continuously stirred tank reactor was investigated by \cite{Zhang2019RTO}. A neural network model was used to model the nonlinear reaction rate in the hybrid plant model.  Similar performances between the first principles and hybrid models were obtained.  However, the hybrid model in practice is more effective due to the challenge of determining the reaction rate of a process based on first principles.
        Different applications of neural networks with MPC for chemical processes are summarized by \cite{Hussain1999} and for biochemical processes by \cite{Mears2017, Mowbray2021}.
        Robust and stochastic MPC methods for uncertainty systems have been reviewed by \cite{Saltık2018}.
    
    \subsubsection{Inverse control using machine learning plant models}
        Inverse model-based methods 
        directly model the inverse process dynamics to predict the desired control actions over a prediction horizon to drive the process to the desired reference state (cf.~Figure~\ref{fig:System_Model_used_for_Process_Control} Case 2) \cite{Thitiyasook2007}.
        The control performance is dependent on the accuracy of these inverse models. Machine learning can be used to obtain the inverse models and relies on the amount and accuracy of the data available to train  \cite{Hussain2014, Thitiyasook2007}.
        Direct inverse control and internal model control are two common types of inverse control methods \cite{Hussain1999, Hussain2014, Ramli2016}.
        The inverse model acts as the controller that is in cascade with the system under control.
        If the inverse model counteracts the system dynamics perfectly, no feedback is needed and the output of the system perfectly follows the desired reference.
        However, additional feedback controllers such as PID controllers are often combined with direct inverse control. 
        The desired outputs (setpoints) along with the past and current process inputs and outputs are then fed into the inverse model that predicts the appropriate control inputs \cite{Hussain1999, Ramli2016}.
        
        A direct inverse control structure, using neural networks, has been combined with a conventional proportional-integral (PI) controller in \cite{Thitiyasook2007} to control the concentration and pH in a steel pickling process.  The PI controller was used to lower setpoint offsets and prevent oscillations of the control variables.  The proposed combined strategy was compared with their respective individual methods for a nominal, disturbance, and model mismatch case.  The PI controller failed to reach the setpoint in the desired time for both the disturbance and model mismatch case while the direct inverse control method was capable of stabilizing the process with a setpoint offset for all three cases.  The combined strategy was capable of tighter control with no setpoint offset.
        In \cite{Lee2009} a recurrent neural network was developed to model the inverse dynamics of the large-scale pulverized coal power plant.  A PID feedback controller was used to eliminate the steady-state error due to model mismatch and disturbances.  They were able to shorten the stabilization time for the process and eliminate the steady-state error (setpoint offset).
        Hybrid modeling was applied in \cite{Andrasik2004} with a direct inverse model-based strategy. The different error components for a PID controller was used as inputs to the neural network along with the states. Better regulatory and tracking performances were achieved with the proposed method compared to normal direct inverse control.
        
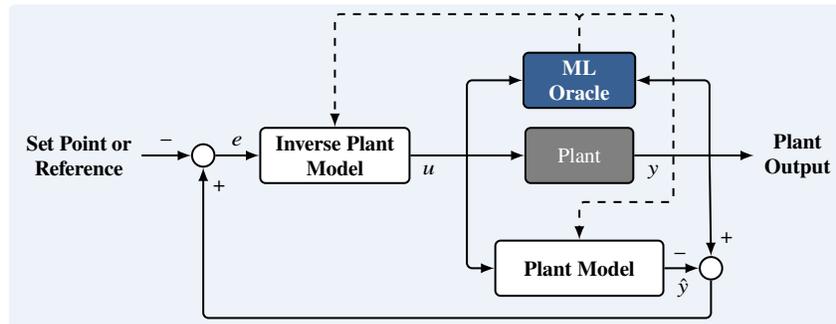
\begin{figure}[hbtp]
            \centering
            \input{images_tot/Inverse_Control_Structure}
            \caption{Internal model control structure.}
            \label{fig:Inverse_Control_Structure}
        \end{figure}

        Internal model control adds a plant model to the direct inverse control strategy to improve the robustness of the controller, (cf.~Figure~\ref{fig:Inverse_Control_Structure})  \cite{Hosen2021, Hussain1999, Thitiyasook2007, Ramli2016}.
        The plant model accounts for mismatches and irregularities that can occur due to equipment degradation and noise \cite{Hussain1999}.
        The inverse model acts as the controller with the plant model minimizing the setpoint offset, even in the presence of noise and disturbances \cite{Hosen2021, Ramli2016}.
        Stability, offset-free tracking, and a parameter free controller are the three major advantages using internal model-based controllers \cite{Hosen2021}.
        The inverse model can be obtained by either inverting the plant model numerically or training the model to identify the inverse \cite{Hussain2014}.
        Internal model control is capable to allow trade-offs between performance and robustness unlike direct inverse control \cite{Seborg2011}.
        
        A neural network was used in an internal model-based control strategy in \cite{Hosen2021}.  The setpoint tracking and disturbance rejection ability was improved by combining it with prediction interval based modeling to add tighter upper and lower bounds to the predictions.
        A neural network was used to develop a direct inverse control and an internal model control strategy in \cite{Ramli2016} for a simulated debutanizer distillation column.  The internal model control strategy had a better performance with faster settling times compared to a conventional PID controller and the direct inverse control strategy.
        Different applications of inverse model-based applications using neural networks are summarized in \cite{Hussain1999}.

    \subsubsection{Outlook for the use of machine learned plant models for process control}
    To enable safe, reliable, and precise process control based on machine-learning models, one of the main topics is availability of high quality models. 
    This quality refers to the accuracy of the machine learning model predictions. Besides, the future of process modeling needs to meet a variety of requirements, including explainability/interpretability, ability to handle uncertainty, and ability to use large quantities of data \cite{MdNor2020}.
    To increase profit and flexibility in operation, focus on adaptive mechanisms for prediction models under frequent operating condition deviations are required. Online learning during control execution can be used to do so. However, safety guarantees for these online adaptations still need to be ensured.
    Robust adaptive controllers or controllers that exploit models of the uncertainties seem promising to tackle these challenges (see also Subsection~\ref{subsec:learning_uncertainty}.
        

%% file: images_tot/Section1_4_1.tex
\tikzset{every picture/.style={line width=0.75pt}} 

\begin{tikzpicture}[node distance=1.5cm]
    \node (T2) at (0,0) [rounded corners=2pt, draw=none, fill=IFATblue3!50]
        {\begin{tikzpicture}[>=latex]
            
            \node (R) at (0.75,1.0) [rounded corners=0pt, draw=none, fill=none]
            {\begin{minipage}[c][0.5cm]{1.25cm}\centering\footnotesize
                \textbf{Reference}
            \end{minipage}
            };
            
            \node (C) [right of=R,  xshift=0.75cm, rounded corners=2pt, draw, fill=white]
            {\begin{minipage}[c][0.5cm]{1.75cm}\centering\footnotesize
                \textbf{Control Layer}
            \end{minipage}
            };
            
            \node (P) [right of=C,  xshift=1.25cm, rounded corners=2pt, draw, fill=black!50]
            {\begin{minipage}[c][0.5cm]{1.25cm}\centering\footnotesize
                \textbf{\textcolor{white}{Plant}}
            \end{minipage}
            };
            
            \node (G) [above of=P, yshift=-0.5cm, rounded corners=2pt, draw, fill=CCPSblue2]
            {\begin{minipage}[c][0.5cm]{1.25cm}\centering\footnotesize
                \textbf{\textcolor{white}{ML Oracle}}
            \end{minipage}
            };
            
            \node (M) [right of=P, xshift=1.1cm, rounded corners=2pt, draw, fill=white]
            {\begin{minipage}[c][0.5cm]{1.4cm}\centering\footnotesize
                \textbf{Monitoring}
            \end{minipage}
            };
            
            \draw [->] (R.east) -- (C.west);
            
            \draw [->] (C.east) -- node[anchor=north, xshift=-0.2cm] {\footnotesize $u$} (P.west);
            \draw [->] ([xshift=0.5cm] C.east) |- (G.west);
            
            \draw [->] (P.east) -- node[anchor=north, xshift=0.0cm] {\footnotesize $y$} (M.west);
            
            \draw [->] ([xshift=0.5cm] P.east) |- (G.east);
            \draw [dashed, ->] (G.north) -- ([yshift=0.25cm] G.north) node[anchor=south, xshift=1.5cm] {\footnotesize Model} -| (M.north);
            
            \draw [->] (M.east) -- node[anchor=south, xshift=0.0cm] {\footnotesize $\hat{x}$} ++(1.0cm, 0.0cm);
            \draw [->] ([xshift=0.5cm] M.east) -- ([shift={(0.5cm,-1cm)}] M.east) -| (C.south);
            
            \draw [->] ([xshift=0.5cm] C.east)  -- ([shift={(0.5cm,-0.75cm)}] C.east) -| (M.south);

        \end{tikzpicture}
    };
\end{tikzpicture}


            
            
            
                
            


%% file: images_tot/Section1_4_2.tex
\tikzset{every picture/.style={line width=0.75pt}} 
\newcommand\sectionfourcontrolwidth{1.25cm}

\begin{tikzpicture}[node distance=1.5cm]
    \node (T2) at (0,0) [rounded corners=2pt, draw=none, fill=IFATblue3!50]
        {\begin{tikzpicture}[>=latex]
            
            \node (R) at (0.75,1.0) [rounded corners=0pt, draw=none, fill=none]
            {\begin{minipage}[c][0.5cm]{1.25cm}\centering\footnotesize
                \textbf{Reference}
            \end{minipage}
            };
            
            \node (C) [right of=R,  xshift=0.75cm, rounded corners=2pt, draw, fill=white]
            {\begin{minipage}[c][0.5cm]{1.75cm}\centering\footnotesize
                \textbf{Control Layer}\\
                \tiny{See Figure~\ref{fig:System_Model_used_for_Process_Control}}
            \end{minipage}
            };
            
            \node (P) [right of=C,  xshift=1.25cm, rounded corners=2pt, draw, fill=black!50]
            {\begin{minipage}[c][0.5cm]{1.25cm}\centering\footnotesize
                \textbf{\textcolor{white}{Plant}}
            \end{minipage}
            };
            
            \node (G) [above of=P, yshift=-0.5cm, rounded corners=2pt, draw, fill=CCPSblue2]
            {\begin{minipage}[c][0.5cm]{1.25cm}\centering\footnotesize
                \textbf{\textcolor{white}{ML Oracle}}
            \end{minipage}
            };
                
            \node (M) [right of=P, xshift=1.1cm, rounded corners=2pt, draw, fill=white]
            {\begin{minipage}[c][0.5cm]{1.4cm}\centering\footnotesize
                \textbf{Monitoring}
            \end{minipage}
            };

            \draw [->] (R.east) -- (C.west);
            
            \draw [->] (C.east) -- node[anchor=north, xshift=-0.2cm] {\footnotesize $u$} (P.west);
            \draw [->] ([xshift=0.5cm] C.east) |- (G.west);
            
            \draw [->] (P.east) -- node[anchor=north, xshift=0.0cm] {\footnotesize $y$} (M.west);
            \draw [->] ([xshift=0.5cm] P.east) |- (G.east);
            
            \draw [dashed, ->] (G.north) -- ([yshift=0.25cm] G.north) node[anchor=south, xshift=-1.5cm] {\footnotesize Provides Model} -| (C.north);
            
            \draw [->] (M.east) -- node[anchor=south, xshift=0.0cm] {\footnotesize $\hat{x}$} ++(1.0cm, 0.0cm);
            \draw [->] ([xshift=0.5cm] M.east) -- ([shift={(0.5cm,-1cm)}] M.east) -| (C.south);
            
             \draw [->] ([xshift=0.5cm] C.east)  -- ([shift={(0.5cm,-0.75cm)}] C.east) -| (M.south);
            
        \end{tikzpicture}
    };
\end{tikzpicture}


            
            
                
            


%% file: images_tot/Section1_4_2_Control_Layer.tex
\tikzset{every picture/.style={line width=0.75pt}} 

\begin{tikzpicture}[node distance=1.0cm]
    \node (T2) at (0,0) [rounded corners=2pt, draw=none, fill=IFATblue3!50]
        {\begin{tikzpicture}[>=latex]
            
            \draw[rounded corners=2pt, draw, inner sep=2pt, fill=white] (2.0, 4.0) rectangle ++(5.5, -3.5);
            \node (C) at (4.0,0.75) [rounded corners=0pt, draw=none, fill=none] {\footnotesize \textbf{Control Layer}};
            
            \node (Adapt) at (3.75, 2.5 ) [rounded corners=2pt, draw, fill=CCPSgray2!50]
            {\begin{minipage}[c][0.5cm]{2.75cm}\centering\footnotesize
                \textbf{Adaptive Control Parameter Mechanism}
            \end{minipage}
            };
            
            \node (R) [below of=Adapt, shift={(-3.0cm, -0.15cm)}, rounded corners=0pt, draw=none, fill=none]
            {\begin{minipage}[c][0.5cm]{1.25cm}\centering\footnotesize
                \textbf{Reference}
            \end{minipage}
            };
            
            \node (Law) [below of=Adapt, xshift=2.0cm,  rounded corners=2pt, draw, fill=CCPSgray2!50]
            {\begin{minipage}[c][0.5cm]{2.75cm}\centering\footnotesize
                \textbf{Control Law}
            \end{minipage}
            };
            
            \node (G) [right of=Adapt, shift={(4.5cm, 1.0cm)}, rounded corners=2pt, draw, fill=CCPSblue2]
            {\begin{minipage}[c][0.5cm]{1.25cm}\centering\footnotesize
                \textbf{\textcolor{white}{ML Oracle}}
            \end{minipage}
            };
            
            \draw [->] (Adapt.south) |- ([yshift=0.15cm]Law.west);
            \draw [->] (R.east) -- ([yshift=-0.15cm] Law.west);
            
            \draw[dashed, -] (G.north) -- ([yshift=0.5cm] G.north) -| ([shift=({-4.5cm, -0.5cm})] G.north);
            \draw[dashed, ->] ([shift=({-4.25cm, -0.5cm})] G.north) -| node[anchor=south, xshift=0.5cm] {\footnotesize Case 1} (Adapt.north);
            \draw[dashed, ->] ([shift=({-4.25cm, -0.5cm})] G.north) -| node[anchor=south, xshift=-0.5cm] {\footnotesize Case 2} (Law.north);
            
            \draw[->] (Law.east) node[anchor=north, xshift=1.0cm] {\footnotesize $u$} -- ++ (3.75cm, 0cm);
            \draw[->] ([xshift=0.75cm] Law.east) |- (G.west);
            
            \draw[->] (11.0cm, 3.5cm) node[anchor=south, xshift=-0.5cm] {\footnotesize $y$} -- (G.east);
            \draw[->] (11.0cm, 0.0cm) node[anchor=south, xshift=-1.5cm] {\footnotesize $\hat{x}$} -| (Law.south);
        \end{tikzpicture}
    };
\end{tikzpicture}

%% file: images_tot/Inverse_Control_Structure.tex
\tikzset{every picture/.style={line width=0.75pt}} 

\begin{tikzpicture}[node distance=1.5cm]
    \node (T2) at (0,0) [rounded corners=2pt, draw=none, fill=IFATblue3!50]
    {\begin{tikzpicture}[>=latex]
        
        \node (setpoint) at (0.85, 3) [rounded corners=2pt, draw=none]
        {\begin{minipage}[c][0.5cm]{1.4cm}\centering\footnotesize
            \textbf{Set Point or Reference}
        \end{minipage}
        };
        
        \node (SP_difference) at (2.5, 3) [circle,inner sep=3pt,fill=white,draw] {};
        
        \node (invmodel) [right of=SP_difference, xshift=0.25cm, rounded corners=2pt,draw, fill=white]
        {\begin{minipage}[c][0.5cm]{1.75cm}\centering\footnotesize
            \textbf{Inverse Plant Model}
        \end{minipage}
        };
        
        \node (plant) [right of=invmodel, xshift=1.75cm, rounded corners=2pt,draw, fill=black!50]
        {\begin{minipage}[c][0.5cm]{1.2cm}\centering\footnotesize
            \textcolor{white}{Plant}
        \end{minipage}
        };
        
        \node (G) [above of=plant, yshift=-0.5cm, rounded corners=2pt, draw, fill=CCPSblue2]
        {\begin{minipage}[c][0.5cm]{1.25cm}\centering\footnotesize
            \textbf{\textcolor{white}{ML Oracle}}
        \end{minipage}
        };
        
        \node (output) [right of=plant, xshift=1.4cm, rounded corners=2pt,draw=none]
        {\begin{minipage}[c][0.5cm]{0.9cm}\centering\footnotesize
            \textbf{Plant Output}
        \end{minipage}
        };
    
        \node (plantmodel) [below of=plant, rounded corners=2pt,draw, fill=white]
        {\begin{minipage}[c][0.5cm]{2.0cm}\centering\footnotesize
            \textbf{Plant Model}
        \end{minipage}
        };
    
        \node (output_difference) [right of=plantmodel, xshift=0.25cm, circle,inner sep=3pt,fill=white,draw] {};
    
        \draw [->] (setpoint.east) -- node[anchor=south] {\footnotesize $-$} (SP_difference.west);
        \draw [->] (SP_difference.east) -- node[anchor=south] {\footnotesize $e$} (invmodel.west);
        
        \draw [->] (invmodel.east) node[anchor=north, xshift = 0.25cm] {\footnotesize $u$} -- (plant.west);
        \draw [->] ([xshift=0.75cm] invmodel.east) |- (plantmodel.west);
        \draw [->] ([xshift=0.75cm] invmodel.east) |- (G.west);
        
        \draw [->] (plant.east) node[anchor=north, xshift=0.25cm] {\footnotesize $y$}-- (output.west);
        \draw [->] ([xshift=1.0cm] plant.east) |- (G.east);
        \draw [->] ([xshift=1.0cm] plant.east) -- (output_difference.north) node[anchor=west, yshift=0.25cm] {\footnotesize $+$};
        
        \draw [->, dashed] (G.north) -- ([yshift=0.5cm] G.north) -| (invmodel.north);
        \draw [->, dashed] (G.north) |- ([shift={(1.25cm,0.5cm)}] G.north) |- ([yshift=0.5cm] plantmodel.north) -- (plantmodel.north);
        
        \draw [->] (plantmodel.east) node[anchor=north, xshift=0.25cm] {\footnotesize $\hat{y}$} -- (output_difference.west) node[anchor=south, xshift=-0.25cm] {\footnotesize $-$};
        
        \draw [->] (output_difference.south) -- ([yshift=-0.5cm] output_difference.south) -| (SP_difference.south) node[anchor=west, yshift=-0.25cm] {\footnotesize $+$};
        
    \end{tikzpicture}};
\end{tikzpicture}

%% file: Sections/5_ML_for_the_control_law.tex
In the previous section, we have showed some possible applications of ML models for estimation, prediction, fault diagnosis, and model-based control. These mainly focus on modeling the plant  using machine learning techniques 
cf.~Figures~\ref{fig:Learning_System_Models_for_Process_Monitoring} and \ref{fig:Learning_System_Models_for_Process_Control}.
In this section, we show that machine learning can also be applied in the control layer of the manufacturing process, which can be used to either improve the control performance of the main baseline controller (for example, an MPC for controlling with a GP for learning the uncertainty), or to replace the baseline controller completely by acting as the main controller of the system.
In Subsection~\ref{subsec:ML_process_control}, the role of machine learning is only to learn the plant model while the control task is achieved by other algorithms. In this section machine learning directly builds or adjusts the control algorithm.
We discuss the adaptation of controllers via machine learning to guarantee safety and constraint satisfaction despite uncertainties as well as the substitution of controllers via machine learning.

\subsection{Learning Uncertainties for Safe Control}
\label{subsec:learning_uncertainty}
There exist different uncertainties  that can effect the control performance of manufacturing systems.  
Uncertainties are present in plant models due to the challenge of accurately modeling real plants.
Moreover, measurement noise and unmodelled external disturbances can limit the control performance.
Significant uncertainties can lead to violations of constraints which, if critical, can cause a fatal failure for the complete system.
The control approach that considers model uncertainty is called \emph{robust} control. Here we are going to focus on robust control approaches that use constraints back-off. This approach is equipped with a representation of the model uncertainty and can provide a certain degree of safety and performance to the plant. Although uncertainty can affect all the components of the control loop, most commonly robust approaches considers model and measurement uncertainties. 
A common approach in constrained control is to build a \emph{nominal} model (uncertainty free) and then back-off the constraints to such an extent that, if the \emph{nominal} model does not violate the shrunk constraints, the real (uncertain) system  will not violate the original constraints. 
Gaussian processes can be used to model the uncertainty, since they naturally provide not only the learned function but also its probability distribution (cf. \ref{sec:gaussian_process}). This information can be used to back-off the constraints and guarantee constraint satisfaction up to a certain probability. In \cite{Bradford2018, Mowbray2021a} this approach was used for lutein production from microalgae. In \cite{Rafiei2018} a Monte Carlo sampling strategy for uncertainty propagation together with a power series expansion to represent the confidence interval of the process constraints was used for constraint back-off in a water treatment plant example. In \cite{Bradford2020} a stochastic MPC using GPs was used to back of the constraints of a batch reactor.
The previous approach provides probabilistic constraint satisfaction, hence it cannot provide constraint satisfaction guarantees for \emph{all} uncertainty realizations. For this purpose, set-based approached can be used. These assume that the uncertainty lies with uniform probability within a finite set and therefore can provide constraint satisfaction guarantees for any uncertainty realization in that set. Some of these approaches use tube-based MPC, where a so-called \emph{ancillary} feed-back controller is used to maintain the system uncertainty in a (possibly small) bounded set and hence to avoid expansion with the prediction horizon \cite{Aswani2013, BETHGE2018517, Soloperto2018}. Nevertheless, these approaches tend to be very conservative, because they consider also uncertainty events with very low probability. 
Other approaches do not consider constraint back-off, but discourage the controller to visit areas where the uncertainty is large, for example, by defining an uncertainty measure that enters the objective function \cite{Teixeira2006, Yang2015a, Morabito2021}. These are, in general, easier to implement in contrast to the constraint back-off method, but do not guarantee robust constraint satisfaction.

\subsection{Substitute Conventional Controllers with Machine learning }

Machine learning methods can act as the controller in the closed loop system. To design the machine learning controllers, imitation learning, reinforcement learning, or iterative learning control can be used, which we outline in the following. 

\subsubsection{Imitation learning with Neural Networks}

        \begin{figure}[hbtp]
            \centering
            \scalebox{1}{\input{images_tot/51-Imitationlearning}}
            \caption{In imitation learning, the neural network, acting as the controller, is trained from  input-output data of a baseline controller.}
            \label{fig:51Imiationlearning}
        \end{figure}
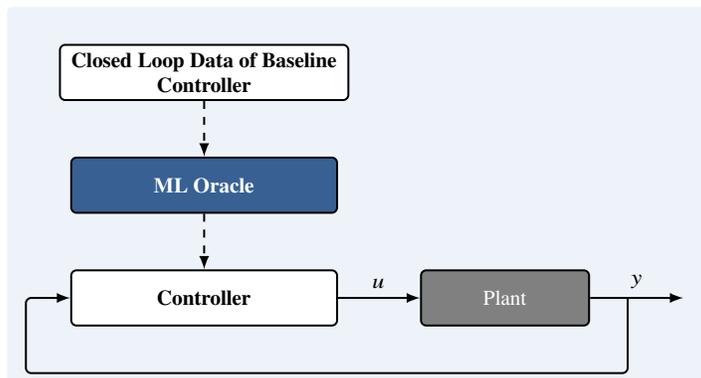

In imitation learning, there exist a baseline controller, which can deliver a desired performance, that is learned and substituted by a machine learning-based controller (see Figure~\ref{fig:51Imiationlearning}). 
This approach can have economical and computational benefits when the learned controller requires less expensive hardware and lower computational power.  
For example, if a model predictive controller is used, a nonlinear optimization problem needs to be solved at each time instant, which is computationally challenging. For linear systems, explicit MPC can help to reduce the computational burden by dividing the state-space into different regions, pre-computing, and storing the control law as an explicit function for each region \cite{Karg2020}. This approach allows rapid and efficient evaluation without solving the optimization problem at each time step. Nevertheless, it suffers storage problems because of the requirement to store all the pre-computed control laws.

A popular approach to solve both problems (fast online computations and large storage demands) is to use a neural network to learn the baseline MPC. There are numerous papers that exploit this approach, for example \cite{Parisini1995, Karg2020, Maddalena2020, Cao2020, Chen2018, Cseko2015, Ponkumar2018}. Other machine learning methods can also be applied.
For example, support vector machines  were used to avoid the problem of storing the explicit MPC for high dimensional systems (\cite{Ankush2017}).
The control inputs and closed-loop data, consisting of the states or outputs of the real or the simulated plant, are generated using the baseline controller. The data is then used to train the machine learning based controller offline.

If we put this into the framework we have mentioned in Section~\ref{sec:Process_setup}, the \MLterm uses the states or measurements as features and the input as labels to learn an explicit control law. Hence, this approach belongs to the category of supervised learning. 
Since the baseline controller needs to be synthesized, at least the nominal mathematical model of the plant is often assumed to be known, which can provide useful information to guarantee safety for the closed-loop system with the machine learning based control.

In general, guaranteeing safety (stability and constraint guarantees) for closed-loop systems with machine learning components is challenging.
However, several approaches have been investigated to guarantee stability for closed-loop systems using neural networks.
The first approach is to exploit the theorems by \cite{Cybenko1989} and \cite{Hornik1991} about the universal approximation ability of multi-layer feed-forward neural networks. They have shown that any continuous mapping over a compact domain can be approximated as accurately as necessary by a feed-forward neural network with one hidden layer. In other words, given any (small) positive number, a neural network with sufficiently large number of nodes can guarantee that the difference between the ground truth function and the approximated function is always smaller than that number for all points in the compact domain. This theorem is combined with small gain theorem to guarantee stability of the closed-loop system in \cite{Akesson2005}.
The second approach exploits the characteristics of the nonlinear activation functions commonly used in deep neural networks. Since these functions often satisfy the sector bound conditions or can be transformed by using loop transform techniques so that the sector bound conditions are satisfied, the stability of the closed-loop system can be reformulated into the form of a \emph{diagonal linear differential inclusion}, which can be guaranteed by available tools from robust control theory such as IQC (Integral quadratic constraint) \cite{Yin2021} or absolute stability of Lure systems \cite{Nguyen2021}.
This approach also allows analyzing the robustness of the system under disturbance and uncertainty.

The work in \cite{Hertneck2018} uses a different approach to obtain the probabilistic guarantees based on Hoeffding's Inequality, where the neural network is used to learn a robust MPC subject to inaccurate inputs within given bounds. The neural network is used to learn this controller via offline samples. Finally, the approximated neural network controller is validated statistically by using Hoeffding's Inequality to obtain a bound on error between the baseline MPC and the learned one. 
More recently, by extending the work of \cite{Hertneck2018}, the authors in \cite{Kumar2021} propose a new neural network architecture to obtain offset-free closed-loop performance for industrial large-scale linear MPC, which are challenging for online QP solvers. Furthermore, the aim of this paper also focuses on setpoint tracking MPC, which is relevant to the manufacturing industry. The input-to-state of the system is guaranteed by using the available results in the MPC literature. The condition under which the neural network controllers are robust to state estimation errors and process disturbances is also established. Some works obtain the guarantee by exploiting a special structure of neural network to impose stability, for example, \cite{ivanov2018,nguyen2020}.

\subsubsection{Reinforcement learning}

        \begin{figure}[hbtp]
            \centering
            \scalebox{1}{\input{images_tot/52-rl}}
            \caption{In reinforcement learning, by using information of States, Actions and Rewards, the \MLterm helps establish an optimal policy }
            \label{fig:52reinforcement}
        \end{figure}
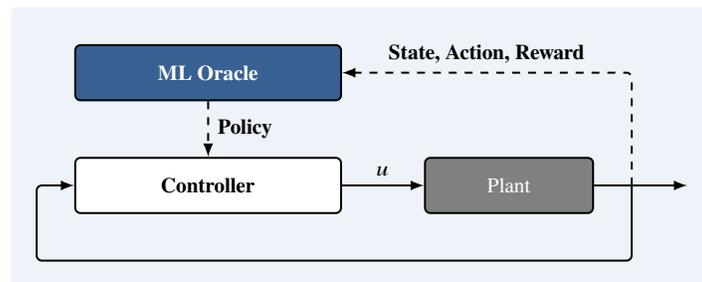

\begin{table}[hbtp]
    \centering
    \caption{Some terminology commonly used in RL, and their control counterparts (adapted from \cite{Bertsekas2019}).}
    \label{tab:RLterms}
    \begin{tabularx}{300pt}{lX}
    \toprule
        Terms in RL & Control-related terms \\ \midrule
         Environment & System \\
         Agent  & Decision maker or controller \\
         Action & Decision or control\\
         Reward of a stage & (Opposite of) Cost of a stage \\
         Value (or reward) function & (Opposite of) Cost function \\
         Deep reinforcement learning & Approximate dynamic programming using value
and/or policy approximation with deep neural networks\\
        Planning & Solving a dynamic programming problem with a known mathematical model\\
        Learning & Solving a dynamic programming problem without using an explicit mathematical model\\
     \bottomrule
    \end{tabularx}

\end{table}

Reinforcement learning (RL) is a data-driven (possibly) model-free control approach, where the mathematical model of the plant is not required and it does not necessarily needs a baseline controller either (Figure~\ref{fig:52reinforcement}).
 
RL is a method to find optimal controllers for nonlinear stochastic systems when the dynamics are unknown or affected by significant uncertainty. 
Although RL has a strong connection with optimal control, there exist differences in the terminology used in the artificial intelligence community and the optimal control community \cite{Bertsekas2019}.
Table~\ref{tab:RLterms} shows some terminology commonly used in RL, and their control counterparts.

If we use the framework from Section~\ref{sec:Process_setup}, we can explain that in RL the \MLterm learns, either offline or online, from samples of transitions and rewards to solve an optimal control problem and obtain a state feedback control law, called policy in RL literature. The \MLterm plays the role of the controller in the closed-loop system.
In order to solve the optimal control problem, the \MLterm has to learn either the optimal value (or action-value) function, the optimal control policy, or both (actor-critic methods) which are impossible to be exactly derived in general.
Therefore, functional approximation must be utilized. A very common type of function approximators, which is extensively used at the moment, are deep neural networks (DNNs). The resulting method is known as deep reinforcement learning (DRL).

There are review/tutorial papers (for example, see \cite{Bertsekas2005}, \cite{GORGES2017}, \cite{Lewis2009}) which draw the connections between RL and two popular control methods: MPC and Adaptive/Approximate dynamic programming (ADP). 
In \cite{Petsagkourakis2020} a reinforcement learning approach was used for batch bioprocesses and compared to NMPC, where extensive offline training was carried out using a simplified model.

Safety for RL, which currently is one of the main drawbacks, is still an open question because stability conditions and constraints cannot be guaranteed naturally during the problem formulation, compared to approaches based on first principles. For example, the constraints and stability conditions are integrated naturally during formulation of the MPC problem. However, this is harder to be integrated for RL during the process of solving for optimal policy. 
The basic problem for RL lies in the fact that RL is based on a finite sampling data set, and without the absolute knowledge of the system, it is impossible to guarantee robust control satisfaction without introducing additional assumptions.
Therefore, there are several works that have combined the advantages of MPC in dealing with stability and safety with RL to achieve some degree of safety (cf. \cite{Hewing2020}). 
One possible combination is proposed in \cite{Zanon2021}, where MPC is used as a function approximator within RL to provide safety and stability guarantees. Also RL is also utilized to tune the MPC parameters, thus improving closed-loop performance in a data-driven fashion.
In order to guarantee stability and safety for RL, \cite{Zhang2020} proposes Q-learning-based MPC for online control of nonlinear systems where an accurate mathematical model is not available. In this approach, two neural networks are implemented with an actor-critic structure. The critic network approximates the Q-function, trained by system input and state measurements, while the actor network is used to approximate a Lyapunov-based MPC to reduce time for solving optimization problem. The stability and safety is guaranteed by the Lyapunov’s second method for stability and constraints integrated in solving the MPC problem.  
In \cite{Khalid2022}, RL is used to learn unknown parameters of chemical processes but an economic MPC is used to guarantee stability.

Several methods that aim at using RL while guaranteeing constraint satisfaction use a \emph{safety filter}, i.e., a backup controller (e.g. MPC) that intervenes in place of the RL when constraint violation cannot be guaranteed \cite{Wabersich2017,Muntwiler2020}. 
In \cite{Wabersich2018} an MPC was used as a \emph{filter} that is entitled to change the RL input as little as possible in such a way that constraints are satisfied.
Besides the approach of combining  MPC with RL for safety guarantees, other theoretical tools in control systems theory are also used, for example Lyapunov design principles in \cite{perkins2002}, or robust control in \cite{jin2020}.
Furthermore, the work \cite{Pan2021} proposes a constrained Q-learning algorithm which can guarantee safety with high probability by using self-tuned constraint tightening.

\subsubsection{Iterative learning control} \label{subsec:Iterative_learning_control} 
Iterative Learning Control (ILC) is an open-loop control technique with the objective to improve the performance of processes that are executed repeatedly as the number of repetitions increases \cite{Bristow2006}. 
In (bio)chemical engineering these processes are (fed)batch processes where the performance is usually measured in terms of a tracking error of a possible time-varying reference \cite{Lee2007}.

\begin{figure}
    \centering
    \input{images_tot/iterative_learning_control2}
    \caption{Depiction of an iterative learning controller for a time-discrete dynamical system.
    The matricies $L$ and $Q$ are the called learning function and $Q$-filter. 
    Figure adapted from \cite{Bristow2006}.}
    \label{fig:iterative_learning_control}
\end{figure}
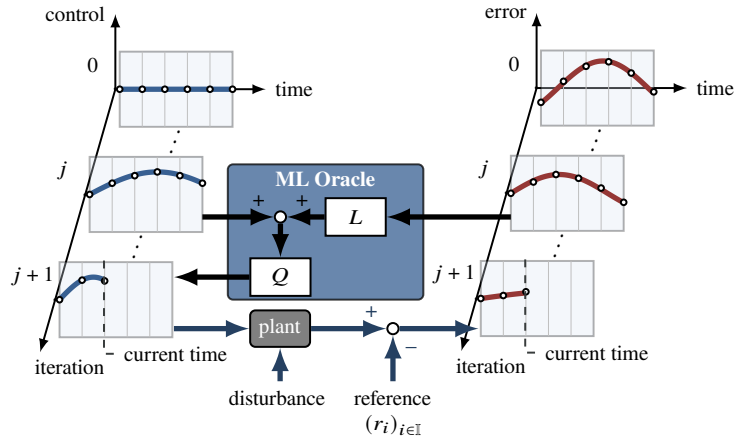

These processes have a fixed operating time, the same initial conditions, are operated repetitively (cf. Subsection~\ref{subsec:Operating_modes_of_plant}).  
Trajectory changes, disturbances, and initialization errors can also occur in these processes to a certain degree \cite{Lee2007}.
Iterative learning control has been developed for more than fifty years, hence a large number of variations have been proposed. 
Here we show only the main idea (refer to the review articles \cite{Moore1992, Bristow2006, Lee2007, Ahn2007, Xu2011} for a more detailed discussion). 
The idea is to learn a so-called \emph{filter} that reduces the tracking error
$\left( {e_k}_i \right)_{i\in\mathbb{I}}$, with ${e_k}_i:= r_i - {y_k}_i$,
between a reference 
$\left( r_i \right)_{i\in\mathbb{I}}$
and the measured plant output sequence
$\left( {y_k}_i \right)_{i\in\mathbb{I}}$
at every iteration $k$ over the set $\mathbb{I}:=\left\{0,1,\ldots,N\right\}$ of sample times.
This means that after each run of the process, the \MLterm obtains the measured output sequence as well as the input sequence
$\left( {u_k}_i \right)_{i\in\mathbb{I}}$
that generated it. 
The idea behind the algorithm of the \MLterm is to update the sequence 
$\left( {u_k}_i \right)_{i\in\mathbb{I}}$ 
such that $ {e_k}_i \rightarrow 0$ for all sample times $i\in\mathbb{I}$ as $k \rightarrow \infty$ (cf. Figure~\ref{fig:iterative_learning_control}).
\\
Using a time-discrete dynamical system in a lifted-system representation (cf. \cite{Markovsky2010,Markovsky2022}), the update algorithm within the \MLterm can also be written explicitly in general form by
\begin{alignat*}{3}
    u_{k+1} = Q \left( u_k + L e_k\right),
\end{alignat*}
where $Q\in\mathbb{R}^{N,N}$ and $L\in\mathbb{R}^{N,N}$ are called the \emph{Q-filter} and the \emph{gain matrix}.
These matrices depend not only on the transfer behavior of the plant but also on the ILC tuning techniques used \cite{Bristow2006}. 
The open-loop control sequence $\left( {u_k}_i \right)_{i\in\mathbb{I}}$ is applied to the plant. 
Often, ILC is coupled with a feedback controller to reject non-repeating disturbances (cf. e.g.\cite{Doh1999}). 
This basic formulation was adapted to a model-based formulation that uses model plant inversion in  \cite{Amann1996,Lee2000}.




ILC is usually applied to linear or linearized models in discrete time, but extensions to continuous-time, nonlinear systems are available (cf.  \cite{Xu2003, Chen1999, Xu2011}). 
Asymptotic stability can be guaranteed with some conditions on the filter matrix and for some cases also in presence of non-repeating perturbations \cite{Norrloef2002}.
For repetitive processes, whose trajectories, dynamics, and disturbances do not change much between iterations, ILC can be a valid control strategy to use in combination with a feedback control strategy. 
Non-repeating disturbances and noise can be detrimental for the performance of the ILC. For this, it is  recommended to couple the ILC control algorithm with a closed-loop controller.



\subsection{OUTLOOK}
Model-based control methods require a model of the system which are often derived from first principles.
The benefit of applying ML for control, especially for the RL case, is that there is no need to build a model of the system, which is often expensive, timely, and requires a lot of effort.
It also means that we do not need prior expert knowledge of the systems if we use a completely model-free or data-driven approach.
However, the drawback of the approach, which is based completely on ML, is that it lacks safety guarantees, robustness analysis, and constraint satisfaction.
Therefore, the future of the research in this direction is to combine both approaches so that we can benefit from the advantages of both approaches.

%% file: images_tot/51-Imitationlearning.tex


\tikzset{every picture/.style={line width=0.75pt}}
\begin{tikzpicture}[node distance=1.5cm]
    \node (T2) at (0,0) [rounded corners=2pt, draw=none, fill=IFATblue3!50]
    {\begin{tikzpicture}[>=latex]
        \draw[step=0.5cm,IFATblue3!50,very thin] (-2.5,0.75) grid (6.5,-2.75);
        \node (baseline) at (0, 0) [rounded corners=2pt, draw, fill=white]
        {\begin{minipage}[c][0.5cm]{3.6cm}\centering\footnotesize
            \textbf{Closed Loop Data of Baseline Controller}
        \end{minipage}
        };
        \node (generator) [below of=baseline, rounded corners=2pt,draw, fill=CCPSblue2]
        {\begin{minipage}[c][0.5cm]{3.3cm}\centering\footnotesize
            \textbf{\textcolor{white}{ML Oracle}}
        \end{minipage}
        };
        \node (controller) [below of=generator, rounded corners=2pt,draw, fill=white]
        {\begin{minipage}[c][0.5cm]{3.3cm}\centering\footnotesize
            \textbf{Controller}
        \end{minipage}
        };
        \node (plant) [right of=controller, xshift=2.5cm, rounded corners=2pt,draw, fill=black!50]
        {\begin{minipage}[c][0.5cm]{2.0cm}\centering\footnotesize
            \textcolor{white}{Plant}
        \end{minipage}
        };
        \node (output) [right of=plant, xshift=1.0cm, rounded corners=2pt,draw=none]
        {\begin{minipage}[c][0.5cm]{1.2cm}\centering\footnotesize
        \end{minipage}
        };
        \draw [->, dashed] (baseline.south) --  (generator.north);
        \draw [->, dashed] (generator.south) --  (controller.north);
        \draw [->] (controller.east) -- node[anchor=south] {\footnotesize$u$}  (plant.west);
        \draw [->] (plant.east) -- node[anchor=south] {\footnotesize $y$} (output.west);
        \draw [->] ([xshift=0.5cm] plant.east) -- ++(0.0,-1.0) -- ++(-8,0.0) -- ++(0.0,1.0) -- (controller.west);
    \end{tikzpicture}};
\end{tikzpicture}

%% file: images_tot/52-rl.tex
\tikzset{every picture/.style={line width=0.75pt}}
\begin{tikzpicture}[node distance=1.5cm]
    \node (T2) at (0,0) [rounded corners=2pt, draw=none, fill=IFATblue3!50]
    {\begin{tikzpicture}[>=latex]
        \draw[step=0.5cm,IFATblue3!50,very thin] (-2.5,0.75) grid (6.5,-2.75);
        \node (baseline) at (0, 0) [rounded corners=2pt, draw, fill=CCPSblue2]
        {\begin{minipage}[c][0.5cm]{3.3cm}\centering\footnotesize
            \textbf{\textcolor{white}{ML Oracle}}
        \end{minipage}
        };

        \node (controller) [below of=baseline, rounded corners=2pt, yshift=0cm, draw, fill=white]
        {\begin{minipage}[c][0.5cm]{3.3cm}\centering\footnotesize
            \textbf{Controller}
        \end{minipage}
        };
        \node (plant) [right of=controller, xshift=2.5cm, rounded corners=2pt,draw, fill=black!50]
        {\begin{minipage}[c][0.5cm]{2.0cm}\centering\footnotesize
            \textcolor{white}{Plant}
        \end{minipage}
        };
        \node (output) [right of=plant, xshift=1.0cm, rounded corners=2pt,draw=none]
        {\begin{minipage}[c][0.5cm]{1.2cm}\centering\footnotesize
        \end{minipage}
        };
        \draw [->, dashed] (baseline.south) --node[anchor=west] {\footnotesize \textbf{Policy}}  (controller.north);
        \draw [->] (controller.east) -- node[anchor=south] {\footnotesize $u$}    (plant.west);
        \draw [->] (plant.east) -- node[anchor=south] {} (output.west);
         \draw [->] ([xshift=0.5cm] plant.east) -- ([shift={(0.5,-1.0)}] plant.east) -- ([shift={(-0.5,-1.0)}] controller.west) -- ([shift={(-0.5,0.0)}] controller.west) -- (controller.west);
         \draw [dashed, ->] ([xshift=0.5cm] plant.east) -- ([shift={(0.5,1.5)}] plant.east) -- node[anchor=south] {\footnotesize \textbf{State, Action, Reward}}(baseline.east);
    \end{tikzpicture}};
\end{tikzpicture}

%% file: images_tot/iterative_learning_control2.tex
\tikzset{every picture/.style={line width=0.75pt}} 

\begin{tikzpicture}>=latex]

\draw[draw=none] (-5.2,-3.5) rectangle ++(10.4,6);

\node (LHS) at (-2.8,0) [draw=none]
    {\begin{tikzpicture}[>=latex]
    
        \draw [->] (0,0) -- ++(2,0) node [right] {\footnotesize time};    
        \draw [->] (0,0) -- ++(0,1) node [left] {\footnotesize control};
        \draw [->] (0,0) -- 
        node[pos=0.0,name=I1,inner sep=0pt]{.}
        node[pos=0.4,name=I2,inner sep=0pt]{.} 
        node[pos=0.8,name=I3,inner sep=0pt]{.} ++(-1,-3.5)
        node [right,xshift=-0.2cm,yshift=-0.2cm] {\footnotesize iteration};

        \node (A1) at (I1.center) [anchor=west,inner sep=0pt]
        {\begin{tikzpicture}[>=latex]
            \draw[fill=IFATblue3,opacity=0.3] (0,0) rectangle ++(1.5,1);
            \foreach \x in {0.3,0.6,...,1.5}
                \draw[IFATgray2,line width=0.25pt] (\x ,0) -- ++(0,1);
            
            \draw[IFATblue2,line width=2pt] (0,0.5) --++ (1.5,0); 
            
            \node at (0.0,0.5) 
            [anchor=center, circle, draw, fill=white, inner sep=0.75pt] {};
            \node at (0.3,0.5) 
            [anchor=center, circle, draw, fill=white, inner sep=0.75pt] {};
            \node at (0.6,0.5) 
            [anchor=center, circle, draw, fill=white, inner sep=0.75pt] {};
            \node at (0.9,0.5) 
            [anchor=center, circle, draw, fill=white, inner sep=0.75pt] {};
            \node at (1.2,0.5) 
            [anchor=center, circle, draw, fill=white, inner sep=0.75pt] {};
            \node at (1.5,0.5) 
            [anchor=center, circle, draw, fill=white, inner sep=0.75pt] {};

         \end{tikzpicture}};
        \node (A1text) at ([xshift=-0.3cm,yshift=-0.2cm]A1.north west) [] {\footnotesize 0}; 
         
        \node (A2) at (I2.center) [anchor=west,inner sep=0pt]
        {\begin{tikzpicture}[>=latex]
            \draw[fill=IFATblue3,opacity=0.3] (0,0) rectangle ++(1.5,1);
            \foreach \x in {0.3,0.6,...,1.5}
                \draw[IFATgray2,,line width=0.25pt] (\x ,0) -- ++(0,1);
                
              \draw[domain=0:1.5, variable=\x, IFATblue2,line width=2pt, smooth] plot ({\x}, {0.3*sin(100*\x)+0.5});
        
            \node at (0.0,0.5) 
            [anchor=center, circle, draw, fill=white, inner sep=0.75pt] {};
            \node at (0.3,0.65) 
            [anchor=center, circle, draw, fill=white, inner sep=0.75pt] {};
            \node at (0.6,0.75) 
            [anchor=center, circle, draw, fill=white, inner sep=0.75pt] {};
            \node at (0.9,0.8) 
            [anchor=center, circle, draw, fill=white, inner sep=0.75pt] {};
            \node at (1.2,0.75) 
            [anchor=center, circle, draw, fill=white, inner sep=0.75pt] {};
            \node at (1.5,0.65) 
            [anchor=center, circle, draw, fill=white, inner sep=0.75pt] {};

             \end{tikzpicture}};
        \node (A2text) at ([xshift=-0.3cm,yshift=-0.2cm]A2.north west) [] {\footnotesize $j$};  
         
        \node (A3) at (I3.center) [anchor=west,inner sep=0pt]
        {\begin{tikzpicture}[>=latex]
            \draw[fill=IFATblue3,opacity=0.3] (0,0) rectangle ++(1.5,1);
            \foreach \x in {0.3,0.6,...,1.5}
                \draw[IFATgray2,line width=0.25pt] (\x ,0) -- ++(0,1);
                
            \draw[domain=0:0.6, smooth, variable=\x, IFATblue2,line width=2pt]  plot ({\x}, {0.3*sin(200*\x)+0.5});
            
            \node at (0.0,0.5) 
            [anchor=center, circle, draw, fill=white, inner sep=0.75pt] {};
            \node at (0.3,0.76) 
            [anchor=center, circle, draw, fill=white, inner sep=0.75pt] {};
            \node at (0.6,0.75) 
            [anchor=center, circle, draw, fill=white, inner sep=0.75pt] {};

         \end{tikzpicture}};
        \node (A3text) at ([xshift=-0.3cm,yshift=-0.2cm]A3.north west) [] {\footnotesize $j+1$};  
        
        \draw[MPItext,dashed] ([xshift=-0.14cm]A3.north) -- ([xshift=-0.14cm,yshift=-0.2cm]A3.south) 
        -- ([xshift=-0.0cm,yshift=-0.2cm]A3.south) node [right,black] {\footnotesize current time};
        
        \node at(0.5+0.3,-0.675) [rotate=70] {\bfseries\footnotesize $\dots$};    
        \node at(0.0+0.3,-2.075) [rotate=70] {\bfseries\footnotesize $\dots$};          
                  
    \end{tikzpicture}};

\node (RHS) at (2.8,0) [draw=none]
    {\begin{tikzpicture}[>=latex]
    
        \draw [->] (0,0) -- ++(2,0) node [right] {\footnotesize time};    
        \draw [->] (0,0) -- ++(0,1) node [left] {\footnotesize error};
        \draw [->] (0,0) -- 
        node[pos=0.0,name=I1,inner sep=0pt]{.}
        node[pos=0.4,name=I2,inner sep=0pt]{.} 
        node[pos=0.8,name=I3,inner sep=0pt]{.} ++(-1,-3.5)
        node [right,xshift=-0.2cm,yshift=-0.2cm] {\footnotesize iteration};
    
        \node (A1) at (I1.center) [anchor=west,inner sep=0pt]
        {\begin{tikzpicture}[>=latex]
            \draw[fill=IFATblue3,opacity=0.3] (0,0) rectangle ++(1.5,1);
            \foreach \x in {0.3,0.6,...,1.5}
                \draw[IFATgray2,line width=0.25pt] (\x ,0) -- ++(0,1);
            
            \draw[domain=0:1.5, variable=\x, IFATred2,line width=2pt, smooth] plot ({\x}, {0.3*sin(150*\x)-0.2*cos(150*\x)+0.5});
            
            \node at (0.0,0.31) 
            [anchor=center, circle, draw, fill=white, inner sep=0.75pt] {};
            \node at (0.3,0.59) 
            [anchor=center, circle, draw, fill=white, inner sep=0.75pt] {};
            \node at (0.6,0.79) 
            [anchor=center, circle, draw, fill=white, inner sep=0.75pt] {};
            \node at (0.9,0.85) 
            [anchor=center, circle, draw, fill=white, inner sep=0.75pt] {};
            \node at (1.2,0.7) 
            [anchor=center, circle, draw, fill=white, inner sep=0.75pt] {};
            \node at (1.5,0.45) 
            [anchor=center, circle, draw, fill=white, inner sep=0.75pt] {};

         \end{tikzpicture}};
        \node (A1text) at ([xshift=-0.3cm,yshift=-0.2cm]A1.north west) [] {\footnotesize 0}; 
         
        \node (A2) at (I2.center) [anchor=west,inner sep=0pt]
        {\begin{tikzpicture}[>=latex]
            \draw[fill=IFATblue3,opacity=0.3] (0,0) rectangle ++(1.5,1);
            \foreach \x in {0.3,0.6,...,1.5}
                \draw[IFATgray2,,line width=0.25pt] (\x ,0) -- ++(0,1);
                
              \draw[domain=0:1.5, variable=\x, IFATred2,line width=2pt, smooth] plot ({\x}, {0.25*sin(140*\x)+0.5});

            \node at (0.0,0.5) 
            [anchor=center, circle, draw, fill=white, inner sep=0.75pt] {};
            \node at (0.3,0.65) 
            [anchor=center, circle, draw, fill=white, inner sep=0.75pt] {};
            \node at (0.6,0.75) 
            [anchor=center, circle, draw, fill=white, inner sep=0.75pt] {};
            \node at (0.9,0.70) 
            [anchor=center, circle, draw, fill=white, inner sep=0.75pt] {};
            \node at (1.2,0.57) 
            [anchor=center, circle, draw, fill=white, inner sep=0.75pt] {};
            \node at (1.5,0.38) 
            [anchor=center, circle, draw, fill=white, inner sep=0.75pt] {};
            
         \end{tikzpicture}};
        \node (A2text) at ([xshift=-0.3cm,yshift=-0.2cm]A2.north west) [] {\footnotesize $j$};  
         
        \node (A3) at (I3.center) [anchor=west,inner sep=0pt]
        {\begin{tikzpicture}[>=latex]
            \draw[fill=IFATblue3,opacity=0.3] (0,0) rectangle ++(1.5,1);
            \foreach \x in {0.3,0.6,...,1.5}
                \draw[IFATgray2,line width=0.25pt] (\x ,0) -- ++(0,1);
                
            \draw[domain=0:0.6, smooth, variable=\x, IFATred2,line width=2pt]  plot ({\x}, {0.1*sin(80*\x)+0.5});
            
            \node at (0.0,0.5) 
            [anchor=center, circle, draw, fill=white, inner sep=0.75pt] {};
            \node at (0.3,0.54) 
            [anchor=center, circle, draw, fill=white, inner sep=0.75pt] {};
            \node at (0.6,0.59) 
            [anchor=center, circle, draw, fill=white, inner sep=0.75pt] {};
            
         \end{tikzpicture}};
        \node (A3text) at ([xshift=-0.3cm,yshift=-0.2cm]A3.north west) [] {\footnotesize $j+1$};  
        \draw[MPItext,dashed] ([xshift=-0.14cm]A3.north) -- ([xshift=-0.14cm,yshift=-0.2cm]A3.south) 
        -- ([xshift=-0.0cm,yshift=-0.2cm]A3.south) node [right,black] {\footnotesize current time};

        \node at(0.5+0.3,-0.675) [rotate=70] {\bfseries\footnotesize $\dots$};    
        \node at(0.0+0.3,-2.075) [rotate=70] {\bfseries\footnotesize $\dots$};           
    \end{tikzpicture}};

\node (GEN) at (-2.8,-0.55) [draw=none,anchor=west]
    {\begin{tikzpicture}[>=latex,anchor=center]

    \draw[fill=CCPSblue2,opacity=0.75,rounded corners=2pt] (-0.7,0.7) rectangle ++(2.6,-1.8);
        \node (H) at(-0.7+1.3,0.5) [] {\footnotesize\bfseries\color{white} ML Oracle};
    
        \node (C1) at(0,0) [circle,inner sep=1.5pt,draw, fill=white] {};
        
        \node (L1) at (1,0)  [inner sep=2pt, draw, fill=white, minimum width=0.8cm, minimum height=0.5cm] {\footnotesize $L$};
        \node (Q1) at (0,-0.8) [inner sep=2pt, draw, fill=white, minimum width=0.8cm, minimum height=0.5cm] {\footnotesize $Q$};

        \draw[->,line width=2pt,black ]  ([xshift=-0.95cm]C1.west) -- (C1.west) node[above,xshift=-0.2cm]{\footnotesize $+$};
        \draw[->,line width=2pt,black ]  ([xshift=1.65cm]L1.east) -- (L1.east);
        \draw[->,line width=2pt,black ]  (Q1.west) -- ([xshift=-1cm]Q1.west);
        
        \draw[->,line width=2pt,black ]  (L1.west) -- (C1.east) node[above,xshift=+0.2cm]{\footnotesize $+$};
        \draw[->,line width=2pt,black ]  (C1.south) -- (Q1.north); 

    \end{tikzpicture}};

\node (GEN) at (-2.8,-2.4) [draw=none,anchor=west]
    {\begin{tikzpicture}[>=latex,anchor=center]

        \node (P1) at (0,-1.6) [rounded corners=2pt,draw,fill=black!50] {\footnotesize\color{white} plant};
        \node (C2) at(1.5,-1.6) [circle,inner sep=1.5pt,draw, fill=white] {};
        
        \node (Out1) at(0,-2.3) [anchor=north] {\footnotesize disturbance \hspace*{0.0cm}};
        \node (Out2) at(1.5,-2.3) [anchor=north] {\begin{minipage}{2cm}\centering\footnotesize 
                                        reference\\ $\left( r_i \right)_{i\in\mathbb{I}}$
                                      \end{minipage}};

        \draw[->,line width=2pt,IFATblue1 ] ([xshift=-1cm]P1.west) -- (P1.west);
        \draw[->,line width=2pt,IFATblue1 ] (P1.east) -- (C2.west) node[above,xshift=-0.2cm]{\footnotesize $+$};
        \draw[->,line width=2pt,IFATblue1 ] (C2.east) -- ([xshift=1.05cm]C2.east);
        
        \draw[->,line width=2pt,IFATblue1 ] (Out1.north) -- (P1.south);
        \draw[->,line width=2pt,IFATblue1 ] (Out2.north) -- (C2.south) node[right,yshift=-0.2cm]{\footnotesize $-$};
    \end{tikzpicture}};

\end{tikzpicture}

%% file: Sections/6_Summary_outlook.tex
\begin{figure}
    \centering
    \input{images_tot/summaryTAB}
    \caption{Selected, non-extensive, overview  of works on machine-learning-supported control. Note that there is no clear-cut classification, as in some cases it is difficult to classify the methods into these categories. }
    \label{fig:overviewML}
\end{figure}

The control of (bio)chemical processes is vital in order to obtain high-quality products and ensure safety, but challenging due to complexity, high-dimensionality, strong nonlinearity, and uncertainties.
To deal with these challenges, the traditional control scheme is often build on mathematical models that are based on first-principle knowledge and using model-based control systems. However, building mathematical models is often difficult, time-consuming, and economically expensive.

In the last decade, there has been a strong, growing interest in machine learning in the (bio)chemical industry due to an increasing amount of available data as well as the breakthroughs in deep learning and reinforcement learning. These techniques are promising, as they can accelerate and facilitate the modeling and control of complex systems such as (bio)chemical processes.
Nevertheless, there are several existing bottlenecks that prevent machine learning from being widely applied in this field, including safety requirements and lack of good-quality data. 

This paper outlined how machine learning can be integrated inside the control systems to enhance the performance. 
We proposed a generic and unified framework for integrating machine learning techniques with closed-loop control systems which is applicable to many machine learning techniques.  In this framework, we characterize the machine learning components as a \MLterm block, which can be described as an abstract map that uses a data set to construct a continuous function to provide a correlation between the feature set and the label set. This abstract concept allows us to describe many machine learning techniques in an unified way.

With this framework, we have reviewed and classified machine learning methods for (bio)chemical systems from two main perspectives. The first perspective discusses machine learning methods that can be used to derive the models of the (bio)chemical plant for identification, analysis, estimation, and monitoring tasks.
The obtained model can then be used for simulation-, optimization-, and model-based controller design; i.e. the control design tasks are done with conventional control theory after obtaining the model from machine learning. This part covers many machine learning techniques that have been used in the control field for decades as well as recent techniques such as Gaussian process.

The second perspective discusses machine learning methods that can be used directly in control algorithms to support the control tasks partially or to replace the controllers completely. This part covers more recent techniques, such as imitation learning using neural networks or reinforcement learning. 
Table \ref{fig:overviewML} summarizes a selected, non-extensive, overview of works on machine-learning-supported control, which have been mentioned in this paper, in the context of the common hierarchical control structure in modern chemical manufacturing processes. 
With regards to machine learning in the context of RTO, some work has proposed to learn the modifiers of a modifier adaptation scheme \cite{Ferreira2018, Marchetti2009, Riochanona2019}.  The technique tries to reach the real plant optimum despite model uncertainties, which is achieved by modifying the objective function and constraints of \eqref{eq:rto} in such a way that the necessary conditions of optimality \emph{of the real plant} are obtained. 

We did not limit ourselves in reviewing only machine learning applications of (bio)chemical processes, but we also mentioned methods that have been used in different fields. For example, to the best of our knowledge, limited work has been done on safety guarantees for machine learning supported control and estimation for (bio)chemical processes. For this reasons, we reviewed approaches that have been developed in other fields, such as robotics. The goal is to bring awareness to the (bio)chemical community about these methods that can be used as inspiration for new (bio)chemical applications.

We also see several research topics as essential for future developments:
The main bottleneck to machine learning supported approaches in (bio)chemical processes is data. For this reason, we believe that extensive research must be done to develop new measurement technologies. Furthermore, to gain the trust of the practitioners, future research should consider the development of robust modeling and control approaches, in other words, approaches that guarantee margins of safety, e.g. on product specifications and or process constraints.

%% file: images_tot/summaryTAB.tex
\tikzset{every picture/.style={line width=0.75pt}} 

\begin{tikzpicture}>=latex]

\draw[draw=none] (-5.2,-3.5) rectangle ++(11,8);

\node (R1) at (-5.1,2) [anchor=west,draw=none,rounded corners=2pt,fill=CCPSblue1]
{\begin{minipage}[c][2cm]{2cm} \bfseries\centering\color{white}\small
    Real-Time\\ Optimization
 \end{minipage}};

\node (R2) at (R1.south) [anchor=north,draw=none,rounded corners=2pt,fill=CCPSblue1]
{\begin{minipage}[c][2cm]{2cm}\bfseries\centering\color{white}\small
    Supervisory\\ Control
 \end{minipage}};
 
 \node (R3) at (R2.south) [anchor=north,draw=none,rounded corners=2pt,fill=CCPSblue1]
{\begin{minipage}[c][2cm]{2cm}\bfseries\centering\color{white}\small
    Regulatory\\ Control
 \end{minipage}};

\node (C1) at (R1.north east) [anchor=south west,draw=none,rounded corners=2pt,fill=CCPSblue1]
{\begin{minipage}{4cm} \bfseries\centering\color{white}\small
    Controller Design\\ supported by\\ Machine Learning
 \end{minipage}};

\node (C2) at (C1.east) [anchor=west,draw=none,rounded corners=2pt,fill=CCPSblue1]
{\begin{minipage}{4cm}\bfseries\centering\color{white}\small
    Controller Design\\ via\\ Machine Learning
 \end{minipage}};
 
\node (D11) at (C1.south) [anchor=north,draw=none,rounded corners=2pt,fill=CCPSgray3]
{\begin{minipage}[t][2cm]{4cm} \footnotesize
    \begin{itemize}
        \item learning modifiers of the modifier adaptation scheme via GPs
                \cite{Andersson2020, Ferreira2018, Riochanona2019}
        \item hybrid modeling for RTO models \cite{Zhang2019}
    \end{itemize}
 \end{minipage}};

\node (D21) at (D11.south) [anchor=north,draw=none,rounded corners=2pt,fill=CCPSblue3]
{\begin{minipage}[t][2cm]{4cm} \footnotesize
    \begin{itemize}
        \item plant models for model-based control \cite{Georgieva2007, Hussain1999, Macmurray1995, Mears2017, Mowbray2021, Saltık2018, Zhang2019}
        \item inverse models to provide control actions \cite{Hosen2021, Hussain1999, Lee2009, Ramli2016}
    \end{itemize}
 \end{minipage}};
 
 \node (D31) at (D21.south) [anchor=north,draw=none,rounded corners=2pt,fill=CCPSgray3]
{\begin{minipage}[t][2cm]{4cm} \footnotesize
    \begin{itemize}
        \item observer for parameter and state estimation \cite{Bishnoi2021, Galvanauskas2006, Georgieva2007, Gharagheizi2011, Nikolaou1993, Psichogios1992}
        \item self-tuning PID controllers \cite{Andrasik2004, Kucherov202186, Lee2020ReinfLearn, Liu2008, Chen20082054, Parlos2001, Thitiyasook2007}
    \end{itemize}
 \end{minipage}};
 
 \node (D12) at (C2.south) [anchor=north,draw=none,rounded corners=2pt,fill=CCPSgray3]
{\begin{minipage}[t][2cm]{4cm} \footnotesize
    \begin{itemize}
        \item solving static optimization of RTO using RL \cite{Powell2020}
    \end{itemize}
 \end{minipage}};
 
 \node (D22) at (D12.south) [anchor=north,draw=none,rounded corners=2pt,fill=CCPSblue3]
{\begin{minipage}[t][2cm]{4cm} \footnotesize
    \begin{itemize}
        \item set-point optimization using RL \cite{Kim2021,Dai20191946}
        \item imitation (supervise) learning \cite{Akesson2005}
        \item RL-based  \cite{Petsagkourakis2020,Khalid2022}
    \end{itemize}
 \end{minipage}};
 
 \node (D32) at (D22.south) [anchor=north,draw=none,rounded corners=2pt,fill=CCPSgray3]
{\begin{minipage}[t][2cm]{4cm} \footnotesize
    \begin{itemize}
        \item RL for tuning PID controller \cite{Dogru2022}
     \end{itemize}
 \end{minipage}};

\end{tikzpicture}

%% file: references.bib
@article{Biegler1998,
    title = {Advances in nonlinear programming concepts for process control},
    author = {Biegler, L.T.},
    journal = {Journal of Process Control},
    volume = {8},
    number = {5},
    pages = {301--311},
    year = {1998},
}

@ARTICLE{Santos2020,
	title = {Improving operation in an industrial MDF flash dryer through physics-based NMPC},
	author = {Santos, P. and Pitarch, J.L. and Vicente, A. and de Prada, C. and García, Á.},
	journal = {Control Engineering Practice},
	volume = {94},
	year = {2020},
}

@ARTICLE{lucia2016predictive,
  title={Predictive control, embedded cyberphysical systems and systems of systems--A perspective},
  author={Lucia, S. and K{\"o}gel, M. and Zometa, P. and Quevedo, D.E. and Findeisen, R.},
  journal={Annual Reviews in Control},
  volume={41},
  pages={193--207},
  year={2016},
}

@ARTICLE{severson2016perspectives,
  title={Perspectives on process monitoring of industrial systems},
  author={Severson, K. and Chaiwatanodom, P. and Braatz, R.D.},
  journal={Annual Reviews in Control},
  volume={42},
  pages={190--200},
  year={2016},
}

@ARTICLE{attia2020closed,
  title={Closed-loop optimization of fast-charging protocols for batteries with machine learning},
  author={Attia, P.M. and Grover, A. and Jin, N. and Severson, K.A. and Markov, T.M. and Liao, Y. and Chen, M.H. and Cheong, B. and Perkins, N. and Yang, Z. and others},
  journal={Nature},
  volume={578},
  number={7795},
  pages={397--402},
  year={2020},
}

@ARTICLE{severson2019data,
  title={Data-driven prediction of battery cycle life before capacity degradation},
  author={Severson, K.A. and Attia, P.M. and Jin, N. and Perkins, N. and Jiang, B. and Yang, Z. and Chen, M.H. and Aykol, M. and Herring, P.K. and Fraggedakis, D. and others},
  journal={Nature Energy},
  volume={4},
  number={5},
  pages={383--391},
  year={2019},
}

@incollection{Paulson2018,
  title={Fast stochastic model predictive control of end-to-end continuous pharmaceutical manufacturing},
  author={Paulson, J. A. and Streif, S. and Findeisen, R. and Braatz, R. D. and Mesbah, A.},
  booktitle={Computer Aided Chemical Engineering},
  volume={41},
  pages={353--378},
  year={2018},
  publisher={Elsevier},
}

@ARTICLE{Darby2011,
  Title                    = {{RTO}: An overview and assessment of current practice},
  Author                   = {Darby, M.L. and Nikolaou, M. and Jones, J. and Nicholson, D.},
  Journal                  = {Journal of Process Control},
  Volume                   = {21},
  Number                   = {6},
  Pages                    = {874--884},
  Year                     = {2011},
}

@ARTICLE{Qin2003,
  Title                    = {A survey of industrial model predictive control technology},
  Author                   = {Qin, S.J. and Badgwell, T.A.},
  Journal                  = {Control Engineering Practice},
  Volume                   = {11},
  Number                   = {7},
  Pages                    = {733--764},
  Year                     = {2003},
}

@BOOK{Skogestad2005,
  Title                    = {{Multivariable Feedback Control: Analysis and Design}},
  Author                   = {Skogestad, S. and Postlethwaite, I.},
  Publisher                = {John Wiley \& Sons, Inc.},
  Year                     = {2005},
}

@ARTICLE{Mayne2000,
  Title                    = {{Constrained model predictive control: Stability and optimality}},
  Author                   = {Mayne, D.Q. and Rawlings, J.B. and Rao, C.V. and Scokaert, P.O.M.},
  Journal                  = {Automatica},
  Volume                   = {36},
  Number                   = {6},
  Pages                    = {789--814},
  Year                     = {2000},
}

@BOOK{Gruene2013,
  Title                    = {{Nonlinear Model Predictive Control: Theory and Algorithms}},
  Author                   = {Gr\"une, L. and Pannek, J.},
  Publisher                = {Springer Publishing Company, Incorporated},
  Year                     = {2013},
}

@BOOK{Rawlings2018,
  Title                    = {{Model Predictive Control: Theory, Computation, and Design}},
  Author                   = {Rawlings, J.B. and Mayne, D.Q. and Diehl, M.M.},
  Publisher                = {Nob Hill Publishing, LLC},
  Year                     = {2018},
  Edition                  = {2nd},
}

@ARTICLE{Dai20191946,
	title = {Multi-rate Layered Optimal Operational Control of Industrial Processes},
	author = {Dai, W. and Lu, W. J. and Fu, J. and Ma, X. P.},
	journal = {Zidonghua Xuebao/Acta Automatica Sinica},
	volume = {45},
	number = {10},
	pages = {1946 – 1959},
	year = {2019},
}

@ARTICLE{Kim2021,
	title = {On-line set-point optimization for intelligent supervisory control and improvement of Q-learning convergence},
	author = {Kim, S.H. and Song, K.R. and Kang, I.Y. and Hyon C.I.},
	journal = {Control Engineering Practice},
	volume = {114},
	year = {2021},
}

@CONFERENCE{Chen20082054,
	title = {Optimal fuzzy pid controller design of an active magnetic bearing system based on adaptive genetic algorithms},
	author = {Chen, H. C.},
	journal = {Proceedings of the 7th International Conference on Machine Learning and Cybernetics, ICMLC},
	volume = {4},
	pages = {2054 – 2060},
	year = {2008},
}

@CONFERENCE{Kucherov202186,
	title = {PID Controller Machine Learning Algorithm Applied to the Mathematical Model of Quadrotor Lateral Motion},
	author = {Kucherov, D. and Kozub, A. and Tkachenko, V. and Rosinska, G. and Poshyvailo, O.},
	journal = {2021 IEEE 6th International Conference on Actual Problems of Unmanned Aerial Vehicles Development, APUAVD 2021 - Proceedings},
	pages = {86 – 89},
	year = {2021},
}

@ARTICLE{Lee2020ReinfLearn,
	title = {Reinforcement learning-based adaptive PID controller for DPS},
	author = {Lee, D. and Lee, S.J. and Yim, S.C.},
	journal = {Ocean Engineering},
	volume = {216},
	year = {2020},
}

@ARTICLE{Liu2008,
	title = {Self-tuning PID controller for a nonlinear system based on support vector machines},
	author = {Liu, H. and Liu, D.},
	journal = {Kongzhi Lilun Yu Yingyong/Control Theory and Applications},
	volume = {25},
	number = {3},
	pages = {468 – 474},
	year = {2008},
}

@Article{Andersson2020,
    title = {Real-time optimization of wind farms using modifier adaptation and machine learning},
    author = {Andersson, L. E. and Imsland, L.},
    journal = {Wind Energy Science},
    volume = {5},
    number = {3},
    pages = {885--896},
    year = {2020},
}

@article{Markovsky2022,
    title   = {Data-driven dynamic interpolation and approximation},
    author = {Markovsky, I. and D\"orfler, F.},
    journal = {Automatica},
    volume  = {135},
    pages   = {110008},
    year    = {2022},
}

@article{Markovsky2010,
    title   = {Closed-loop data-driven simulation},
    author  = {Markovsky, I},
    journal = {International Journal of Control},
    volume  = {83},
    number  = {10},
    pages   = {2134--2139},
    year    = {2010},
}

@article{Chung2014empirical,
  title={Empirical evaluation of gated recurrent neural networks on sequence modeling},
  author={Chung, J. and Gulcehre, C. and Cho, K. H. and Bengio, Y.},
  journal={arXiv preprint arXiv:1412.3555},
  year={2014},
}

@article{Sherstinsky2020,
  title={Fundamentals of recurrent neural network (RNN) and long short-term memory (LSTM) network},
  author={Sherstinsky, A.},
  journal={Physica D: Nonlinear Phenomena},
  volume={404},
  pages={132306},
  year={2020},
}

@article{BETHGE202014356,
    title = {Modelling Human Driving Behavior for Constrained Model Predictive Control in Mixed Traffic at Intersections},
    author = {Bethge, J. and Morabito, B. and Rewald, H. and Ahsan, A. and Sorgatz, S. and Findeisen, R.},
    journal = {IFAC-PapersOnLine},
    volume = {53},
    number = {2},
    pages = {14356-14362},
    year = {2020},
}

@ARTICLE{Mohanty2009991,
    title={Artificial neural network based system identification and model predictive control of a flotation column},
    author={Mohanty, S.},
    journal={Journal of Process Control},
    volume={19},
    number={6},
    pages={991-999},
    year={2009},
}

@BOOK{Rasmussen2006,
  title = {Gaussian Processes for Machine Learning},
  author = {Rasmussen, C. E. and Williams, C. K. I.},
  publisher = {MIT Press},
  series = {Adaptive computation and machine learning},
  year = {2006},
}

@ARTICLE{Dahunsi2010,
    title={Neural network-based identification and approximate predictive control of a servo-hydraulic vehicle suspension system},
    author={Dahunsi, O.A. and Pedro, J.O.},
    journal={Engineering Letters},
    volume={18},
    number={4},
    year={2010},
}

@ARTICLE{Ou2002195,
    title={Grouped-neural network modeling for model predictive control},
    author={Ou, J. and Rhinehart, R.R.},
    journal={ISA Transactions},
    volume={41},
    number={2},
    pages={195-202},
    year={2002},
}

@ARTICLE{Mjalli2006539,
    title={Adaptive and predictive control of liquid-liquid extractors using neural-based instantaneous linearization technique},
    author={Mjalli, F.S.},
    journal={Chemical Engineering and Technology},
    volume={29},
    number={5},
    pages={539-549},
    year={2006},
}

@CONFERENCE{Varshney2009543,
    title={ANN based IMC scheme for CSTR},
    author={Varshney, T. and Varshney, R. and Sheel, S.},
    booktitle={Proceedings of the International Conference on Advances in Computing, Communication and Control, ICAC3'09},
    pages={543-546},
    year={2009},
}

@ARTICLE{VanDenBoom2005639,
    title={Design of an analytic constrained predictive controller using neural networks},
    author={Van Den Boom, T.J.J. and Botto, M.A. and Hoekstra, P.},
    journal={International Journal of Systems Science},
    volume={36},
    number={10},
    pages={639-650},
    year={2005},
}

@ARTICLE{Sarali2019,
    title={An Improved Design for Neural-Network-Based Model Predictive Control of Three-Phase Inverters},
    author={Sarali, D.S. and Agnes Idhaya Selvi, V. and Pandiyan, K.},
    journal={2019 International Conference on Clean Energy and Energy Efficient Electronics Circuit for Sustainable Development, INCCES 2019},
    year={2019},
}

@ARTICLE{Shao2014717,
    title={An internal model controller for three-phase APF based on LS-extreme learning machine},
    author={Shao, Z. and Chen, T. and Chen, L.-A. and Tian, H.},
    journal={Open Electrical and Electronic Engineering Journal},
    volume={8},
    pages={717-722},
    year={2014},
}

@CONFERENCE{Chen20213273,
    title={Cyber-Security of Decentralized and Distributed Control Architectures with Machine-Learning Detectors for Nonlinear Processes},
    author={Chen, S. and Wu, Z. and Christofides, P.D.},
    booktitle={Proceedings of the American Control Conference},
    pages={3273-3280},
    year={2021},
}

@CONFERENCE{Kheirabadi20213077,
    title={Real-time Relocation of Floating Offshore Wind Turbines for Power Maximization Using Distributed Economic Model Predictive Control},
    author={Kheirabadi, A.C. and Nagamune, R.},
    journal={Proceedings of the American Control Conference},
    volume={2021-May},
    pages={3077-3081},
    year={2021},
}

@CONFERENCE{Embaby2020109,
    title={Adaptive Nonlinear Model Predictive Control algorithm for blood glucose regulation in type 1 diabetic patients},
    author={Embaby, A.A. and Nossair, Z. and Badr, H.},
    booktitle={2nd Novel Intelligent and Leading Emerging Sciences Conference, NILES 2020},
    pages={109-115},
    year={2020},
}

@CONFERENCE{Samek2005335,
    title={MPC using adaline},
    author={Samek, D. and Dostal, P.},
    booktitle={Annals of DAAAM and Proceedings of the International DAAAM Symposium},
    pages={335-336},
    year={2005},
}

@ARTICLE{Pan20123089,
    title={Model predictive control of unknown nonlinear dynamical systems based on recurrent neural networks},
    author={Pan, Y. and Wang, J.},
    journal={IEEE Transactions on Industrial Electronics},
    volume={59},
    number={8},
    pages={3089-3101},
    year={2012},
}

@ARTICLE{Yan2012717,
    title={Model predictive control for tracking of underactuated vessels based on recurrent neural networks},
    author={Yan, Z. and Wang, J.},
    journal={IEEE Journal of Oceanic Engineering},
    volume={37},
    number={4},
    pages={717-726},
    year={2012},
}

@ARTICLE{Lu20081366,
    title={Adaptive predictive control with recurrent neural network for industrial processes: An application to temperature control of a variable-frequency oil-cooling machine},
    author={Lu, C.-H. and Tsai, C.-C.},
    journal={IEEE Transactions on Industrial Electronics},
    volume={55},
    number={3},
    pages={1366-1375},
    year={2008},
}

@ARTICLE{Kittisupakorn2009579,
    title={Neural network based model predictive control for a steel pickling process},
    author={Kittisupakorn, P. and Thitiyasook, P. and Hussain, M.A. and Daosud, W.},
    journal={Journal of Process Control},
    volume={19},
    number={4},
    pages={579-590},
    year={2009},
}

@ARTICLE{AlSeyab2008568,
    title={Nonlinear system identification for predictive control using continuous time recurrent neural networks and automatic differentiation},
    author={Al Seyab, R.K. and Cao, Y.},
    journal={Journal of Process Control},
    volume={18},
    number={6},
    pages={568-581},
    year={2008},
}

@ARTICLE{Yan2012746,
    title={Model predictive control of nonlinear systems with unmodeled dynamics based on feedforward and recurrent neural networks},
    author={Yan, Z. and Wang, J.},
    journal={IEEE Transactions on Industrial Informatics},
    volume={8},
    number={4},
    pages={746-756},
    year={2012},
}

@ARTICLE{Thuruthel2019127,
    title={Model-Based Reinforcement Learning for Closed-Loop Dynamic Control of Soft Robotic Manipulators},
    author={Thuruthel, T.G. and Falotico, E. and Renda, F. and Laschi, C.},
    journal={IEEE Transactions on Robotics},
    volume={35},
    number={1},
    pages={127-134},
    year={2019},
}

@ARTICLE{Dalamagkidis2011818,
    title={Nonlinear model predictive control with neural network optimization for autonomous autorotation of small unmanned helicopters},
    author={Dalamagkidis, K. and Valavanis, K.P. and Piegl, L.A.},
    journal={IEEE Transactions on Control Systems Technology},
    volume={19},
    number={4},
    pages={818-831},
    year={2011},
}

@ARTICLE{Patan20151147,
    title={Neural Network-Based Model Predictive Control: Fault Tolerance and Stability},
    author={Patan, K.},
    journal={IEEE Transactions on Control Systems Technology},
    volume={23},
    number={3},
    pages={1147-1155},
    year={2015},
}

@ARTICLE{Zarkogianni20112467,
    title={An insulin infusion advisory system based on autotuning nonlinear model-predictive control},
    author={Zarkogianni, K. and Vazeou, A. and Mougiakakou, S.G. and Prountzou, A. and Nikita, K.S.},
    journal={IEEE Transactions on Biomedical Engineering},
    volume={58},
    number={9},
    pages={2467-2477},
    year={2011},
}

@ARTICLE{Temeng199519,
    title={Model predictive control of an industrial packed bed reactor using neural networks},
    author={Temeng, K.O. and Schnelle, P.D. and McAvoy, T.J.},
    journal={Journal of Process Control},
    volume={5},
    number={1},
    pages={19-27},
    year={1995},
}

@ARTICLE{Zhang2008322,
    title={Model predictive control of water management in PEMFC},
    author={Zhang, L. and Pan, M. and Quan, S.},
    journal={Journal of Power Sources},
    volume={180},
    number={1},
    pages={322-329},
    year={2008},
}

@ARTICLE{Yan20164377,
    title={Tube-Based Robust Model Predictive Control of Nonlinear Systems via Collective Neurodynamic Optimization},
    author={Yan, Z. and Le, X. and Wang, J.},
    journal={IEEE Transactions on Industrial Electronics},
    volume={63},
    number={7},
    pages={4377-4386},
    year={2016},
}

@ARTICLE{Pan20081685,
    title={Two neural network approaches to model predictive control},
    author={Pan, Y. and Wang, J.},
    journal={Proceedings of the American Control Conference},
    pages={1685-1690},
    year={2008},
}

@ARTICLE{Chen20198461,
    title={Human-Centered Trajectory Tracking Control for Autonomous Vehicles with Driver Cut-In Behavior Prediction},
    author={Chen, Y. and Hu, C. and Wang, J.},
    journal={IEEE Transactions on Vehicular Technology},
    volume={68},
    number={9},
    pages={8461-8471},
    year={2019},
}

@ARTICLE{Pan2009683,
    title={Model predictive control for nonlinear affine systems based on the simplified dual neural network},
    author={Pan, Y. and Wang, J.},
    journal={Proceedings of the IEEE International Conference on Control Applications},
    pages={683-688},
    year={2009},
}

@ARTICLE{Wu202074,
    title={Process structure-based recurrent neural network modeling for model predictive control of nonlinear processes},
    author={Wu, Z. and Rincon, D. and Christofides, P.D.},
    journal={Journal of Process Control},
    volume={89},
    pages={74-84},
    year={2020},
}

@ARTICLE{Wu20202275,
    title={Real-Time Adaptive Machine-Learning-Based Predictive Control of Nonlinear Processes},
    author={Wu, Z. and Rincon, D. and Christofides, P.D.},
    journal={Industrial and Engineering Chemistry Research},
    volume={59},
    number={6},
    pages={2275-2290},
    year={2020},
}

@ARTICLE{Wu2020,
    title={A predictive energy management strategy for multi-mode plug-in hybrid electric vehicles based on multi neural networks},
    author={Wu, Y. and Zhang, Y. and Li, G. and Shen, J. and Chen, Z. and Liu, Y.},
    journal={Energy},
    volume={208},
    year={2020},
}

@ARTICLE{Núñez20202859,
    title={Neural Network-Based Model Predictive Control of a Paste Thickener over an Industrial Internet Platform},
    author={Núñez, F. and Langarica, S. and Díaz, P. and Torres, M. and Salas, J.C.},
    journal={IEEE Transactions on Industrial Informatics},
    volume={16},
    number={4},
    pages={2859-2867},
    year={2020},
}

@ARTICLE{Wang201929,
    title={Predicting plug loads with occupant count data through a deep learning approach},
    author={Wang, Z. and Hong, T. and Piette, M.A.},
    journal={Energy},
    volume={181},
    pages={29-42},
    year={2019},
}

@ARTICLE{Atuonwu20101418,
    title={Identification and predictive control of a multistage evaporator},
    author={Atuonwu, J.C. and Cao, Y. and Rangaiah, G.P. and Tadé, M.O.},
    journal={Control Engineering Practice},
    volume={18},
    number={12},
    pages={1418-1428},
    year={2010},
}

@ARTICLE{Pan20101597,
    title={A neurodynamic optimization approach to nonlinear model predictive control},
    author={Pan, Y. and Wang, J.},
    journal={Conference Proceedings - IEEE International Conference on Systems, Man and Cybernetics},
    pages={1597-1602},
    year={2010},
}

@ARTICLE{Yang2021,
    title={Experiment study of machine-learning-based approximate model predictive control for energy-efficient building control},
    author={Yang, S. and Wan, M.P. and Chen, W. and Ng, B.F. and Dubey, S.},
    journal={Applied Energy},
    volume={288},
    year={2021},
}

@ARTICLE{Pereira20213213,
    title={Nonlinear Model Predictive Control for the Energy Management of Fuel Cell Hybrid Electric Vehicles in Real Time},
    author={Pereira, D.F. and Lopes, F.D.C. and Watanabe, E.H.},
    journal={IEEE Transactions on Industrial Electronics},
    volume={68},
    number={4},
    pages={3213-3223},
    year={2021},
}

@ARTICLE{Huang2015256,
    title={A hybrid model predictive control scheme for energy and cost savings in commercial buildings: Simulation and experiment},
    author={Huang, H. and Chen, L. and Hu, E.},
    journal={Proceedings of the American Control Conference},
    volume={2015-July},
    pages={256-261},
    year={2015},
}

@ARTICLE{Zhang2019524,
    title={Model predictive control for electrochemical impedance spectroscopy measurement of fuel cells based on neural network optimization},
    author={Zhang, L. and Zhou, Z. and Chen, Q. and Long, R. and Quan, S.},
    journal={IEEE Transactions on Transportation Electrification},
    volume={5},
    number={2},
    pages={524-534},
    year={2019},
}

@ARTICLE{Kusiak20201594,
    title={Convolutional and generative adversarial neural networks in manufacturing},
    author={Kusiak, A.},
    journal={International Journal of Production Research},
    volume={58},
    number={5},
    pages={1594-1604},
    year={2020},
}

@ARTICLE{Zhou20171229,
    title={Review of Convolutional Neural Network},
    author={Zhou, F.-Y. and Jin, L.-P. and Dong, J.},
    journal={Jisuanji Xuebao/Chinese Journal of Computers},
    volume={40},
    number={6},
    pages={1229-1251},
    year={2017},
}

@article{Oliveira2004,
    title = {{Combining first principles modelling and artificial neural networks: A general framework}},
    author = {Oliveira, R.},
    journal = {Computers and Chemical Engineering},
    volume = {28},
    number = {5},
    pages = {755--766},
    year = {2004},
}

@article{McCulloch1943,
    title = {{A logical calculus of the ideas immanent in nervous activity}},
    author = {McCulloch, W.-S. and Pitts, W.},
    journal = {Bulletin of Mathematical Biophysics},
    volume = {5},
    number = {4},
    pages = {115--133},
    year = {1943},
}

@ARTICLE{Powell2020,
    title={Real-time optimization using reinforcement learning},
    author={Powell, B.K.M. and Machalek, D. and Quah, T.},
    journal={Computers and Chemical Engineering},
    volume={143},
    year={2020},
}

@article{Riochanona2019,
    title = {Modifier-Adaptation Schemes Employing Gaussian Processes and Trust Regions for Real-Time Optimization⁎⁎The first two authors contributed equally to the paper},
    author = {del {Rio Chanona}, E.A. and Graciano, J.E.A. and Bradford, E. and Chachuat, B.},
    journal = {IFAC-PapersOnLine},
    volume = {52},
    number = {1},
    pages = {52-57},
    year = {2019},
}

@INPROCEEDINGS{Ferreira2018,
    booktitle={2018 European Control Conference (ECC)}, 
    author={de Avila Ferreira, T. and Shukla, H. A. and Faulwasser, T. and Jones, C. N. and Bonvin, D.},
    title={Real-Time optimization of Uncertain Process Systems via Modifier Adaptation and Gaussian Processes}, 
    volume={},
    number={},
    pages={465-470},
    year={2018},
}

@ARTICLE{Pan2021,
    title={Constrained model-free reinforcement learning for process optimization},
    author={Pan, E. and Petsagkourakis, P. and Mowbray, M. and Zhang, D. and Rio-Chanona, E.A.D.},
    journal={Computers and Chemical Engineering},
    volume={154},
    year={2021},
}

@article{Marchetti2009,
    title={Modifier-adaptation methodology for real-time optimization},
    author={Marchetti, A. and Chachuat, B. and Bonvin, D.},
    journal={Industrial \& engineering chemistry research},
    volume={48},
    number={13},
    pages={6022--6033},
    year={2009},
}

@article{Hewing2020,
    title = {Learning-Based Model Predictive Control: Toward Safe Learning in Control},
    author = {Hewing, L. and Wabersich, K. P. and Menner, M. and Zeilinger, M. N.},
    journal = {Annual Review of Control, Robotics, and Autonomous Systems},
    volume = {3},
    number = {1},
    pages = {269-296},
    year = {2020},
}

@ARTICLE{Xu2011,
    title={A survey on iterative learning control for nonlinear systems},
    author={Xu, J.-X.},
    journal={International Journal of Control},
    volume={84},
    number={7},
    pages={1275-1294},
    year={2011},
}

@ARTICLE{Ahn2007,
    journal={IEEE Transactions on Systems, Man, and Cybernetics, Part C (Applications and Reviews)}, 
    author={Ahn, H.-S. and Chen, Y. Q. and Moore, K. L.},
    title={Iterative Learning Control: Brief Survey and Categorization}, 
    volume={37},
    number={6},
    pages={1099-1121},
    year={2007},
}

@article{Moore1992,
    title={Iterative learning control: A survey and new results},
    author={Moore, K. L. and Dahleh, M. and Bhattacharyya, S.P.},
    journal={Journal of Robotic Systems},
    volume={9},
    number={5},
    pages={563--594},
    year={1992},
}

@book{Chen1999,
    title={Iterative learning control: convergence, robustness and applications},
    author={Chen, Y. Q. and Wen, C.},
    publisher={Springer London},
    year={1999},
}

@book{Xu2003,
    title={Linear and nonlinear iterative learning control},
    author={Xu, J.-X. and Tan, Y.},
    volume={291},
    publisher={Springer},
    year={2003},
}

@article{Norrloef2002,
    title = {Time and frequency domain convergence properties in iterative learning control},
    author = {Norrlöf, M. and Gunnarsson, S.},
    journal = {International Journal of Control},
    volume = {75},
    number = {14},
    pages = {1114-1126},
    year  = {2002},
}

@article{Doh1999,
    title = {Robust iterative learning control with current feedback for uncertain linear systems},
    author = {Doh, T.-Y.},
    journal = {International Journal of Systems Science},
    volume = {30},
    number = {1},
    pages = {39-47},
    year  = {1999},
}

@article{Bristow2006,
    title={A survey of iterative learning control},
    author={Bristow, D. A. and Tharayil, M. and Alleyne, A. G.},
    journal={IEEE control systems magazine},
    volume={26},
    number={3},
    pages={96--114},
    year={2006},
}

@article{Lee2000,
    title={Model-based iterative learning control with a quadratic criterion for time-varying linear systems},
    author={Lee, J. H. and Lee, K. S. and Kim, W.},
    journal={Automatica},
    volume={36},
    pages={641-657},
    year={2000},
}

@article{Amann1996,
    journal={IEE Proceedings - Control Theory and Applications}, 
    author={Amann, N. and Owens, D. and Rogers, E.},
    title={Iterative learning control for discrete-time systems with exponential rate of convergence}, 
    volume={143},
    number={2},
    pages={217-224},
    year={1996},
}

@article{Lee2007,
    title = {{Iterative learning control applied to batch processes: An overview}},
    author = {Lee, J. H. and Lee, K. S.},
    journal = {Control Engineering Practice},
    volume = {15},
    number = {10 SPEC. ISS.},
    pages = {1306--1318},
    year = {2007},
}

@article{Yang2015a,
    title = {{Risk-Sensitive Model Predictive Control with Gaussian Process Models}},
    author = {Yang, X. and Maciejowski, J.},
    journal = {IFAC-PapersOnLine},
    volume = {48},
    number = {28},
    pages = {374--379},
    year = {2015},
}

@ARTICLE{Soloperto2018,
    title = {{Learning-Based Robust Model Predictive Control with State-Dependent Uncertainty}},
    author = {Soloperto, R. and M{\"{u}}ller, M. A. and Trimpe, S. and Allg{\"{o}}wer, F.},
    journal={IFAC-PapersOnLine},
    pages = {442--447},
    volume={51},
    number={20},
    pages={442--447},
    year = {2018},
}

@article{BETHGE2018517,
    title={Multi-mode learning supported model predictive control with guarantees},
    author={Bethge, J. and Morabito, B. and Matschek, J. and Findeisen, R.},
    journal={IFAC-PapersOnLine},
    volume={51},
    number={20},
    pages={517--522},
    year={2018},
}

@article{Morabito2021,
    title={Towards Risk-aware Machine Learning Supported Model Predictive Control and Open-loop Optimization for Repetitive Processes},
    author={Morabito, B. and Pohlodek, J. and Matschek, J. and Savchenko, A. and Carius, L. and Findeisen, R.},
    journal={IFAC-PapersOnLine},
    volume={54},
    number={6},
    pages={321--328},
    year={2021},
}

@article{Aswani2013,
    journal = {Automatica},
    author = {Aswani, A. and Gonzalez, H. and Sastry, S.S. and Tomlin, C.},
    title = {{Provably safe and robust learning-based model predictive control}},
    volume = {49},
    number = {5},
    pages = {1216--1226},
    year = {2013},
}

@article{Wabersich2018,
    title = {{A predictive safety filter for learning-based control of constrained nonlinear dynamical systems}},
    author = {Wabersich, K. P. and Zeilinger, M. N.},
    journal={arXiv},
    year = {2018},
}

@article{Muntwiler2020,
    title = {{Distributed model predictive safety certification for learning-based control}},
    author = {Muntwiler, S. and Wabersich, K. P. and Carron, A. and Zeilinger, M. N.},
    journal = {IFAC-PapersOnLine},
    volume = {53},
    number = {2},
    pages = {5258--5265},
    year = {2020},
}

@article{Wabersich2017,
    title = {{Scalable synthesis of safety certificates from data with application to learning-based control}},
    author = {Wabersich, K. P. and Zeilinger, M. N.},
    journal = {arXiv},
    pages = {1691--1697},
    year = {2017},
}

@article{Bradford2020,
    title = {{Stochastic data-driven model predictive control using gaussian processes}},
    author = {Bradford, E. and Imsland, L. and Zhang, D. and {del Rio Chanona}, E. A.},
    journal = {Computers and Chemical Engineering},
    volume = {139},
    year = {2020},
}

@article{Rafiei2018,
    title = {Stochastic Back-Off Approach for Integration of Design and Control Under Uncertainty},
    author = {Rafiei, M. and Ricardez-Sandoval, L. A.},
    journal = {Industrial \& Engineering Chemistry Research},
    volume = {57},
    number = {12},
    pages = {4351-4365},
    year = {2018},
}

@article{Karniadakis2021,
    title = {Physics-informed machine learning},
    author = {Karniadakis, G. and Kevrekidis, Y. and Lu, L. and Perdikaris, P. and Wang, S. and Yang, L.},
    journal={Nature Reviews Physics},
    volume={3},
    pages={422-440},
    year = {2021},
}

@article{Dobbelaere2021,
    title = {Machine Learning in Chemical Engineering: Strengths, Weaknesses, Opportunities, and Threats},
    author = {Dobbelaere, M. R. and Plehiers, P. P. and {Van de Vijver}, R. and Stevens, C. V. and {Van Geem}, K. M.},
    journal = {Engineering},
    year = {2021},
}

@article{Mowbray2021,
    title = {{Machine learning for biochemical engineering: A review}},
    author = {Mowbray, M. and Savage, T. and Wu, C. and Song, Z. and Cho, B. A. and {Del Rio-Chanona}, E. A. and Zhang, D.},
    journal = {Biochemical Engineering Journal},
    volume = {172},
    pages = {108054},
    year = {2021},
}

@ARTICLE{Henriques2017,
    title={Data-driven reverse engineering of signaling pathways using ensembles of dynamic models},
    author={Henriques, D. and Villaverde, A.F. and Rocha, M. and Saez-Rodriguez, J. and Banga, J.R.},
    journal={PLoS Computational Biology},
    volume={13},
    number={2},
    year={2017},
}

@ARTICLE{Teixeira2007,
    title={Hybrid semi-parametric mathematical systems: Bridging the gap between systems biology and process engineering},
    author={Teixeira, A.P. and Carinhas, N. and Dias, J.M.L. and Cruz, P. and Alves, P.M. and Carrondo, M.J.T. and Oliveira, R.},
    journal={Journal of Biotechnology},
    volume={132},
    number={4},
    pages={418-425},
    year={2007},
}

@CONFERENCE{Lee2021,
    title={A Hybrid Mechanistic Data-Driven Approach for Modeling Uncertain Intracellular Signaling Pathways},
    author={Lee, D. and Jayaraman, A. and Kwon, J.S.-I.},
    booktitle={Proceedings of the American Control Conference},
    volume={2021-May},
    pages={1903-1908},
    year={2021},
}

@ARTICLE{Lee2020,
    title={Development of a hybrid model for a partially known intracellular signaling pathway through correction term estimation and neural network modeling},
    author={Lee, D. and Jayaraman, A. and Kwon, J.S.},
    journal={PLoS Computational Biology},
    volume={16},
    number={12},
    year={2020},
}

@article{DelRio-Chanona2016,
    title = {{Dynamic modeling and optimization of cyanobacterial C-phycocyanin production process by artificial neural network}},
    author = {del Rio-Chanona, E. A. and Manirafasha, E. and Zhang, D. and Yue, Q. and Jing, K.},
    journal = {Algal Research},
    pages = {7--15},
    volume = {13},
    year = {2016},
}

@article{Garcia-Camacho2016,
    title = {{Artificial neural network modeling for predicting the growth of the microalga Karlodinium veneficum}},
    author = {Garc{\'{i}}a-Camacho, F. and L{\'{o}}pez-Rosales, L. and S{\'{a}}nchez-Mir{\'{o}}n, A. and Belarbi, E. H. and Chisti, Yusuf and Molina-Grima, E.},
    journal = {Algal Research},
    pages = {58--64},
    volume = {14},
    year = {2016},
}

@article{Zhang2019,
    title = {{Hybrid physics-based and data-driven modeling for bioprocess online simulation and optimization}},
    author = {Zhang, D. and {Del Rio-Chanona}, E. A. and Petsagkourakis, P. and Wagner, J.},
    journal = {Biotechnology and Bioengineering},
    volume = {116},
    number = {11},
    pages = {2919--2930},
    year = {2019},
}

@article{Mowbray2021a,
    title={Safe Chance Constrained Reinforcement Learning for Batch Process Control},
    author={Mowbray, M. and Petsagkourakis, P. and del Rio-Chanona, E. A. and Zhang, D.},
    journal={arXiv},
    year={2021},
}

@article{Bradford2018,
    title = {Dynamic modeling and optimization of sustainable algal production with uncertainty using multivariate Gaussian processes},
    author = {Bradford, E. and Schweidtmann, A. M. and Zhang, D. and Jing, K. and del Rio-Chanona, E. A.},
    journal = {Computers \& Chemical Engineering},
    volume = {118},
    pages = {143-158},
    year = {2018},
}

@article{Petsagkourakis2020,
    title = {Reinforcement learning for batch bioprocess optimization},
    author = {Petsagkourakis, P. and Sandoval, I.O. and Bradford, E. and Zhang, D. and del Rio-Chanona, E. A.},
    journal = {Computers \& Chemical Engineering},
    volume = {133},
    pages = {106649},
    year = {2020},
}

@article{Teixeira2006,
    title={Bioprocess iterative batch-to-batch optimization based on hybrid parametric/nonparametric models},
    author={Teixeira, A. P. and Clemente, J. J. and Cunha, A. E. and Carrondo, M. J. T. and Oliveira, R.},
    journal={Biotechnology progress},
    volume={22},
    number={1},
    pages={247--258},
    year={2006},
}

@article{VonStosch2014,
    title={Hybrid semi-parametric modeling in process systems engineering: Past, present and future},
    author={Von Stosch, M. and Oliveira, R. and Peres, J. and de Azevedo, S. F.},
    journal={Computers \& Chemical Engineering},
    volume={60},
    pages={86--101},
    year={2014},
}

@article{Ponkumar2018,
    title = {A Deep Learning Architecture for Predictive Control},
    author = {Pon Kumar, S. S. and Tulsyan, A. and Gopaluni, B. and Loewen, P.},
    journal = {IFAC-PapersOnLine},
    volume = {51},
    number = {18},
    pages = {512-517},
    year = {2018},
}

@ARTICLE{Cseko2015,
    title={Explicit MPC-Based RBF Neural Network Controller Design With Discrete-Time Actual Kalman Filter for Semiactive Suspension}, 
    author={Csekő, L. H. and Kvasnica, M. and Lantos, B.},
    journal={IEEE Transactions on Control Systems Technology}, 
    volume={23},
    number={5},
    pages={1736-1753},
    year={2015},
}

@INPROCEEDINGS{Chen2018,  
    title={Approximating Explicit Model Predictive Control Using Constrained Neural Networks},   
    author={Chen, S. and Saulnier, K. and Atanasov, N. and Lee, D. D. and Kumar, V. and Pappas, G. J. and Morari, M.},  
    booktitle={2018 Annual American Control Conference (ACC)},   
    pages={1520-1527},
    year={2018},
}

@article{Cao2020,
    title = {Deep Neural Network Approximation of Nonlinear Model Predictive Control},
    author = {Cao, Y. and Gopaluni, R. B.},
    journal = {IFAC-PapersOnLine},
    volume = {53},
    number = {2},
    pages = {11319-11324},
    year = {2020},
}

@article{Maddalena2020,
    title = {A Neural Network Architecture to Learn Explicit MPC Controllers from Data},
    author = {Maddalena, E.T. and {da S. Moraes}, C.G. and Waltrich, G. and Jones, C.N.},
    journal = {IFAC-PapersOnLine},
    volume = {53},
    number = {2},
    pages = {11362-11367},
    year = {2020},
}

@article{Parisini1995,
    title = {A receding-horizon regulator for nonlinear systems and a neural approximation},
    author = {Parisini, T. and Zoppoli, R.},
    journal = {Automatica},
    volume = {31},
    number = {10},
    pages = {1443-1451},
    year = {1995},
}

@article{Karg2020,
    title={Efficient representation and approximation of model predictive control laws via deep learning},
    author={Karg, B. and Lucia, S.},
    journal={IEEE Transactions on Cybernetics},
    volume={50},
    number={9},
    pages={3866--3878},
    year={2020},
}

@article{Andrasik2004,
    title = {{On-line tuning of a neural PID controller based on plant hybrid modeling}},
    author = {Andr{\'{a}}{\v{s}}ik, A. and M{\'{e}}sz{\'{a}}ros, A. and {De Azevedo}, S. F.},
    journal = {Computers and Chemical Engineering},
    volume = {28},
    number = {8},
    pages = {1499--1509},
    year = {2004},
}

@article{Basheer2000,
    title = {{Artificial neural networks: fundamentals, computing, design, and application}},
    author = {Basheer, I.A and Hajmeer, M},
    journal = {Journal of Microbiological Methods},
    volume = {43},
    number = {1},
    pages = {3--31},
    year = {2000},
}

@article{Bhutani2006,
    title = {{First-Principles, Data-Based, and Hybrid Modeling and Optimization of an Industrial Hydrocracking Unit}},
    author = {Bhutani, N. and Rangaiah, G. P. and Ray, A. K.},
    journal = {Industrial {\&} Engineering Chemistry Research},
    volume = {45},
    number = {23},
    pages = {7807--7816},
    year = {2006},
}

@article{Bishnoi2021,
    title = {{Scalable Gaussian processes for predicting the optical, physical, thermal, and mechanical properties of inorganic glasses with large datasets}},
    author = {Bishnoi, S. and Ravinder, R. and Grover, H. S. and Kodamana, H. and Krishnan, N. M. A.},
    journal = {Materials Advances},
    volume = {2},
    number = {1},
    pages = {477--487},
    year = {2021},
}

@article{Chai2014,
    title = {{Optimal operational control for complex industrial processes}},
    author = {Chai, T. and Qin, S. J. and Wang, H.},
    journal = {Annual Reviews in Control},
    volume = {38},
    number = {1},
    pages = {81--92},
    year = {2014},
}

@article{Curteanu2006,
    title = {{Hybrid Neural Network Models Applied to a Free Radical Polymerization Process}},
    author = {Curteanu, S. and Leon, F.},
    journal = {Polymer-Plastics Technology and Engineering},
    volume = {45},
    number = {9},
    pages = {1013--1023},
    year = {2006},
}

@article{Fezai2020,
    title = {{Online reduced Gaussian process regression based generalized likelihood ratio test for fault detection}},
    author = {Fezai, R and Mansouri, M and Abodayeh, K and Nounou, H and Nounou, M},
    journal = {Journal of Process Control},
    volume = {85},
    pages = {30--40},
    year = {2020},
}

@inproceedings{Galvanauskas2006,
    booktitle = {The 2006 IEEE International Joint Conference on Neural Network Proceedings},
    author = {Galvanauskas, V. and Georgieva, P. and de Azevedo, S.F.},
    title = {{Dynamic Optimisation of Industrial Sugar Crystallization Process based on a Hybrid (mechanistic+ANN) Model}},
    pages = {2728--2735},
    publisher = {IEEE},
    year = {2006},
}

@article{Ge2011,
    title = {{Quality prediction for polypropylene production process based on CLGPR model}},
    author = {Ge, Z. and Chen, T. and Song, Z.},
    journal = {Control Engineering Practice},
    volume = {19},
    number = {5},
    pages = {423--432},
    year = {2011},
}

@article{Ge2017a,
    title = {{Data Mining and Analytics in the Process Industry: The Role of Machine Learning}},
    author = {Ge, Z. and Song, Z. and Ding, S. X. and Huang, B.},
    journal = {IEEE Access},
    volume = {5},
    pages = {20590--20616},
    year = {2017},
}

@article{Georgieva2007,
    title = {{Neural Network-Based Control Strategies Applied to a Fed-Batch Crystallization Process}},
    author = {Georgieva, P. and de Azevedo, S. F.},
    journal = {International Journal of Chemical and Molecular Engineering},
    volume = {1},
    number = {12},
    pages = {145--154},
    year = {2007},
}

@article{Gharagheizi2011,
    title = {{Use of artificial neural network-group contribution method to determine surface tension of pure compounds}},
    author = {Gharagheizi, F. and Eslamimanesh, A. and Mohammadi, A. H. and Richon, D.},
    journal = {Journal of Chemical and Engineering Data},
    volume = {56},
    number = {5},
    pages = {2587--2601},
    year = {2011},
}

@article{Himmelblau2008,
    title = {{Accounts of Experiences in the Application of Artificial Neural Networks in Chemical Engineering}},
    author = {Himmelblau, D. M.},
    journal = {Industrial {\&} Engineering Chemistry Research},
    volume = {47},
    number = {16},
    pages = {5782--5796},
    year = {2008},
}

@article{Hosen2021,
    title = {{NN-based Prediction Interval for Nonlinear Processes Controller}},
    author = {Hosen, M. A. and Khosravi, A. and Kabir, H. M. D and Johnstone, M. and Creighton, D. and Nahavandi, S. and Shi, P.},
    journal = {International Journal of Control, Automation and Systems},
    volume = {19},
    number = {9},
    pages = {3239--3252},
    year = {2021},
}

@article{Hussain1999,
    title = {{Review of the applications of neural networks in chemical process control — simulation and online implementation}},
    author = {Hussain, M. A.},
    journal = {Artificial Intelligence in Engineering},
    volume = {13},
    number = {1},
    pages = {55--68},
    year = {1999},
}

@article{Hussain2014,
    title = {{Neural network inverse model control strategy: Discrete-time stability analysis for relative order two systems}},
    author = {Hussain, M. A. and {Mohd Ali}, J. and Khan, M. J.H.},
    journal = {Abstract and Applied Analysis},
    volume = {2014},
    year = {2014},
}

@article{Kano2009,
    title = {{The State of the Art in Advanced Chemical Process Control in Japan}},
    author = {Kano, M. and Ogawa, M.},
    journal = {IFAC Proceedings Volumes},
    volume = {42},
    number = {11},
    pages = {10--25},
    year = {2009},
}

@incollection{Landau2003,
    title = {{Controls, Adaptive Systems}},
    author = {Landau, I.D.},
    booktitle = {Encyclopedia of Physical Science and Technology},
    pages = {649--658},
    publisher = {Elsevier},
    year = {2003},
    doi = {10.1016/B0-12-227410-5/00142-3},
}

@article{Lee2009,
    title = {{Inverse Dynamic Neuro-Controller for superheater steam temperature control of a large-scale ultra-supercritical (USC) boiler unit}},
    author = {Lee, K. Y. and Ma, L. and Boo, C. J. and Jung, W. H. and Kim, S. H.},
    journal = {IFAC Proceedings Volumes (IFAC-PapersOnline)},
    volume = {42},
    number = {9},
    pages = {107--112},
    year = {2009},
}

@article{Macmurray1995,
    title = {{Modeling and control of a packed distillation column using artificial neural networks}},
    author = {Macmurray, J. C. and Himmelblau, D. M.},
    journal = {Computers and Chemical Engineering},
    volume = {19},
    number = {10},
    pages = {1077--1088},
    year = {1995},
}

@article{MdNor2020,
    title = {{A review of data-driven fault detection and diagnosis methods: applications in chemical process systems}},
    author = {{Md Nor}, N. and {Che Hassan}, C. R. and Hussain, M. A.},
    journal = {Reviews in Chemical Engineering},
    volume = {36},
    number = {4},
    pages = {513--553},
    year = {2020},
}

@article{Mears2017,
    title = {{A review of control strategies for manipulating the feed rate in fed-batch fermentation processes}},
    author = {Mears, L. and Stocks, S. M. and Sin, G. and Gernaey, K. V.},
    journal = {Journal of Biotechnology},
    volume = {245},
    pages = {34--46},
    year = {2017},
}

@article{MohdAli2015,
    title = {{Artificial Intelligence techniques applied as estimator in chemical process systems – A literature survey}},
    author = {{Mohd Ali}, J. and Hussain, M.A. and Tade, M.O. and Zhang, J.},
    journal = {Expert Systems with Applications},
    volume = {42},
    pages = {5915--5931},
    number = {14},
    year = {2015},
}

@article{Nentwich2019,
    title = {{Surrogate Modeling of Fugacity Coefficients Using Adaptive Sampling}},
    author = {Nentwich, C. and Winz, J. and Engell, S.},
    journal = {Industrial and Engineering Chemistry Research},
    volume = {58},
    number = {40},
    pages = {18703--18716},
    year = {2019},
}

@article{Nikolaou1993,
    title = {{Control of nonlinear dynamical systems modeled by recurrent neural networks}},
    author = {Nikolaou, M. and Hanagandi, V.},
    journal = {AIChE Journal},
    volume = {39},
    number = {11},
    pages = {1890--1894},
    year = {1993},
}

@article{Panerati2019,
    journal = {The Canadian Journal of Chemical Engineering},
    author = {Panerati, J. and Schnellmann, M. A. and Patience, C. and Beltrame, G. and Patience, G. S.},
    title = {{Experimental methods in chemical engineering: Artificial neural networks–ANNs}},
    volume = {97},
    number = {9},
    pages = {2372--2382},
    year = {2019},
}

@article{Pantelides2013,
    title = {{The online use of first-principles models in process operations: Review, current status and future needs}},
    author = {Pantelides, C.C. and Renfro, J.G.},
    journal = {Computers \& Chemical Engineering},
    volume = {51},
    pages = {136--148},
    year = {2013},
}

@article{Parlos2001,
    title = {{Neuro-predictive process control using on-line controller adaptation}},
    author = {Parlos, A. G. and Parthasarathy, S. and Atiya, A. F.},
    journal = {IEEE Transactions on Control Systems Technology},
    volume = {9},
    number = {5},
    pages = {741--755},
    year = {2001},
}

@article{Pirdashti2013,
    title = {{Artificial neural networks: applications in chemical engineering}},
    author = {Pirdashti, M. and Curteanu, S. and Kamangar, M. H. and Hassim, M. H. and Khatami, M. A.},
    journal = {Reviews in Chemical Engineering},
    volume = {29},
    number = {4},
    pages = {205--239},
    year = {2013},
}

@article{Psichogios1992,
    title = {{A hybrid neural network‐first principles approach to process modeling}},
    author = {Psichogios, D. C. and Ungar, L. H.},
    journal = {AIChE Journal},
    volume = {38},
    number = {10},
    pages = {1499--1511},
    year = {1992},
}

@article{Raissi2018,
    title = {{Hidden Fluid Mechanics: A Navier-Stokes Informed Deep Learning Framework for Assimilating Flow Visualization Data}},
    author = {Raissi, M. and Yazdani, A. and Karniadakis, G. E.},
    journal = {arXiv},
    year = {2018},
}

@article{Ramli2016,
    title = {{Multivariable control of a debutanizer column using equation based artificial neural network model inverse control strategies}},
    author = {Ramli, N. M. and Hussain, M. A. and Jan, B. M.},
    journal = {Neurocomputing},
    volume = {194},
    pages = {135--150},
    year = {2016},
}

@article{Saltık2018,
    title = {{An outlook on robust model predictive control algorithms: Reflections on performance and computational aspects}},
    author = {Saltık, M. B. and {\"{O}}zkan, L. and Ludlage, J. H. A. and Weiland, S. and {Van den Hof}, P. M. J.},
    journal = {Journal of Process Control},
    volume = {61},
    pages = {77--102},
    year = {2018},
}

@article{Sansana2021,
    title = {{Recent trends on hybrid modeling for Industry 4.0}},
    author = {Sansana, J. and Joswiak, M. N. and Castillo, I. and Wang, Z. and Rendall, R. and Chiang, L. H. and Reis, M. S.},
    journal = {Computers and Chemical Engineering},
    volume = {151},
    pages = {107365},
    year = {2021},
}

@book{Seborg2011,
    title = {{Process Dynamics and Control}},
    author = {Seborg, D. E. and Edgar, T. F. and Mellichamp, D. A. and Doyle, F. J.},
    edition = {3rd Editio},
    isbn = {978-0-470-64610-6},
    pages = {464},
    year = {2011},
    publisher = {Wiley},
}

@article{Shang2019,
    title = {{Data Analytics and Machine Learning for Smart Process Manufacturing: Recent Advances and Perspectives in the Big Data Era}},
    author = {Shang, C. and You, F.},
    journal = {Engineering},
    volume = {5},
    number = {6},
    pages = {1010--1016},
    year = {2019},
}

@article{Shohei2020,
    title = {{Fault detection and diagnosis for heat source system using convolutional neural network with imaged faulty behavior data}},
    author = {Shohei, M. and Jongyeon, L. and Yasunori, A. and Yasuhiro, K. and Katsuhiko, T.},
    journal = {Science and Technology for the Built Environment},
    volume = {26},
    number = {1},
    pages = {52--60},
    year = {2020},
}

@article{Subramanian2021,
    title = {{White-box Machine learning approaches to identify governing equations for overall dynamics of manufacturing systems: A case study on distillation column}},
    author = {Subramanian, R. and Moar, R. R. and Singh, S.},
    journal = {Machine Learning with Applications},
    volume = {3},
    pages = {100014},
    year = {2021},
}

@article{Thitiyasook2007,
    title = {{Dual-mode control with neural network based inverse model for a steel pickling process}},
    author = {Thitiyasook, P. and Kittisupakorn, P. and Hussain, M. A.},
    journal = {Asia-Pacific Journal of Chemical Engineering},
    volume = {2},
    number = {6},
    pages = {536--543},
    year = {2007},
}

@article{Venkatasubramanian2019,
    title = {{The promise of artificial intelligence in chemical engineering: Is it here, finally?}},
    author = {Venkatasubramanian, V.},
    journal = {AIChE Journal},
    volume = {65},
    number = {2},
    pages = {466--478},
    year = {2019},
}

@article{Willard2003,
    title = {{Integrating Physics-Based Modeling with Machine Learning: A Survey}},
    author = {Willard, J. and Jia, X. and Xu, S. and Steinbach, M. and Kumar, V.},
    journal = {CoRR},
    volume = {2003},
    number = {04919},
    pages = {271--278},
    year = {2020},
}

@article{Zendehboudi2018,
    title = {{Applications of hybrid models in chemical, petroleum, and energy systems: A systematic review}},
    author = {Zendehboudi, S. and Rezaei, N. and Lohi, A.},
    journal = {Applied Energy},
    volume = {228},
    number = {December 2017},
    pages = {2539--2566},
    year = {2018},
}

@article{Zhang2019RTO,
    title = {{Real-Time Optimization and Control of Nonlinear Processes Using Machine Learning}},
    author = {Zhang, Z. and Wu, Z. and Rincon, D. and Christofides, P.},
    journal = {Mathematics},
    volume = {7},
    number = {10},
    pages = {890},
    year = {2019},
}

@ARTICLE{Ankush2017,
    journal={IEEE Transactions on Automatic Control}, 
    author={Chakrabarty, A. and Dinh, V. and Corless, M. J. and Rundell, A. E. and Żak, S. H. and Buzzard, G. T.},
    title={Support Vector Machine Informed Explicit Nonlinear Model Predictive Control Using Low-Discrepancy Sequences}, 
    volume={62},
    number={1},
    pages={135-148},
    year={2017},
 }

@ARTICLE{Hertneck2018,
    journal={IEEE Control Systems Letters}, 
    author={Hertneck, M. and Köhler, J. and Trimpe, S. and Allgöwer, F.},
    title={Learning an Approximate Model Predictive Controller With Guarantees}, 
    volume={2},
    number={3},
    pages={543-548},
    year={2018},
 }

@article{Kumar2021,
    title = {Industrial, large-scale model predictive control with structured neural networks},
    author = {Kumar, P. and Rawlings, J.B. and Wright, S.J.},
    journal = {Computers and Chemical Engineering},
    volume = {150},
    pages = {107291},
    year = {2021},
}

@ARTICLE{Yin2021,
    title={Stability Analysis using Quadratic Constraints for Systems with Neural Network Controllers}, 
    author={Yin, H. and Seiler, P. and Arcak, M.},
    journal={IEEE Transactions on Automatic Control}, 
    year={2021},
}

@INPROCEEDINGS{Nguyen2021,
    title={Stability Certificates for Neural Network Learning-based Controllers using Robust Control Theory}, 
    author={Nguyen, H. H. and Zieger, T. and Wells, S. C. and Nikolakopoulou, A. and Braatz, R. D. and Findeisen, R.},
    booktitle={2021 American Control Conference (ACC)}, 
    year={2021},
    pages={3564-3569},
}

@article{Akesson2005,
    title = {Neural network approximation of a nonlinear model predictive controller applied to a pH neutralization process},
    author = {\r{A}kesson, B.M. and Toivonen, H.T. and Waller, J.B. and Nyström, R.H.},
    journal = {Computers and Chemical Engineering},
    volume = {29},
    number = {2},
    pages = {323-335},
    year = {2005},
}

@article{Zhang2020,
    title = {Q-Learning-Based Model Predictive Control for Nonlinear Continuous-Time Systems},
    author = {Zhang, H. and Li, S. and Zheng, Y.},
    journal = {Industrial \& Engineering Chemistry Research},
    volume = {59},
    number = {40},
    pages = {17987-17999},
    year = {2020},
}

@article{Khalid2022,
    title = {A reinforcement learning-based economic model predictive control framework for autonomous operation of chemical reactors},
    author = {Alhazmi, K. and Albalawi, F. and Sarathy, S.M.},
    journal = {Chemical Engineering Journal},
    volume = {428},
    pages = {130993},
    year = {2022},
}

@ARTICLE{Zanon2021,
    journal={IEEE Transactions on Automatic Control},
    author={Zanon, M. and Gros, S.},
    title={Safe Reinforcement Learning Using Robust MPC},
    volume={66},
    number={8},
    pages={3638-3652},
    year={2021},
}

@article{Bertsekas2005,
    title = {Dynamic Programming and Suboptimal Control: A Survey from ADP to MPC},
    author = {Dimitri P. B.},
    journal = {European Journal of Control},
    volume = {11},
    number = {4},
    pages = {310-334},
    year = {2005},
}

@article{GORGES2017,
    title = {Relations between Model Predictive Control and Reinforcement Learning},
    author = {G\"{o}rges, D.},
    journal = {IFAC-PapersOnLine},
    volume = {50},
    number = {1},
    pages = {4920-4928},
    year = {2017},
}

@ARTICLE{Lewis2009,
    title={Reinforcement learning and adaptive dynamic programming for feedback control}, 
    author={Lewis, F. L. and Vrabie, D.},
    journal={IEEE Circuits and Systems Magazine}, 
    volume={9},
    number={3},
    pages={32-50},
    year={2009},
}

@ARTICLE{Cybenko1989,
    title={Approximation by superpositions of a sigmoidal function}, 
    author={Cybenko, G.},
    journal={Mathematics of Control, Signals and Systems}, 
    volume={2},
    pages={303–314},
    year={1989},
}

@article{Hornik1991,
    title = {Approximation capabilities of multilayer feedforward networks},
    author = {Hornik, K.},
    journal = {Neural Networks},
    volume = {4},
    number = {2},
    pages = {251-257},
    year = {1991},
}

@ARTICLE{ivanov2018,
    title={Verisig: verifying safety properties of hybrid systems with neural network controllers}, 
    author={Ivanov, R. and Weimer, J. and Alur, R. and Pappas, G.J. and Lee, I.},
    journal = {arXiv},
    year={2018},
}

@INPROCEEDINGS{nguyen2020,
    title={Towards nominal stability certification of deep learning-based controllers}, 
    author={Nguyen, H.H. and Matschek, J. and Zieger, T. and Savchenko, A. and Noroozi, N. and Findeisen, R.},
    booktitle={2020 American Control Conference (ACC)}, 
    year={2020},
    pages={3886-3891},
}

@article{perkins2002,
    title={Lyapunov design for safe reinforcement learning},
    author={Perkins, T. J. and Barto, A. G.},
    journal={Journal of Machine Learning Research},
    volume={3},
    pages={803--832},
    year={2002},
}

@article{jin2020,
    title={Stability-certified reinforcement learning: A control-theoretic perspective},
    author={Jin, M. and Lavaei, J.},
    journal={IEEE Access},
    volume={8},
    pages={229086--229100},
    year={2020},
}

@book{Bertsekas2019,
  title={Reinforcement Learning and Optimal Control},
  author={Bertsekas, D. P.},
  year={2019},
  publisher={Athena Scientific},
}

@InProceedings{Kocijan2003,
  Title                    = {Predictive control with {G}aussian {P}rocess models},
  Author                   = {Kocijan, J. and Murray-Smith, R. and Rasmussen, C. E. and Likar, B.},
  Booktitle                = {Eurocon 2003: The International Conference on Computer as a Tool},
  Organization             = {IEEE},
  Volume                   = {1},
  Pages                    = {352--356},
  Year                     = {2003},
}

@InProceedings{Kocijan2004,
  Title                    = {{G}aussian {P}rocess model based predictive control},
  Author                   = {Kocijan, J. and Murray-Smith, R. and Rasmussen, C. E. and Girard, A.},
  Booktitle                = {Proceedings of the 2004 American Control Conference},
  Organization             = {IEEE},
  Volume                   = {3},
  Pages                    = {2214--2219},
  Year                     = {2004},
}

@Article{Murray2003,
  Title                    = {Adaptive, cautious, predictive control with {G}aussian {P}rocess priors},
  Author                   = {Murray-Smith, R. and Sbarbaro, D. and Rasmussen, C. E. and Girard, A.},
  Journal                  = {IFAC Proceedings Volumes},
  Volume                   = {36},
  Number                   = {16},
  Pages                    = {1155--1160},
  Year                     = {2003},
}

@Article{Likar2007,
  Title                    = {Predictive control of a gas--liquid separation plant based on a {G}aussian {P}rocess model},
  Author                   = {Likar, B. and Kocijan, J.},
  Journal                  = {Computers \& chemical engineering},
  Volume                   = {31},
  Number                   = {3},
  Pages                    = {142--152},
  Year                     = {2007},
}

@inproceedings{Grancharova2007,
  title={Explicit stochastic nonlinear predictive control based on {G}aussian {P}rocess models},
  author={Grancharova, A. and Kocijan, J. and Johansen, T. A.},
  booktitle={European Control Conference (ECC)},
  pages={2340--2347},
  year={2007},
  organization={IEEE},
}

@article{Grancharova2008,
  title={Explicit stochastic predictive control of combustion plants based on {G}aussian {P}rocess models},
  author={Grancharova, A. and Kocijan, J. and Johansen, T. A.},
  journal={Automatica},
  volume={44},
  number={6},
  pages={1621--1631},
  year={2008},
}

@inproceedings{Cao2016,
  title={{G}aussian {P}rocess based {M}odel {P}redictive {C}ontrol for linear time varying systems},
  author={Cao, G. and Lai, E.M.K and Alam, F.},
  booktitle={14th International Workshop on Advanced Motion Control (AMC)},
  pages={251--256},
  year={2016},
  organization={IEEE},
}

@article{Cao2017b,
    title={{G}aussian {P}rocess {M}odel {P}redictive {C}ontrol of unknown non-linear systems},
    author={Cao, G. and Lai, E.M.K. and Alam, F.},
    journal={IET Control Theory \& Applications},
    volume={11},
    number={5},
    pages={703--713},
    year={2017},
}

@inproceedings{Nghiem2017,
  title={Data-driven demand response modeling and control of buildings with {G}aussian processes},
  author={Nghiem, T. X. and Jones, C. N.},
  booktitle={American Control Conference (ACC)},
  pages={2919--2924},
  year={2017},
  organization={IEEE},
}

@inproceedings{Maiworm2018b,
  title={Stability of {G}aussian {P}rocess {L}earning {B}ased {O}utput {F}eedback {M}odel {P}redictive {C}ontrol},
  author={Maiworm, M. and Limon, D. and Manzano, J. M. and Findeisen, R.},
  booktitle = {6th IFAC Conference on Nonlinear Model Predictive Control},
  pages={551-557},
  year={2018},
}

@article{Hewing2017,
  title={Cautious {M}odel {P}redictive {C}ontrol using {G}aussian {P}rocess regression},
  author={Hewing, L. and Zeilinger, M. N},
  journal={arXiv},
  year={2017},
}

@Article{Yang2015b,
  Title                    = {Fault tolerant control using {G}aussian {P}rocesses and {M}odel {P}redictive {C}ontrol},
  Author                   = {Yang, X. and Maciejowski, J. M.},
  Journal                  = {International Journal of Applied Mathematics and Computer Science},
  Volume                   = {25},
  Number                   = {1},
  Pages                    = {133--148},
  Year                     = {2015},
}

@article{maiworm2019online,
    title={Online Gaussian Process learning-based Model Predictive Control with Stability Guarantees},
    author={Maiworm, M. and Limon, D. and Findeisen, R.},
    journal={International Journal of Robust and Nonlinear Control},
    year={2021},
}

@incollection{bradford2021hybrid,
    title={Hybrid Gaussian Process Modeling Applied to Economic Stochastic Model Predictive Control of Batch Processes},
    author={Bradford, E. and Imsland, L. and Reble, M. and del Rio-Chanona, E. A.},
    booktitle={Recent Advances in Model Predictive Control},
    pages={191--218},
    year={2021},
    publisher={Springer},
}

@inproceedings{caldwell2021towards,
    title={Towards Efficient Learning-Based Model Predictive Control via Feedback Linearization and Gaussian Process Regression},
    author={Caldwell, J. and Marshall, J. A},
    booktitle={2021 IEEE/RSJ International Conference on Intelligent Robots and Systems (IROS)},
    pages={4306--4311},
    year={2021},
    organization={IEEE},
}

@article{li2021adaptive,
    title={Adaptive stochastic model predictive control of linear systems using Gaussian process regression},
    author={Li, F. and Li, H. and He, Y.},
    journal={IET Control Theory \& Applications},
    volume={15},
    number={5},
    pages={683--693},
    year={2021},
}

@book{Goodfellow2016,
    title={Deep Learning},
    author={Goodfellow, I. and Bengio, Y. and Courville, A.},
    publisher={MIT Press},
    year={2016},
}

@book{Sandro2018,
    title={Introduction to Deep Learning From Logical Calculus to Artificial Intelligence},
    author={Skansi, S.},
    publisher={Springer},
    year={2018},
}

@article{Dogru2022,
    title = {Reinforcement learning approach to autonomous PID tuning},
    author = {Dogru, O. and Velswamy, K. and Ibrahim, F. and Wu, Y. and Sundaramoorthy, A.S. and Huang, B. and Xu, S. and Nixon, M. and Bell, N.},
    journal = {Computers \& Chemical Engineering},
    volume = {161},
    pages = {107760},
    year = {2022},
}
